\documentclass[10pt,letterpaper]{article}

\usepackage{jheppub_mod}
\usepackage{slashed}

\newcommand{\cA}{\mathcal{A}}
\newcommand{\cC}{\mathcal{C}}
\newcommand{\cD}{\mathcal{D}}
\newcommand{\cF}{\mathcal{F}}
\newcommand{\cG}{\mathcal{G}}
\newcommand{\cH}{\mathcal{H}}
\newcommand{\cJ}{\mathcal{J}}
\newcommand{\cL}{\mathcal{L}}
\newcommand{\cM}{\mathcal{M}}
\newcommand{\cN}{\mathcal{N}}
\newcommand{\cO}{\mathcal{O}}
\newcommand{\cQ}{\mathcal{Q}}
\newcommand{\cR}{\mathcal{R}}
\newcommand{\cS}{\mathcal{S}}
\newcommand{\cV}{\mathcal{V}}
\newcommand{\cW}{\mathcal{W}}
\newcommand{\cY}{\mathcal{Y}}

\newcommand{\uD}{\mathrm{D}}
\newcommand{\uH}{\mathrm{H}}
\newcommand{\uI}{\mathrm{I}}
\newcommand{\uL}{\mathrm{L}}
\newcommand{\uR}{\mathrm{R}}
\newcommand{\uT}{\mathrm{T}}

\newcommand{\ud}{\mathrm{d}}
\newcommand{\ue}{\mathrm{e}}
\newcommand{\ui}{\mathrm{i}}
\newcommand{\um}{\mathrm{m}}
\newcommand{\us}{\mathrm{s}}
\newcommand{\ut}{\mathrm{t}}

\newcommand{\II}{\mathbb{I}}

\newcommand{\Ga}{\Gamma}
\newcommand{\Lam}{\Lambda}
\newcommand{\Sig}{\Sigma}

\newcommand{\ep}{\epsilon}
\newcommand{\ga}{\gamma}
\newcommand{\ka}{\kappa}
\newcommand{\lam}{\lambda}
\newcommand{\sig}{\sigma}

\newcommand{\ve}{\varepsilon}
\newcommand{\vp}{\varphi}
\newcommand{\vt}{\vartheta}

\newcommand{\vol}{\mathrm{vol}}
\newcommand{\YM}{\mathrm{YM}}

\newcommand{\SO}[1]{\mathrm{SO}\left(#1\right)}
\newcommand{\SU}[1]{\mathrm{SU}\left(#1\right)}
\newcommand{\U}[1]{\mathrm{U}\left(#1\right)}

\newcommand{\abs}[1]{\left\vert#1\right\rvert}
\newcommand{\ul}[1]{\underline{#1}}
\newcommand{\bs}[1]{\boldsymbol{#1}}

\newcommand{\nn}{\!\!}

\title{Hidden-sector current-current correlators in\\ holographic
  gauge mediation}

\author{Paul McGuirk}

\affiliation{Laboratory for Elementary-Particle Physics, Cornell
  University, Ithaca, New York, 14853, USA}

\affiliation{Department of Physics, University of Wisconsin-Madison,
  Madison, Wisconsin, 53706, USA}

\emailAdd{mcguirk@cornell.edu}

\date{\today}

\preprint{MAD-TH-11-08}

\abstract{We discuss gauge mediation in the case where the hidden
  sector is strongly coupled but, via the gauge-gravity
  correspondence, admits a weakly-coupled description in terms of a
  warped higher-dimensional spacetime.  In this framework, known as
  holographic gauge mediation, the visible-sector gauge group is
  realized in the gravitational description by probe D-branes and the
  non-supersymmetric state by normalizable perturbations to the
  geometry.  Using the formalism of general gauge mediation,
  supersymmetry-breaking soft terms in the visible sector can be
  related to the two-point functions of the hidden-sector current
  superfield that couples to the visible-sector gauge group.  Such
  correlation functions cannot be directly calculated in the strongly
  coupled field theory but can be determined using the gauge-gravity
  correspondence and holographic renormalization.  We explore this
  procedure by considering a toy geometry where such two-point
  functions can be explicitly calculated.  Unlike previous
  implementations of holographic gauge mediation where sfermion masses
  were not calculable directly in a purely holographic framework, such
  terms are readily obtained via these correlators, while (due to the
  simplicity of the geometry considered) the visible-sector gauginos
  remain massless to leading order in the visible-sector coupling.

}

\begin{document}
\setcounter{tocdepth}{2}

\setcounter{page}{0}

\maketitle

\section{Introduction}

The search for physics beyond the standard model is motivated in large
part by the search for naturalness.  In particular, the electroweak
scale in the standard model is tied directly to the mass of a
fundamental scalar and since the former exhibits an exponential
hierarchy compared to the Planck scale ($m_{Z}\sim
10^{-16}m_{\mathrm{P}}$) while the latter is quadratically sensitive
to short-wavelength physics, one is left either with the acceptance of
an unnatural amount of fine tuning of classical effects against
quantum effects or the acceptance of the existence of new physics.

One possibility for new physics is supersymmetry (SUSY) which
addresses the electroweak hierarchy problem by tempering the quantum
corrections to certain operators such as scalar masses.  If the universe
were in a supersymmetric state, then for every fermionic field there
would be a bosonic field with the same charge and (in a Minkowski
spacetime) same mass and vice-versa.  The lack of discovery of such
superpartners indicates that our universe is not in such a state.
Nevertheless, if supersymmetry exists as a spontaneously broken
symmetry, then the protection from quantum effects is to a large
extent preserved.

In order for such a scenario to be phenomenologically viable, the
spontaneous breaking of supersymmetry must not occur in the visible
sector, e.g. the minimally supersymmetric standard model (MSSM), but within
another set of fields.  In a theory with finite $m_{\mathrm{P}}$, this
breaking will be communicated to the visible sector via quantum
effects related to the superconformal
anomaly~\cite{Randall:1998uk,*Giudice:1998xp,*Bagger:1999rd} and
classically by irrelevant operators which are generically expected to
be
present~\cite{Chamseddine:1982jx,*Barbieri:1982eh,*Hall:1983iz,*Soni:1983rm,*Nilles:1983ge,*Brignole:1993dj,*Brignole:1995fb,*Brignole:1997dp}.

Whether or not gravity is present, the breaking of supersymmetry can
also be mediated to the standard model via so-called messenger fields
which transform under the visible-sector gauge group and couple to the
sector in which supersymmetry is
broken~\cite{Dine:1981za,*Dimopoulos:1981au,*Dine:1981gu,*Nappi:1982hm,*AlvarezGaume:1981wy,*AlvarezGaume:1981wy,*Dine:1993yw,*Dine:1994vc,*Dine:1995ag}
(see also~\cite{Giudice:1998bp} for a review).  This mechanism, known
as gauge mediation, has the advantage that the effects of supersymmetry
breaking respect the flavor structure of the visible sector, unlike
mediation from irrelevant operators which will generically result in
unacceptable flavor-changing neutral currents\footnote{Although
  anomaly mediation also respects this flavor structure, it leads to
  spontaneous breaking of $\U{1}_{\mathrm{em}}$ unless supplemented by
  comparable contributions from gauge
  mediation~\cite{Pomarol:1999ie,*Rattazzi:1999qg}, mediation from
  irrelevant operators~\cite{Choi:2004sx,*Choi:2005ge}, or
  both~\cite{Everett:2008qy,*Everett:2008ey,*Altunkaynak:2010xe,*Altunkaynak:2010tn}.
}.

In any of these scenarios, the effects of the breaking of
supersymmetry on the visible sector can be captured by so-called soft
terms: operators in the effective Lagrangian of the visible-sector
that do not reintroduce quadratic sensitivity on ultraviolet physics
even though they do not respect supersymmetry.  Since supersymmetry,
if it exists as an underlying symmetry, is broken, all of the
phenomenological implications of supersymmetry result from these soft
terms (see~\cite{Martin:1997ns,*Chung:2003fi} for reviews) and it is
thus of clear importance to study them in various scenarios especially
in light of the increasing experimental constraints on simple models.

In this work, we will focus on gauge mediation of supersymmetry
breaking.  Even within this class of scenarios, there are a number of
possibilities based on the nature of the messenger sector.  The
minimal scenario involves messengers that are neutral under any gauge
group of the SUSY-breaking sector, but couple to the
SUSY-breaking-sector fields through superpotential operators.  In
direct gauge mediation scenarios, such as those
in~\cite{Affleck:1984xz,*Poppitz:1996fw,*ArkaniHamed:1997jv,*Murayama:1997pb}
the messenger fields are charged under SUSY-breaking-sector gauge
groups, and indeed there is little distinction between the
SUSY-breaking sector and the messenger sector.  Semi-direct models are
a compromise between these two scenarios in which the messenger
sectors couple to the SUSY-breaking sector (in the original
semi-direct proposal~\cite{Seiberg:2008qj}, the messengers had only
gauge couplings to the SUSY-breaking sector) but do not participate in
the breaking of supersymmetry.

In~\cite{Meade:2008wd} (see
also~\cite{Buican:2008ws,Komargodski:2008ax,Dumitrescu:2010ha}), a
general framework for gauge mediation scenarios was presented.  In the
limit where the visible-sector gauge couplings are taken to zero, the
visible-sector gauge group is realized as a global symmetry of the
messenger and SUSY-breaking sectors, which together compromise the
hidden sector.  In models of gauge mediation, the hidden sector and
visible sector decouple in this limit.  The conserved hidden-sector
current $j_{\mu}$ corresponding to this global symmetry is a component
field of a linear superfield $\cJ$ which contains also a scalar
component and a spinor component.  It was shown in~\cite{Meade:2008wd}
that once the visible-sector gauge group becomes weakly gauged,
visible-sector soft terms arise and can be given in terms of two-point
functions of these currents\footnote{An important exception to this in
  the MSSM is the Lagrangian-level operator $B\mu H_{u}H_{d}$ (where
  $H_{u}$ and $H_{d}$ are the two Higgs doublets of the MSSM and $\mu$
  appears in the analogous superpotential coupling), which must be
  treated differently as in~\cite{Komargodski:2008ax}.}.

Although the couplings between the visible and hidden sectors are
small, the hidden sector itself may be strongly coupled and such
current-current correlators cannot be directly calculated.  However,
certain strongly coupled gauge theories admit a weakly coupled dual
description in terms of a classical gravitational theory on a curved
spacetime of higher
dimension~\cite{Maldacena:1997re,*Witten:1998qj,*Gubser:1998bc}
(see~\cite{Aharony:1999ti,*D'Hoker:2002aw} for reviews).  This
duality, known as (non-)AdS/(non-)CFT or the gauge-gravity
correspondence, is the best understood example of the holographic
principle~\cite{'tHooft:1993gx,*Susskind:1994vu}.  If the gauge theory
is supersymmetric, then a non-supersymmetric state can be constructed
by considering particular perturbations to the geometry.  For example,
it was argued in~\cite{Kachru:2002gs} that the addition of a small
number of anti-branes to the geometry of Klebanov and
Strassler~\cite{Klebanov:2000hb} is dual to the preparation of a
metastable non-supersymmetric state in a particular\footnote{$\cN_{D}$
  denotes the amount of supersymmetry in $D$ spacetime
  dimensions. Hence, $\cN_{4}=1$ has four supercharges while both
  $\cN_{4}=2$ and $\cN_{5}=1$ have eight supercharges.}  $\cN_{4}=1$
gauge theory\footnote{See, however,~\cite{Bena:2009xk} for possible
  concerns with this procedure.}.

In the limit of vanishing visible-sector gauge coupling, the
visible-sector gauge group becomes a global symmetry.  If the hidden
sector admits a dual gravity description, this global symmetry can be
realized by D-branes, known as flavor branes, that extend along the
holographic direction~\cite{Karch:2002sh}.  According to the AdS/CFT
dictionary, some of the open-string excitations of these D-branes are
dual to the components of the current superfield $\cJ$.  The
calculation of the classical two-point functions of these components
thus corresponds to the calculation of the current-current correlators
in the strongly-coupled dual field theory.

This paper will explore this procedure of calculating hidden-sector
current-current correlators using holographic techniques.  Such
holographic models of supersymmetry breaking were first considered
in~\cite{Benini:2009ff} and further studied
in~\cite{McGuirk:2009am,Fischler:2011xd} (see
also~\cite{Gabella:2007cp}) where certain soft terms (namely the mass
of the visible-sector gaugino) were deduced via dimensional reduction
to 4d with additional soft terms following from gaugino
mediation~\cite{Kaplan:1999ac,*Chacko:1999mi}. The approach here
differs from this previous work in that we make use of the formalism
of general gauge mediation~\cite{Meade:2008wd} to calculate soft terms
in terms of current-current correlators.  A drawback of this procedure
is that it is difficult to precisely calculate such correlators in the
types of geometries considered in~\cite{Benini:2009ff,McGuirk:2009am}
and so in order to be able to calculate explicitly, we consider a toy
geometry described below.  We emphasize that the barrier to explicitly
calculate two-point functions in the gravity picture is of an entirely
different nature than the barrier in the direct gauge picture; in the
former the complication is the inability to analytically solve in
curved spacetime classical equations of motion, while in the latter
the barrier is the inapplicability of perturbative techniques in a
quantum theory.  As a consequence of the simple geometry however, the
visible-sector gauginos will remain massless in the construction we
consider here, and we leave the analysis of more phenomenologically
viable geometries for future work.

Our paper is organized as follows.  The formalism of general gauge
mediation is reviewed in section~\ref{sec:ggm}.  In
section~\ref{sec:geometry}, we summarize the framework of holographic
gauge mediation and introduce the geometry that we consider in this
work.  In section~\ref{sec:5deft} we deduce the classical 5d effective
field theory (EFT) describing the open string fluctuations that is
dual to the generating functional for current correlators in the dual
field theory.  We deduce this EFT in two different ways: in
section~\ref{sec:on_shell} we find the on-shell action by dimensional
reduction from the well-known action of a $\uD$-brane, and in
section~\ref{sec:off_shell} we find the off-shell action by making use
of the known off-shell action in the 5d Minkowski spacetime $R^{4,1}$.
In section~\ref{sec:SUSY_currents}, we calculate the current-current
correlators in the supersymmetric case for the both the case of
massless and massive messengers.  This is done using the techniques of
holographic renormalization which we also briefly review.  In
section~\ref{sec:non_SUSY_currents}, we extend this calculation to a
non-supersymmetric example and in doing so we effectively determine
the visible-sector soft terms which are the main subject of interest
of this work.  Section~\ref{sec:conclusions} contains some concluding
remarks and our conventions are presented in appendix~\ref{app:conv}.

We note also that general gauge mediation has been considered together
with warped geometries elsewhere in the
literature~\cite{McGarrie:2010yk}.  The essential difference
between~\cite{McGarrie:2010yk} and the work below is that in the
former, the SUSY-breaking sector is realized in an entirely field
theoretic way in the warped geometry, while here the SUSY-breaking
sector is realized by the geometry itself.

\section{\label{sec:ggm}General gauge mediation}

As discussed in the introduction, general gauge
mediation~\cite{Meade:2008wd} relates visible-sector soft terms to
hidden-sector current-current correlators.  The underlying assumption
in the formalism is that in the limit that the visible-sector gauge
coupling $g_{\mathrm{vis}}$ vanishes, the visible sector and hidden
sector decouple (this implicitly requires $m_{\mathrm{P}}\to\infty$).
For simplicity of presentation, we consider the case in which the
visible-sector gauge group is $\U{1}$.  The hidden sector then possess
a conserved current $j^{\mu}$, e.g. a real vector satisfying the
condition (here we are working on the Minkowski spacetime $R^{3,1}$)
\begin{equation}
  \partial^{\mu}j_{\mu}=0.
\end{equation}
In an $\cN_{4}=1$ theory, this is a component of a linear superfield
$\cJ$ which in $\cN_{4}=1$ superspace takes the form
\begin{equation}
  \label{eq:define_current_superfield}
  \cJ=J+\ui\theta j-\ui\bar{\theta}\bar{j}
  -\theta\sig^{\mu}\bar{\theta}j_{\mu}
  +\frac{1}{2}\theta^{2}\bar{\theta}\bar{\sig}^{\mu}\partial_{\mu}j
  -\frac{1}{2}\bar{\theta}^{2}\theta\sig^{\mu}\partial_{\mu}\bar{j}
  -\frac{1}{4}\theta^{2}\bar{\theta}^{2}\partial^{2}J,
\end{equation}
in which $j$ is a two-component spinor and $J$ is a real scalar.  With
these conditions, $\cJ$ satisfies $D^{2}\cJ=\bar{D}^{2}\cJ=0$ where
$D$ is the usual supercovariant derivative
\begin{equation}
  D_{\alpha}=\frac{\partial}{\partial\theta^{\alpha}}
  +\ui\sig^{\mu}_{\alpha\dot{\alpha}}\bar{\theta}^{\dot{\alpha}}\partial_{\mu},
  \notag\quad
  \bar{D}_{\dot{\alpha}}=-\frac{\partial}{\partial\bar{\theta}^{\dot{\alpha}}}
  -\ui\theta^{\alpha}\sig^{\mu}_{\alpha\dot{\alpha}}\partial_{\mu}.
\end{equation}

Upon weakly gauging the visible sector, this superfield couples to the
visible-sector vector superfield which, in the Wess-Zumino gauge,
takes the form
\begin{equation}
  \cV=-\theta\sig^{\mu}\bar{\theta}A_{\mu}
  +\ui\theta^{2}\bar{\theta}\bar{\lam}
  -\ui\bar{\theta}^{2}\theta\lam
  +\frac{1}{2}\theta^{2}\bar{\theta}^{2}D.
\end{equation}
The coupling between the current and vector superfields is
\begin{equation}
  2g_{\mathrm{vis}}\int_{R^{3,1}}\nn\nn\nn\ud^{4}x\,
  \int\ud^{4}\theta\, \cV\cJ
  =g_{\mathrm{vis}}\int_{R^{3,1}}\nn\nn\nn\ud^{4}x\,
  \bigl\{DJ-\lam j-\bar{\lam}\bar{j}-A^{\mu}j_{\mu}\bigr\},
\end{equation}
in which $g_{\mathrm{vis}}$ is the visible-sector gauge coupling.

It is convenient to cast the two-point correlators as~\cite{Meade:2008wd}
\begin{align}
  \bigl\langle J\bigl(x\bigr)J\bigr(0\bigr)\bigr\rangle=&
  \frac{1}{x^{4}}C_{0}\bigl(x^{2}M^{2}\bigr),\notag\\
  \label{eq:correlator_functions}
  \bigl\langle j_{\alpha}\bigl(x\bigr)\bar{j}_{\dot{\alpha}}\bigl(0\bigr)
  \bigr\rangle=&
  -\ui\sig^{\mu}_{\alpha\dot{\alpha}}\partial_{\mu}
  \biggl(\frac{1}{x^{4}}C_{1/2}\bigl(x^{2}M^{2}\bigr)\biggr),\\
  \bigl\langle j_{\mu}\bigl(x\bigr)j_{\nu}\bigl(0\bigr)\bigr\rangle
  =&
  \bigl(\eta_{\mu\nu}\partial^{2}
  -\partial_{\mu}\partial_{\nu}\bigr)
  \biggl(\frac{1}{x^{4}}C_{1}\bigl(x^{2}M^{2}\bigr)\biggr),\notag\\
  \bigl\langle j_{\alpha}\bigl(x\bigr)j_{\beta}\bigl(0\bigr)\bigr\rangle
  =&\ep_{\alpha\beta}\frac{1}{x^{5}}B_{1/2}\bigl(x^{2}M^{2}\bigr),\notag
\end{align}
where $M$ is some characteristic mass scale.  In the supersymmetric
limit~\cite{Meade:2008wd},
\begin{equation}
  C_{0}=C_{1/2}=C_{1},\quad B_{1/2}=0.
\end{equation}

The Fourier transforms take the form
\begin{align}
  \label{eq:general_correlator}
  \bigl\langle J\bigl(k\bigr)J\bigl(q\bigr)\bigr\rangle=&
  C_{0}\bigl(k^{2}/M^{2}\bigr),\notag \\
  \bigl\langle j_{\alpha}\bigl(k\bigr)\bar{j}_{\dot{\alpha}}\bigl(q\bigr)
  \bigr\rangle=&
  -\sig^{\mu}_{\alpha\dot{\alpha}}k_{\mu}
  C_{1/2}\bigl(k^{2}/M^{2}\bigr),\\
  \bigl\langle j_{\mu}\bigl(k\bigr)j_{\nu}\bigl(q\bigr)\bigr\rangle
  =&-\bigl(k^{2}\eta_{\mu\nu}-k_{\mu}k_{\nu}\bigr)
  C_{1}\bigl(k^{2}/M^{2}\bigr),\notag\\
  \bigl\langle j_{\alpha}\bigl(k\bigr)
  j_{\beta}\bigl(q\bigr)\bigr\rangle
  =&\ep_{\alpha\beta}MB_{1/2}\bigl(k^{2}/M^{2}\bigr),\notag
\end{align}
where we have, and will in what follows, suppressed the
momentum-conserving delta function
\begin{equation}
  \bigl(2\pi\bigr)^{4}\delta^{4}\bigl(k+q\bigr).
\end{equation}
We use the same notation to denote the functions $C_{a}$ and $B$ and
their Fourier transforms
\begin{align}
  C_{a}\bigl(k^{2}/M^{2}\bigr)=&
  \int_{R^{3,1}}\nn\nn\nn\ud^{4}x\, \ue^{\ui k\cdot x}
  \frac{1}{x^{4}}C_{a}\bigl(x^{2}M^{2}\bigr),\notag\\
  B_{1/2}\bigl(k^{2}/M^{2}\bigr)=&
  \int_{R^{3,1}}\nn\nn\nn\ud^{4}x\, \ue^{\ui k\cdot x}
  \frac{1}{Mx^{5}}B_{1/2}\bigl(x^{2}M^{2}\bigr).
\end{align}
In general, these integrals require the introduction of a UV cutoff
$\Lam$, the dependence on which is suppressed in the above formulae.

A central result of~\cite{Meade:2008wd} is that the visible-sector
soft masses (except for $B\mu$-like terms) can be expressed to
leading order in $g_{\mathrm{vis}}$ in terms of these two-point
functions.  For the visible-sector gaugino corresponding to the
partner of the $\U{1}$ gauge-boson,
\begin{equation}
  m_{1/2}=g_{\mathrm{vis}}^{2}MB_{1/2}\bigl(0\bigr).
\end{equation}
For the sfermion masses, we now suppose that the visible-sector gauge
group takes the form $G_{\mathrm{vis}}=\bigotimes_{i}G_{i}$ and that
the sfermion transforms under the representations $r_{i}$ for each of
the $G_{i}$.  Then,
\begin{equation}
  m_{\tilde{f}}^{2}=\sum_{i}g_{i}^{4}c_{2}\bigl(r_{i}\bigr)\Ga_{i},
\end{equation}
in which $g_{i}$ is the coupling for $G_{i}$, $c_{2}\bigl(r_{i}\bigr)$
is the quadratic Casimir for the representation $r_{i}$ of the group
$G_{i}$ and $\Ga_{i}$ is built from the current-current correlators for
the corresponding group
\begin{equation}
  \label{eq:sfermion_mass}
  \Ga_{i}=-\frac{M^{2}}{16\pi^{2}}\int_{0}^{\infty}\ud y\,
  \bigl\{3C_{1}\bigl(y\bigr)
  -4C_{1/2}\bigl(y\bigr)+C_{0}\bigl(y\bigr)\bigr\}.
\end{equation}
In the event of a vacuum expectation value for the scalar component of
one the vector superfields (which does not violate any symmetries when the
group is Abelian), there is an additional contribution which we will
not consider here.

\section{\label{sec:geometry}Geometric setup}

We will now consider a special class of hidden sectors, namely those
for which the SUSY-breaking sector is in the Maldacena
limit~\cite{Maldacena:1997re}.  In the simplest case of $\cN_{4}=4$
$\SU{N}$ super Yang-Mills, this limit is obtained by first holding the
't~Hooft coupling $\lam_{\ut}=g_{\YM}^{2}N$ fixed and then taking the
number of colors $N$ to infinity (which of course requires taking the
Yang-Mills coupling of the hidden sector $g_{\YM}$ to zero) and then
taking $\lam_{\ut}$ to be large.  The gauge theory in this limit is
dual to classical type-IIB supergravity on the space $AdS^{5}\times
S^{5}$ where the 10d metric takes the form
\begin{equation}
  \ud s_{10}^{2}=\frac{r^{2}}{L^{2}}\eta_{\mu\nu}\ud x^{\mu}\ud x^{\nu}
  +\frac{L^{2}}{r^{2}}\ud r^{2}+L^{2}\ud\Omega_{5}^{2},
\end{equation}
in which $\ud\Omega_{5}^{2}=\hat{g}_{\phi\psi}\ud y^{\phi}\ud
y^{\psi}$ is the metric for a unit $S^{5}$ and $L$ is set by the
't~Hooft coupling,
\begin{equation}
  L^{4}=4\pi\ell_{\us}^{4}g_{\us}N,
\end{equation}
where $g_{\us}=g_{\YM}^{2}$ is the string coupling and $\ell_{\us}$ is
the string length.  The geometry is supported by a $5$-form flux
\begin{equation}
  F^{\left(5\right)}=\bigl(1+\ast_{10}\bigr)\cF^{\left(5\right)},
\end{equation}
in which $\ast_{10}$ is the 10d Hodge-$\ast$ and
$\cF^{\left(5\right)}=\ud C^{\left(4\right)}$ with
\begin{equation}
  C^{\left(4\right)}=\frac{r^{4}}{g_{\us}L^{4}}\ud\vol_{R^{3,1}},
\end{equation}
where $\ud\vol_{R^{3,1}}$ is the volume element of $R^{3,1}$.  The
duality can be motivated by considering a stack of $N$ $\uD 3$-branes
in Minkowski spacetime $R^{9,1}$ which in this limit has an
open-string description in terms of the gauge theory and a
closed-string description in terms of this geometry.

A less symmetric example is the Klebanov-Strassler (KS)
theory~\cite{Klebanov:2000hb}.  Although we will consider the simpler
$AdS^{5}\times S^{5}$ geometry in what follows, the breaking of
supersymmetry has been recently studied in this geometry and so we
will discuss it as an illustration of geometries suitable for
holographic gauge mediation.  The geometry is found by considering a
collection of $M$ fractional $\uD 3$-branes (i.e. $\uD 5$-branes
wrapping a collapsing $2$-cycle) at a conifold point.  The geometry is
similar to the above case in that it is a warped geometry
\begin{equation}
  \ud s_{10}^{2}=\ue^{2A}\eta_{\mu\nu}\ud x^{\mu}\ud x^{\nu}
  +\ue^{-2A}\ud s_{6}^{2},
\end{equation}
in which $\ud s_{6}^{2}$ is the Ricci-flat metric for a particular
Calabi-Yau manifold over which the warp factor $\ue^{4A}$ varies
non-trivially.  The conifold point additionally becomes deformed so
that instead of there being a singularity, there is now a finite-sized
$S^{3}$.  In addition to the $5$-form flux, the geometry is supported
by an imaginary-self-dual $3$-form flux.  The dual gauge theory is no
longer conformal but instead is an $\cN_{4}=1$ theory that can be
described by a series of Seiberg dualities~\cite{Seiberg:1994pq}.  If
a number of $\overline{\uD 3}$-branes are present on the finite
$S^{3}$ where the conifold has been deformed, the geometry will no
longer be supersymmetric.  However, so long as the number $p$ of
$\overline{\uD 3}$ branes is small compared to the amount of
background flux, the geometry will be metastable~\cite{Kachru:2002gs};
without any $\uD 3$-branes to directly annihilate against, the
$\overline{\uD 3}$s will decay only non-perturbatively, first puffing
up via the Myers effect~\cite{Myers:1999ps} to
$\overline{\mathrm{NS}5}$-branes which will then dissolve into flux
and $\uD 3$-branes.  The influence of the $\overline{\uD 3}$-branes on
the geometry is
involved~\cite{DeWolfe:2008zy,Bena:2009xk,McGuirk:2009xx,*Bena:2010ze,*Dymarsky:2011pm,*Bena:2011wh},
but by considering the geometry at distances far away from the tip
where the geometry simplifies~\cite{Klebanov:2000nc} it was argued
in~\cite{DeWolfe:2008zy} that the perturbation to the geometry is such
that the dual theory is in a non-supersymmetric state of the original
theory, rather than a perturbation to the theory itself.  More
precisely, the duality states that for every operator $\cO$ in the
gauge theory, there is a corresponding operator $\Phi$ on the gravity
side such that, if we imagine the gauge theory living on the boundary
of, for example, $AdS^{5}$, the coupling of the bulk field to the
field theory is
\begin{equation}
  \int_{\delta AdS^{5}}\nn\nn\nn\nn\nn\ud^{4}x\,\sqrt{{h}}\cO\Phi,
\end{equation}
in which $h$ is the metric induced on the boundary.  $\Phi$
will satisfy a second-order differential equation and as $r\to\infty$
will behave as
\begin{equation}
  \label{eq:general_asymptote}
  \Phi\sim \phi_{1}r^{-\Delta}+\phi_{2}r^{\Delta-4},
\end{equation}
in which $\Delta$ is the mass dimension of $\cO$.  Solutions involving
$\phi_{2}$ are not normalizable and correspond to deformations of the
Lagrangian in the gauge theory, $\delta\cL\sim \phi_{2}\cO$, while
those involving just $\phi_{1}$ are normalizable and correspond to a
vacuum expectation value, $\bigl\langle\cO\bigr\rangle\sim \phi_{1}$.
The large-radius solution of~\cite{DeWolfe:2008zy} has only
normalizable perturbations implying that the addition of
$\overline{\uD 3}$-branes produces a particular metastable state and
does not change the underlying theory\footnote{We again note the
  possible objections raised in~\cite{Bena:2009xk}.}.

A global flavor group can be added to the gauge theory by adding a
number of D-branes into the geometry~\cite{Karch:2002sh}.  For the
warped geometries of type-IIB that we are considering here, the
appropriate type of brane to add is a $\uD 7$-brane that fills
$R^{3,1}$ and wraps a non-compact $4$-cycle in the transverse space.
A stack of $K$ such branes will produce an $\SU{K}$ flavor group.  In
addition, the matter content of the dual gauge theory will be modified
by the addition of quarks: matter that transforms under a
bifundamental of the flavor group and the dual gauge
group\footnote{For clarity, we emphasize that the terms ``flavor'' and
  ``quark'' are used only in analogy with the standard model and not
  related to the corresponding concepts in the visible sector.}.  In
the brane picture, these correspond to open strings that stretch from
the (fractional or elementary) $\uD 3$-branes that produce the
geometry and the $\uD 7$-branes so that the mass of the quarks is set
by the position of the $\uD 7$-branes.  In the case when the number of
flavor branes is much smaller than the number of color branes, the
$\uD 7$-branes may be considered in the probe approximation where the
backreaction of the $\uD 7$-branes can be neglected, an approximation
which we make here.  In addition to the gauge couplings, the quarks
may possess superpotential couplings to other matter in the hidden
sector~\cite{Ouyang:2003df,*Kuperstein:2004hy}.  One of the
excitations of the $\uD 7$-branes is the $1$-form $A_{\mu}$ that acts
as the 4d gauge field once the flavor group is weakly gauged and thus
couples to a current on a boundary theory via $\cL\sim
j^{\mu}A_{\mu}$.  That is, if we identify this flavor group as the
visible-sector gauge group, the open-string field $A_{\mu}$ is dual to
the current $j^{\mu}$ discussed in section~\ref{sec:ggm}.

We now have the ingredients to put together a dual gravity description
of gauge mediation as in~\cite{Benini:2009ff}:
\begin{enumerate}
\item Begin with a theory of matter and gauge group $G_{\mathrm{hid}}$
  that admits a geometric description via the gauge-gravity
  correspondence.  This theory will function as the SUSY-breaking
  sector.
\item Add $\uD 7$-branes\footnote{Of course, in other classes of
    solutions, different sorts of branes would need to be used here.}
  to the geometry, giving rise to a visible-sector group
  $G_{\mathrm{vis}}$ and quarks that transform under
  $G_{\mathrm{vis}}$ and $G_{\mathrm{hid}}$.  These quarks (or rather
  their bound states) which will serve as messengers.  The messengers
  and SUSY-breaking sector together constitute the hidden sector.  At
  this point, $G_{\mathrm{vis}}$ is a global symmetry in the dual
  theory and there is a corresponding conserved current $j_{\mu}$
  constructed from hidden-sector fields.
\item Prepare a SUSY-breaking state in the hidden sector.  In the
  geometry, this corresponds to a SUSY-breaking normalizable
  perturbations to the geometry from the addition of non-SUSY sources.
  This state should be metastable, though we will not address this
  issue here.
\item Calculate the classical two-point functions of $A_{\mu}$ and
  other fields related to it via supersymmetry.  This is equivalent to
  calculating the two-point functions for $j_{\mu}$ and its related
  fields in the dual theory.
\item Weakly gauge $G_{\mathrm{vis}}$ and introduce the visible-sector
  matter.  On the gravity side, this requires gluing the warped
  geometry into a compact space and introducing (for example) a
  network of intersecting $7$-branes.  Fortunately, all that is
  necessary for calculating the soft terms considered here is
  knowledge of the representations and visible-sector gauge couplings.
  Such soft terms can be determined from the current-current
  correlators as in section~\ref{sec:ggm}.
\end{enumerate}
In the cases studied in~\cite{Benini:2009ff,McGuirk:2009am}, the $\uD
7$s were taken in the probe approximation and the breaking of
supersymmetry occurs whether are not they are added.  Such modes are
thus closely related to models of semi-direct gauge
mediation~\cite{Seiberg:2008qj}: the messengers couple to the hidden
sector via gauge and superpotential couplings but are not involved in
the participation of the breaking of supersymmetry.  As in
section~\ref{sec:ggm}, we will take the case of a single $\uD
7$-brane, the extension to larger rank being a straightforward
generalization.

A significant barrier to this procedure is the non-trivial geometries
involved.  In particular, even if the background corresponding to the
SUSY-breaking sector is known before the breaking, the addition of the
non-SUSY sources will backreact on the geometry and fluxes in a manner
that is non-supersymmetric and difficult to compute.  Once this is
known, the equations of motion for the open-string modes have to be
solved.  Although the behavior along the radial direction will be
under relatively good control, many of the relevant fields will
transform non-trivially under the isometries of the angular space and
so will not be constant along the internal directions (even for the
lowest-lying state) and so the corresponding Laplace-Beltrami equation
is difficult to solve\footnote{Note that this remains true even in the
  large-radius region of the KS solution where the internal Calabi-Yau
  is a cone over the homogeneous space
  $T^{1,1}$~\cite{Candelas:1989js,Klebanov:1998hh,Klebanov:2000nc}.
  Although a procedure exists for a harmonic analysis for such
  manifolds, the angular space wrapped by the $\uD 7$ will not be as
  symmetric.}.  For the sake of calculability, we will model the
warped geometry as\footnote{Note that more generally we could have
  different factors multiplying the $\ud r^{2}$ piece and the
  $\ud\Omega_{5}^{2}$ piece, but by a redefinition of $r$ they can be
  set equal.}
\begin{equation}
  \ud s_{10}^{2}=\ue^{2A}\eta_{\mu\nu}\ud x^{\mu}\ud x^{\nu}
  +\ue^{-2\bar{B}}\bigl(\ud r^{2}+r^{2}\ud\Omega_{5}^{2}\bigr),
\end{equation}
supported by the RR-potential
\begin{equation}
  C^{\left(4\right)}=g_{\us}^{-1}\ue^{4C}\ud\vol_{R^{3,1}}.
\end{equation}
The dilaton will be taken to be a constant, $\Phi=\log g_{\us}$, and
the remaining closed string fields to vanish.  The functions $A$,
$\bar{B}$, and $C$, are taken to be functions of $r$ alone. We will in
addition consider the case in which the warp factor takes the
$AdS$-form $A=\log r/L$ before supersymmetry breaking.  In this case
the gauge theory is conformal before the addition of the flavor branes
and, as argued in~\cite{DeWolfe:2008zy}, such a theory cannot
spontaneously break supersymmetry while preserving Lorentz symmetry.
However, we will find below that the correlators of interest do not
obey the relationships expected from supersymmetry, suggesting that
SUSY is broken after the $\uD 7$ is added.

Finally, we note that the dual theory will have extended
supersymmetry.  Our interest is only in the coupling to the visible
sector which we will take to be only $\cN_{4}=1$.  We will therefore
only couple part of the hidden sector to the visible sector, namely
though the operator $\sim\int\ud^{4}\theta\,\cJ\cV$ where $\cV$ is an
$\cN_{4}=1$ vector multiplet.

Explicitly, we take the coordinates on the $S^{5}$ to be
\begin{align}
  x^{4}=&r
  \sin\left(\vp_{5}\right)
  \sin\left(\vp_{4}\right)
  \sin\left(\vp_{3}\right)
  \sin\left(\vp_{2}\right)
  \sin\left(\vp_{1}\right),\notag\\
  x^{5}=&r
  \cos\left(\vp_{5}\right)
  \sin\left(\vp_{4}\right)
  \sin\left(\vp_{3}\right)
  \sin\left(\vp_{2}\right)
  \sin\left(\vp_{1}\right),\notag\\
  x^{6}=&r
  \cos\left(\vp_{4}\right)
  \sin\left(\vp_{3}\right)
  \sin\left(\vp_{2}\right)
  \sin\left(\vp_{1}\right),\\
  x^{7}=&r
  \cos\left(\vp_{3}\right)
  \sin\left(\vp_{2}\right)
  \sin\left(\vp_{1}\right),\notag\\
  x^{8}=&r
  \cos\left(\vp_{2}\right)
  \sin\left(\vp_{1}\right),\notag\\
  x^{9}=&r\cos\left(\vp_{1}\right)\notag,
\end{align} so that $\vp_{5}\in\left[0,2\pi\right)$ while $\vp_{i\neq
  5}\in\left[0,\pi\right)$.  We place the $\uD 7$ at a radial distance
$r=\mu$ which we arrange by taking $x^{8}=0$, $x^{9}=\mu$.  The metric
induced onto the $\uD 7$-brane is
\begin{align}
  \ud s_{8}^{2}=&\ue^{2A}\eta_{\mu\nu}
  +\ue^{-2B}\frac{L^{2}}{\rho^{2}}\ud\rho^{2}+\ue^{-2B}L^{2}\ud\Omega_{3}^{2}
  \notag\\
  =&\tilde{g}_{mn}\ud x^{m}\ud x^{n}
  +\ue^{-2B}L^{2}\breve{g}_{\phi\psi}\ud y^{\phi}\ud y^{\psi},
  \label{eq:inducedmetric}
\end{align}
in which $\ud\Omega_{3}^{2}$ is the line element for a unit $S^{3}$,
$\rho$ is defined by the relationship $r^{2}=\rho^{2}+\mu^{2}$, and
$B=\bar{B}+\log L/\rho$.  We denote the non-compact 5d part of the
worldvolume by $\cM$.

\section{\label{sec:5deft}5d effective field theory}

Our goal is to calculate the two-point correlation functions of the
component fields of the current
superfield~\eqref{eq:define_current_superfield}.  The duality relates
the generating functional on the gauge theory side to the classical
action on the gravity side.  We will determine the latter in two ways.
In section~\ref{sec:on_shell} we perform a dimensional reduction of
the 8d action that describes the low-energy excitations of the $\uD
7$-brane.  The resulting 5d theory will be useful in that it is valid
whether or not supersymmetry is broken.  In
section~\ref{sec:off_shell}, the off-shell action is determined and
written in $\cN_{4}=1$ superspace language, using the flat spacetime
result as a bootstrap.  This method will only be effective when the
closed string background is supersymmetric, since only the open-string
modes are taken off-shell.  However, this action is needed since the
scalar component of the chiral superfield couples to the scalar
component of the $\cN_{4}=1$ vector superfield and the latter is an
auxiliary field.  Note that if we did not need to make use of the
off-shell 5d theory, we could the 8d equations of motion and need not
perform the intermediate dimensional reduction to get a 5d action.

\subsection{\label{sec:on_shell}On-shell theory from dimensional reduction}

The starting place for the on-shell action is the Dirac-Born-Infeld
(DBI) and Chern-Simons (CS) action describing the long-wavelength
dynamics of a $\uD p$-brane
\begin{equation}
  S_{\uD p}^{\mathrm{DBI}}=S_{\uD p}^{\mathrm{DBI}}+S_{\uD p}^{\mathrm{CS}}.
\end{equation}
In the 10d Einstein frame, the DBI action takes the form
\begin{equation}
  S_{\uD p}^{\mathrm{DBI}}=-\tau_{\uD p}\int_{\cW}\nn\ud^{p+1}\xi\,
  \bigl(g_{\us}^{-1}\ue^{\Phi}\bigr)^{\frac{p-3}{4}}
  \sqrt{\abs{\det\left(M_{\alpha\beta}\right)}},
\end{equation}
in which
\begin{equation}
  M_{\alpha\beta}
  =\mathrm{P}\bigl[g_{\alpha\beta}-g_{\us}^{1/2}\ue^{-\Phi/2}
  B_{\alpha\beta}\bigr]
  +\lam g_{\us}^{1/2}\ue^{-\Phi/2}F_{\alpha\beta}.
\end{equation}
$\mathrm{P}$ denotes the pullback from the 10d spacetime on to the
worldvolume $\cW$ of the $\uD p$ brane,
\begin{equation}
  \mathrm{P}\bigl[v_{\alpha}\bigr]=v_{M}\frac{\partial x^{M}}
  {\partial\xi^{\alpha}},
\end{equation}
where $\xi^{\alpha}$ are coordinates on $\cW$ and are in general
dynamic. In what follows, we take the static gauge
$\xi^{\alpha}=x^{\alpha}$.  $g_{MN}$ is the 10d metric and
$B^{\left(2\right)}$ the NS-NS $2$-form which vanishes for the
backgrounds that we consider here.  $F_{\alpha\beta}$ are the
components of the field strength for the $\U{1}$
$\left(p+1\right)$-dimensional vector potential living on the brane,
$F^{\left(2\right)}=\ud A^{\left(1\right)}$.  The tension of a $\uD
p$-brane is given by $\tau_{\uD
  p}^{-1}=\left(2\pi\right)^{p}\ell_{\us}^{\left(p+1\right)}g_{\us}$
and we have $\lambda=2\pi\ell_{\us}^{2}$.  The Chern-Simons action is
\begin{equation}
  S_{\uD p}^{\mathrm{CS}}
  =\tau_{\uD p}g_{\us}\int_{\cW}\nn
  \mathrm{P}\biggl[\cC\wedge\ue^{-B^{\left(2\right)}}\biggr]
  \wedge\ue^{\lam F^{\left(2\right)}},
\end{equation}
in which $\cC$ is the formal sum of all of the RR-potentials.  The
non-Abelian generalization of this action is more
intricate~\cite{Myers:1999ps}; however, to leading order in $\ell_{\us}$
and to quadratic order in the open string fields, it can be obtained
by promoting $A^{\left(1\right)}$ and the fluctuations of the position
to adjoint-valued fields and taking a trace over gauge indices.

To leading order in $\ell_{\us}$, the action for the gauge field on $\uD
7$ in the above background can be found via a Taylor expansion
\begin{equation}
  S=-\frac{1}{4g_{8}^{2}}
  \int_{\cW}\nn\ud^{8}x\sqrt{g}
  \biggl\{g^{\alpha\beta}g^{\ga\delta}F_{\alpha\ga}F_{\beta\delta}
  -\frac{g_{\us}}{2\cdot 4!\sqrt{g}}\ep^{\alpha\beta\ga\delta\ep\eta\zeta\theta}
  C_{\alpha\beta\ga\delta}F_{\ep\eta}F_{\zeta\theta}\biggr\},
\end{equation}
in which $g_{8}^{2}=8\pi^{3}\ell_{\us}^{4}$ and $\ep^{0\cdots 7}=+1$.  After
integrating by parts, this can be written as
\begin{align}
  S=\frac{L^{3}}{g_{8}^{2}}\int_{\cW}\nn\ud^{8}x\,
  \sqrt{\tilde{g}}\ue^{-3B}\sqrt{\breve{g}}\biggl\{&
  -\frac{1}{4}\tilde{g}^{mn}\tilde{g}^{st}F_{ms}F_{nt}
  -\frac{\ue^{2B}}{2L^{2}}\tilde{g}^{mn}\breve{g}^{\phi\psi}
  \partial_{m}A_{\phi}\partial_{n}A_{\psi}\notag\\
  &+\frac{\ue^{2B}}{2L^{2}}\tilde{g}^{mn}\breve{g}^{\phi\psi}
  \breve{\nabla}_{\phi}\breve{\nabla}_{\psi}A_{m}A_{n}
  -\frac{2\rho C'}{L^{4}}
  \ue^{4C+4B-4A}\breve{\ve}^{\phi\psi\zeta}
  A_{\phi}\breve{\nabla}_{\psi}A_{\zeta}\notag\\
  &+\frac{\ue^{4B}}{2L^{4}}
  \breve{g}^{\phi\psi}\breve{g}^{\zeta\xi}
  \bigl(\breve{\nabla}_{\phi}\breve{\nabla}_{\psi}A_{\zeta}
  -\breve{\nabla}_{\zeta}\breve{\nabla}_{\phi}A_{\psi}
  -\breve{R}_{\phi\zeta}A_{\psi}\bigr)A_{\xi}\notag\\
  &+\frac{\ue^{2B}}{L^{2}}
  \tilde{g}^{mn}\breve{g}^{\phi\psi}
  \tilde{\nabla}_{m}A_{n}\breve{\nabla}_{\phi}A_{\psi}
  -\frac{\rho^{2}B'\ue^{4B}}{L^{4}}\breve{g}^{\phi\psi}A_{\rho}
  \breve{\nabla}_{\phi}A_{\psi}
  \biggr\},
\end{align}
in which $\tilde{\nabla}_{m}$ is the covariant derivative built from
the metric $\tilde{g}_{mn}$ for the 5d space $\cM$.  Similarly,
$\breve{\nabla}_{\phi}$ is the covariant derivative on $S^{3}$, and
the associated Ricci tensor is $\breve{R}_{\phi\psi}$, and
$\breve{\ve}^{\phi\psi\zeta}$ is the antisymmetric tensor on a unit
$S^{3}$. $'$ denotes a derivative with respect to $\rho$.
  
The components of the connection with the legs on $\cM$ transform as
scalars under rotations of the $S^{3}$ and thus can be expanded in
terms of scalar spherical harmonics
\begin{equation}
  A_{m}=\sum^{\infty}_{l=0}A_{m}^{\left(l\right)}\bigl(x^{m}\bigr)
  \cY_{l}\bigl(y^{\theta}\bigr),
\end{equation}
where
\begin{equation}
  \breve{\nabla}^{2}\cY_{l}=-l\bigl(l+2\bigr)\cY_{l}.
\end{equation}
The harmonics satisfy the orthogonality relationship
\begin{equation}
  \int_{S^{3}}\nn\ud\vol_{S^{3}}\cY_{l}\cY_{l'}= \cV_{S^{3}}\delta_{ll'},
\end{equation}
in which $\cV_{S^{3}}=2\pi^{2}$ is the volume of a unit $S^{3}$.  We
impose the gauge-fixing condition
\begin{equation}
  \tilde{g}^{mn}\tilde{\nabla}_{m}A_{n}=0\Rightarrow
  \tilde{g}^{mn}\tilde{\nabla}_{m}A^{\left(l\right)}_{n}=0.
\end{equation}
Similarly, the angular components are expanded into the $1$-form
harmonics\footnote{These are related to the more familiar vector
  spherical harmonics by contraction with the metric.  See,
  e.g.~\cite{Higuchi:1986wu, Ceresole:1999ht} for discussions of
  tensor spherical harmonics.}
\begin{equation}
  A_{\phi}=\sum_{l=0}^{\infty}B^{\left(l\right)}\bigl(x^{m}\bigr)
  \breve{\nabla}_{\phi}\cY_{l}\bigl(y^{\theta}\bigr)+
  L\sum^{\infty}_{l=1}\biggl\{
  a^{\left(l,+\right)}\bigl(x^{m}\bigr)
  \cY_{\phi,l}^{+}\bigl(y^{\theta}\bigr)
  +a^{\left(l,-\right)}\bigl(x^{m}\bigr)
  \cY_{\phi,l}^{-}\bigl(y^{\theta}\bigr)\biggr\},
\end{equation}
where the $\cY_{\phi,l}^{\pm}$ satisfy
\begin{equation}
  \breve{\nabla}^{2}\cY_{\phi,l}^{\pm}
  -2\cY_{\phi,l}^{\pm}
  =-\left(l+1\right)^{2}\cY_{\phi,l}^{\pm},\quad
  \breve{\nabla}^{\phi}\cY_{\phi,l}^{\pm}=0,\quad
  \ve^{\phi\psi\zeta}\breve{\nabla}_{\psi}\cY^{\pm}_{\zeta,l}
  =\pm\bigl(l+1\bigr)\breve{g}^{\phi\xi}\cY^{\pm}_{\xi,l},
\end{equation}
and the orthogonality relationship
\begin{equation}
  \int_{S^{3}}\nn\ud\vol_{S^{3}}
  \breve{g}^{\phi\psi}
  \cY^{\ep}_{\phi,l}\cY^{\ep'}_{\psi,l'}
  \propto \delta_{ll'}\delta^{\ep\ep'}.
\end{equation}

Owing to the various orthogonality relationships, the harmonics of
different types decouple from each other.  For the scalar harmonics,
we get, after integrating over the $S^{3}$,
\begin{align}
  S=\frac{1}{g_{5}^{2}}\int_{\cM}\nn\!\ud^{5}x\,
  \sqrt{{\tilde{g}}}\ue^{-3B}\sum_{l=0}^{\infty}\biggl\{&-\frac{1}{4}
  \tilde{g}^{mn}\tilde{g}^{st}F^{\left(l\right)}_{ms}F^{\left(l\right)}_{nt}
  -\frac{l\left(l+2\right)\ue^{2B}}{2L^{2}}
  \tilde{g}^{mn}\partial_{m}B^{\left(l\right)}\partial_{n}B^{\left(l\right)}\notag\\
  &-\frac{l\left(l+2\right)\ue^{2B}}{2L^{2}}
  \tilde{g}^{mn}A^{\left(l\right)}_{m}A^{\left(l\right)}_{n}
  +\frac{l\left(l+2\right)\rho^{2}\ue^{4B}}{L^{4}}A_{\rho}^{\left(l\right)}
  B^{\left(l\right)}\biggr\}.
\end{align}
in which $g_{5}^{2}=4\pi\ell_{\us}^{4}L^{-3}$ is the 5d gauge coupling.
For the $1$-form sector,
\begin{align}
  S=\frac{1}{g_{5}^{2}}\int_{\cM}\nn\!\ud^{5}x\,\sqrt{\tilde{g}}
  \sum^{\infty}_{l=1}\biggl\{&
  -\frac{\ue^{-B}}{2}\tilde{g}^{mn}\partial_{m}a^{\left(l,\pm\right)}
  \partial_{n}a^{\left(l,\pm\right)}\notag\\
  &-\frac{\ue^{B}}{2L^{2}}\bigl[\left(l+1\right)^{2}\pm 4\left(l+1\right)
  \rho C'\ue^{4C-4A}\bigr]a^{\left(l,\pm\right)}a^{\left(l,\pm\right)}\biggr\}.
\end{align}

The remaining bosonic degrees of freedom are the transverse
fluctuations of the $\uD 7$-brane.  To leading order in
$\ell_{\us}$ they enter only through the pullback of the metric
in this background
\begin{equation}
  \mathrm{P}\bigl[g_{\alpha\beta}\bigr]
  =g_{\alpha\beta}+\lambda^{2}\frac{L^{2}}{\rho^{2}}
  \ue^{-2B}\delta_{ij}\partial_{\alpha}\Phi^{i}\partial_{\beta}\Phi^{j},
\end{equation}
where $\Phi^{i=1,2}$ are related to the position of the $\uD 7$ brane
by
\begin{equation}
  x^{8}=\lam\Phi^{1},\quad
  x^{9}=\mu+\lam\Phi^{2}.
\end{equation}
The action is
\begin{equation}
  S=-\frac{1}{2g_{8}^{2}}\int_{\cW}\nn\ud^{8}x\,\sqrt{g}
  g^{\alpha\beta}\frac{\rho^{2}}{L^{2}}\ue^{-2B}\delta_{ij}
  \partial_{\alpha}\Phi^{i}\partial_{\beta}\Phi^{j}.
\end{equation}
Expanding in scalar spherical harmonics and integrating over the $S^{3}$,
\begin{equation}
  S=\frac{1}{g_{5}^{2}}\int_{\cM}\nn\!\ud^{5}x\,\sqrt{\tilde{g}}
  \sum_{l=0}^{\infty}\biggl\{
  -\frac{L^{2}\ue^{-5B}}{2\rho^{2}}\tilde{g}^{mn}\partial_{m}\Phi^{i\left(l\right)}
  \partial_{m}\Phi^{i\left(l\right)}
  -\frac{l\left(l+2\right)\ue^{-3B}}{2\rho^{2}}\Phi^{i\left(l\right)}
    \Phi^{i\left(l\right)}\biggr\}.
\end{equation}

For the fermionic degrees of freedom, we begin with the Dirac-like
action of~\cite{Marolf:2003ye,*Marolf:2003vf,*Martucci:2005rb}, which in
the Einstein frame reads~\cite{Marchesano:2008rg}
\begin{equation}
  S^{\mathrm{F}}_{\uD p}=-\frac{\ui}{g_{8}^{2}}
  \int_{\cW}\nn\ud^{8}x\,\bigl(g_{\us}^{-1}\ue^{\Phi}\bigr)^{\frac{p-3}{4}}
  \sqrt{\abs{\det\left(M_{\alpha\beta}\right)}}
  \bar{\Theta}P_{-}^{\uD p}
  \biggl\{
  \bigl(\cM^{-1}\bigr)^{\alpha\beta}
  \mathrm{P}\biggl[\Ga_{\beta}\bigl(\cD_{\alpha}
  +\frac{1}{4}\Ga_{\alpha}\Delta\bigr)\biggr]
  -\Delta\biggr\}\Theta,
\end{equation}
in which $\Theta$ is the bispinor
\begin{equation}
  \Theta=\begin{pmatrix}\theta_{1} \\ \theta_{2}\end{pmatrix},
\end{equation}
where $\theta_{1,2}$ are 10d Majorana-Weyl spinors\footnote{Our
  fermionic conventions are presented in appendix~\ref{app:conv}.},
$P_{-}^{\uD p}$ is the projection operator
\begin{equation}
  P_{\pm}^{\uD p}=\frac{1}{2}\bigl(1\pm \Ga_{\uD p}\bigr)
  =\frac{1}{2}\begin{pmatrix}
    1 & \pm \breve{\Ga}_{\uD p}^{-1} \\
    \pm \breve{\Ga}_{\uD p} & 1
  \end{pmatrix},
\end{equation}
where
\begin{equation}
  \breve{\Ga}_{\uD p}=\ui^{\left(p-2\right)\left(p-3\right)}
  \Ga_{\uD p}^{\left(0\right)}L\bigl(\cF\bigr),
\end{equation}
with
\begin{align}
  \Ga_{\uD p}^{\left(0\right)}
  =&\frac{1}{\left(p+1\right)!}
  \ve_{\alpha_{1}\cdots\alpha_{p+1}}\Ga^{\alpha_{1}\cdots\alpha_{p+1}},\notag\\
  L\bigl(\cF\bigr)=&
  \frac{\sqrt{\abs{\det\left(\mathrm{P}\left[g\right]\right)}}}
  {\sqrt{\abs{\det\left(M\right)}}}
  \sum_{q}\frac{\bigl(g_{\us}\ue^{-\Phi}\bigr)^{q/2}}{q!2^{q}}
  \cF_{\alpha_{1}\alpha_{2}}\cdots\cF_{\alpha_{2q-1}\alpha_{2q}}
  \Ga^{\alpha_{1}\cdots\alpha_{2q}},
\end{align}
with
$\cF^{\left(2\right)}=-\mathrm{P}\bigl[B^{\left(2\right)}\bigr]+\lam
F^{\left(2\right)}$ and $\ve_{\alpha_{1}\cdots\alpha_{p+1}}$ is the
antisymmetric tensor.  The operators $\cD_{M}$ and $\Delta$ are
involved in the SUSY-variations of the Einstein-frame gravitini and
dilatini as in appendix~\ref{app:typeIIB}.

The action above is subject to a gauge redundancy known as
$\ka$-symmetry where we make the identification
\begin{equation}
  \Theta\sim \Theta+\Ga_{-}^{\uD p}\ka,
\end{equation}
in which $\ka$ is an arbitrary 10d Majorana-Weyl bispinor.  We choose the gauge
\begin{equation}
  \Theta=\begin{pmatrix} \theta \\ 0\end{pmatrix},
\end{equation}
and then in the above background, the action becomes
\begin{equation}
  S_{\uD 7}^{\mathrm{F}}=-\frac{\ui}{2g_{8}^{2}}
  \int_{\cW}\nn\ud^{8}x\,\sqrt{g}\,\bar{\theta}\biggl\{g^{\alpha\beta}
  \Ga_{\alpha}\nabla_{\beta}
  +\frac{g_{\us}}{16}g^{\alpha\beta}
  \breve{\Ga}_{\uD 7}^{-1}\Ga_{\alpha}\slashed{F}^{\left(5\right)}\Ga_{\beta}
  \biggr\}\theta,
\end{equation}
with
\begin{equation}
  \breve{\Ga}_{\uD 7}
  =-\ui\sig^{\ul{3}}\otimes\II_{4}\otimes\sig^{\ul{3}}\otimes\II_{2}.
\end{equation}
$\theta$ is a 10d Majorana-Weyl spinor and thus can be written as
\begin{equation}
  \theta=\begin{pmatrix}1 \\ 0\end{pmatrix}
  \otimes\lambda\otimes\chi\otimes\psi
  -\ui\begin{pmatrix} 1 \\ 0\end{pmatrix}
  \otimes\tilde{B}_{5}\lambda^{\ast}\otimes\sig^{\ul{1}}\chi^{\ast}
  \otimes\ui\sig^{\ul{2}}\psi^{\ast},
\end{equation}
where $\lambda$, $\chi$, and $\psi$ are $\SO{4,1}$, $\SO{2}$, and
$\SO{3}$ Dirac spinors respectively.  That is, $\lambda$ is a spinor
on $\cM$ and $\psi$ is a spinor on the $S^{3}$ wrapped by the $\uD
7$-brane.  We additionally take the ansatz that $\lambda$ depends only on
the coordinates on $\cM$, $\psi$ depends only on the 3-cycle
coordinates, and $\chi$ is a constant spinor.  We have
\begin{equation}
  g^{\alpha\beta}\Ga_{\alpha}\slashed{F}^{\left(5\right)}\Ga_{\beta}
  =4\slashed{\cF}^{\left(5\right)},
\end{equation}
where we have used the self-duality of $F^{\left(5\right)}$ and that
this operator is acting on a 10d Weyl spinor. In this setup, we can
take the aichtbein for the metric~\ref{eq:inducedmetric} to be
\begin{equation}
  e_{\alpha}^{\phantom{\alpha}\ul{\beta}}
  =\begin{pmatrix}
    \ue^{A}\delta_{\mu}^{\phantom{\mu}\ul{\nu}} & & \\
    & \ue^{-B}\frac{L}{\rho} & \\
    & & \ue^{-B}L\breve{e}_{\ul{\theta}}^{\phantom{\theta}\ul{\phi}}
  \end{pmatrix},
\end{equation}
where underlined indices denote the non-coordinate frame and $\breve{e}_{\ul{\theta}}^{\phantom{\theta}\ul{\phi}}$ is the dreibein for a unit $S^{3}$.  Then
\begin{equation}
  4\slashed{\cF}
  =\frac{16\ui C'}{g_{\us}L}\ue^{4C-4A+B}
  \sig^{\ul{1}}\otimes\II_{4}\otimes\II_{2}\otimes\II_{2},
\end{equation}
where we have again made use of the fact that it acts on 10d Weyl
spinor.

The covariant derivative can be written in terms of the spin
connection
\begin{equation}
  \nabla_{\alpha}=\partial_{\alpha}+
  \frac{1}{4}\omega_{\alpha}^{\phantom{\alpha}\ul{\beta}\ul{\ga}}
  \Ga_{\ul{\beta}\ul{\ga}},
\end{equation}
where
\begin{equation}
  \omega_{{\alpha}}^{\phantom{\alpha}\ul{\beta}\ul{\ga}}
  =\frac{1}{2}e_{\alpha}^{\phantom{\alpha}\ul{\delta}}
  \bigl(T^{\phantom{\delta}\ul{\beta}\ul{\ga}}_{\ul{\delta}}
  -T^{\ul{\beta}\ul{\ga}\phantom{\delta}}_{\phantom{\beta\ga}\ul{\delta}}
  -T^{\ul{\ga}\phantom{\delta}\ul{\beta}}_{\phantom{\ga}\ul{\delta}\phantom{\beta}}\bigr),
  \quad
  T^{\ul{\alpha}}_{\phantom{\alpha}\ul{\beta}\ul{\ga}}
  =\bigl(
  e^{\beta}_{\phantom{\beta}\ul{\beta}}e^{\ga}_{\phantom{\ga}\ul{\ga}}-
  e^{\ga}_{\phantom{\ga}\ul{\beta}}e^{\beta}_{\phantom{\beta}\ul{\ga}}\bigr)
  \partial_{\ga}e_{\beta}^{\phantom{\beta}\ul{\alpha}}.
\end{equation}
For the above choice of aichtbein, the non-vanishing components of the
spin connection are
\begin{equation}
  \label{eq:D7_spin_connection}
  \omega_{\phi}^{\phantom{\phi}\ul{\psi}\ul{\zeta}}=
  \breve{\omega}_{\phi}^{\phantom{\phi}\ul{\psi}\ul{\zeta}},\quad
  \omega_{\phi}^{\phantom{\phi}\ul{\psi}\ul{4}}
  =-\rho B'\delta_{\phi}^{\ul{\psi}},\quad
  \omega_{\mu}^{\phantom{\mu}\ul{4}\ul{\pi}}
  =-\frac{\rho A'}{L}\ue^{A+B}\delta_{\mu}^{\ul{\pi}}.
\end{equation}
To leading order in $\ell_{\us}$, the bosonic and fermionic
fluctuations decouple and the action takes the form
\begin{multline}
  S=-\frac{\ui L^{3}}{2g_{8}^{2}}
  \int_{\cW}\nn\ud^{8}x\,\sqrt{\tilde{g}}\ue^{-3B}\sqrt{\breve{g}}
  \biggl\{
  \bigl(\bar{\lambda}\tilde{\slashed{\nabla}}\lambda\bigr)
  \bigl(\chi^{\dagger}\chi\bigr)
  \bigl(\psi^{\dagger}\psi\bigr)
  +\frac{\rho C'}{L}\ue^{4C-4A+B}
  \bigl(\bar{\lambda}\lambda\bigr)
  \bigl(\chi^{\dagger}\sig^{3}\chi\bigr)
  \bigl(\psi^{\dagger}\psi\bigr)
  \\
  -\frac{3\rho B'}{2L}\ue^{B}
  \bigl(\bar{\lambda}\ga_{\left(4\right)}\lambda\bigr)
  \bigl(\chi^{\dagger}\chi\bigr)
  \bigl(\psi^{\dagger}\psi\bigr)
  +\frac{\ui}{L}\ue^{B}
  \bigl(\bar{\lambda}\lambda\bigr)
  \bigl(\chi^{\dagger}\sig^{3}\chi\bigr)
  \bigl(\psi^{\dagger}\breve{\slashed{\nabla}}\psi\bigr)
  \biggr\}
  +\mathrm{c.c.}
\end{multline}
Under $\SO{5}\to\SO{2}\times\SO{3}$, a Dirac spinor decomposes as
(see, e.g.~\cite{Wetterich:1982ed})
\begin{equation}
  \eta\to \chi_{+}\otimes\psi_{+}+\chi_{-}\otimes\psi_{-},
\end{equation}
where $\chi_{\pm}$ are $\SO{2}$ Weyl spinors while $\psi_{\pm}$ are
$\SO{3}$ Dirac spinors.  As an $\SO{3}$ spinor, $\psi$ can be expanded
in spinor spherical harmonics\footnote{See
  e.g.~\cite{Ceresole:1999ht}.} which satisfy
\begin{equation}
  \breve{\slashed{\nabla}}\cY_{l,\pm}=\pm\ui\left(l+\frac{3}{2}\right)
  \cY_{l,\pm},
\end{equation}
where $l=0,1,2,\ldots$.  They again satisfy the orthogonality condition
\begin{equation}
  \int_{S^{3}}\nn\nn\ud\vol_{S^{3}} \cY_{l,\ep}^{\dagger}
  \cY_{l',\ep}\propto \delta_{ll'}\delta_{\ep\ep'}.
\end{equation}
Then we take the expansion
\begin{equation}
  \theta=\sum_{l,\sig,\ep}
  \begin{pmatrix}1 \\ 0\end{pmatrix}\otimes
  \lam^{\left(l,\sigma,\ep\right)}\otimes\chi_{\sigma}\otimes
  \cY_{l,\ep}+\cdots,
\end{equation}
where $\cdots$ indicates the terms necessary to ensure that $\theta$
is Majorana.  With this expansion,
\begin{align}
  S=\frac{1}{g_{5}^{2}}\int_{\cM}\nn\!\ud^{5}x\,
  \sqrt{\tilde{g}}\ue^{-3B}\sum_{l=0}^{\infty}\sum_{\ep,\sig}
  \biggl\{&
  -\ui\bar{\lam}^{l,\sig,\ep}\tilde{\slashed{\nabla}}\lam^{l,\sig,\ep}
  -\frac{3\ui\rho B'\ue^{B}}{2L}\bar{\lam}^{l,\sig,\ep}\ga_{\left(4\right)}
  \lam^{l,\sig,\ep}\notag\\
  &+\frac{\ui\sig\ue^{B}}{L}
  \biggl(\ep\bigl(l+\frac{3}{2}\bigr)-\rho C'\ue^{4C-4A}\biggr)
  \bar{\lam}^{l,\sig,\ep}\lam^{l,\sig,\ep}
  \biggr\}
\end{align}

The degrees of freedom should be able to be organized into $\cN_{5}=1$
super multiplets and our interest is in the lightest vector multiplet.
We can identify this multiplet by comparing against the known results
in the $AdS^{5}$ case which is recovered by taking $A=\bar{B}=C=\log
r/L$ and $\mu=0$.  In this case, the lightest (i.e. most negative
$m^{2}$) comes from the $\left(1,-\right)$ sector of the scalar
descending from $A_{\phi}$, which saturates the BF
bound~\cite{Breitenlohner:1982jf,*Breitenlohner:1982bm},
\begin{equation}
  m^{2}=-\frac{4}{L^{2}}.
\end{equation}
This is the expected mass for the scalar in the $\cN_{5}=1$ massless
vector hypermultiplet~\cite{Kim:1985ez,Shuster:1999zf}.  The
corresponding Dirac fermion has mass $\frac{1}{2L}$ and comes from the
$\left(l,\ep,\sig\right)=\left(0,-,+\right)$ mode of the fermions.
The vector component of this multiplet comes from the $l=0$ mode of
$A_{m}$ and so the action for these modes is
\begin{align}
  \label{eq:non_canonical_on_shell_action}
  S=-\frac{1}{g_{5}^{2}}\int_{\cM}\nn\!\ud^{5}x\,\sqrt{\tilde{g}}\ue^{-3B}
  \biggl\{&\frac{1}{4}\tilde{g}^{mn}\tilde{g}^{st}F_{ms}F_{nt}
  +\ui\bar{\lam}\tilde{\ga}^{m}\tilde{\nabla}_{m}\lam
  +\frac{\ue^{2B}}{2}\tilde{g}^{mn}\partial_{m}a\partial_{n}a\notag\\
  &+\ui m_{\lam}\bar{\lam}\lam
  +\ui \alpha\bar{\lam}\ga_{\left(4\right)}\lam
  +\frac{1}{2}m_{a}^{2}a^{2}\biggr\},
\end{align}
in which
\begin{align}
  L^{2}m_{a}^{2}=4\ue^{4B}\bigl(1-2\rho C'\ue^{4C-4A}\bigr),\quad
  Lm_{\lam}=\ue^{B}\bigl(\frac{3}{2}-\rho C'\ue^{4C-4A}\bigr),\quad
  L\alpha=-\frac{3\rho B'}{2}\ue^{B}.
\end{align}
The higher modes give rise to $\cN_{5}=1$ hyper multiplets and massive
vector multiplets.  Note that although not real by itself, when
$\ui\alpha\bar{\lam}\ga_{\left(4\right)}\lam$ added to the kinetic term for the
fermion, the entire action is real.

Since these are all components of a single supermultiplet, it will be
convenient to perform a field redefinition so that they all have the
same kinetic terms.  Changing the kinetic term of $A_{m}$ would spoil
manifest gauge invariance, and so we redefine the scalar to match the
gauge kinetic term.  Defining $\Sig=\ue^{A}a$, we get
\begin{align}
  \label{eq:on_shell_action}
  S=-\frac{1}{g_{5}^{2}}\int_{\cM}\nn\!\ud^{5}x\,\sqrt{\tilde{g}}\ue^{-3B}
  \biggl\{&\frac{1}{4}\tilde{g}^{mn}\tilde{g}^{st}F_{ms}F_{nt}
  +\ui\bar{\lam}\tilde{\ga}^{m}\tilde{\nabla}_{m}\lam
  +\frac{1}{2}\tilde{g}^{mn}\partial_{m}\Sig\partial_{n}\Sig\\
  &+\ui m_{\lam}\bar{\lam}\lam
  +\ui \alpha\bar{\lam}\ga_{\left(4\right)}\lam
  +\frac{1}{2}m_{\Sig}^{2}\Sig^{2}\biggr\},\notag
\end{align}  
in which
\begin{equation}
  L^{2}m_{\Sig}^{2}=\ue^{2B}\bigl({\rho^{2}B''}-{\rho^{2}B'^{2}}
  +4\rho^{2}A'B'+{\rho B'}-8\rho C'\ue^{4C-4A}+4\bigr).
\end{equation}

\subsection{\label{sec:off_shell}Off-shell theory from flat space}

$\cN_{5}=1$ supersymmetry is generated by eight real supercharges that
can be arranged into a pair of symplectic-Majorana spinors
$\cR^{i=1,2}$.  In addition to the connection $A^{\left(1\right)}$ and
the gaugino (the degrees of freedom of which can again be expressed
as a pair of symplectic-Majorana spinors $\lambda^{1,2}$) the
off-shell theory contains three real auxiliary scalar fields
$X^{I=1,2,3}$ (see e.g.~\cite{West:1990tg}).  Under the $\SU{2}$
R-symmetry that rotates the $\cR^{i}$ into each other, the gauge field
is inert, while the fermionic degrees transform as a $\mathbf{2}$ and
the auxiliary fields as a $\mathbf{3}$.  Under a SUSY transformation
parametrized by $\eta^{i}$, the fields transform in flat space as
\begin{align}
\label{eq:5d_flatspace_SUSY}
  \delta_{\eta}A_{m}=&\ui\bar{\eta}_{i}\tilde{\gamma}_{m}\lambda^{i},\notag\\
  \delta_{\eta}\Sigma=&\bar{\eta}_{i}\lambda^{i},\\
  \delta_{\eta}X^{I}=&\bar{\eta}_{i}
  \bigl(\sig^{I}\bigr)^{i}_{\ j}
  \tilde{\ga}^{m}\partial_{m}\lambda^{j},\notag\\
  \delta_{\eta}\lambda^{i}=&
  -\frac{1}{2}F_{mn}\tilde{\ga}^{mn}\eta^{i}
  +\ui\tilde{\ga}^{m}\partial_{m}\Sig\eta^{i}
  +\ui X^{I}\bigl(\sig^{I}\bigr)^{i}_{\ j}\eta^{j}\notag,
\end{align}
where $\bigl(\sig^{I}\bigr)^{i}_{\ j}$ are the components of the
usual Pauli matrices.  The algebra closes in the sense that
\begin{equation}
  \bigl[\delta_{\eta},\delta_{\xi}\bigr]=2\ui\bar{\xi}_{i}\tilde{\ga}^{m}
  \partial_{m}\eta^{i}\partial_{m},
\end{equation}
except when acting on the gauge field which closes only up to a gauge
transformation.  The off-shell action is
\begin{equation}
  \label{eq:offshell_maxwell_action}
  S=-\frac{1}{g_{5}^{2}}\int_{R^{4,1}}\nn\nn\nn\ud^{5}x\,
  \biggl\{\frac{1}{4}F_{mn}F^{mn}
  +\frac{1}{2}\partial_{m}\Sig\partial^{m}\Sig
  +\frac{\ui}{2}
  \bar{\lam}_{i}\tilde{\ga}^{m}\partial_{m}\lam^{i}
  -\frac{1}{2}X^{I}X^{I}\biggr\}.
\end{equation}
The vector multiplet can be written in $\cN_{4}=1$ language as a
vector superfield and a neutral chiral
superfield~\cite{Mirabelli:1997aj,*Hebecker:2001ke}.  In particular, we
can embed an $\cN_{4}=1$ into the higher supersymmetry by considering
the transformations~\eqref{eq:5d_flatspace_SUSY} and taking
$\eta_{\uR}=0$.  Under this restricted set, the fields transform as
\begin{align}
  \delta_{\eta_{\uL}}A_{\mu}=&
  \ui\bar{\eta}_{\uL}\bar{\sig}_{\mu}\lambda_{\uL}
  +\ui\eta_{\uL}\sig_{\mu}\bar{\lambda}_{\uL},\notag\\
  \delta_{\eta_{\uL}}A_{4}=&
  \bar{\eta}_{\uL}\bar{\lam}_{\uR}
  +\eta_{\uL}\lam_{\uR},\notag\\
  \delta_{\eta_{\uL}}\Sig=&
  -\ui\bar{\eta}_{\uL}\bar{\lam}_{\uR}
  +\ui\eta_{\uL}\lam_{\uR},\\
  \delta_{\eta_{\uL}}X^{1}=&\eta_{\uL}\sig^{\mu}\partial_{\mu}\bar{\lam}_{\uR}
  +\ui\eta_{\uL}\partial_{4}\lam_{\uL}
  -\bar{\eta}_{\uL}\bar{\sig}^{\mu}\partial_{\mu}\lam_{\uR}
  -\ui\bar{\eta}_{\uL}\partial_{4}\bar{\lam}_{\uR}
  ,\notag\\
  \delta_{\eta_{\uL}}X^{2}
  =&\ui\eta_{\uL}\sig^{\mu}\partial_{\mu}\bar{\lam}_{\uR}
  -\eta_{\uL}\partial_{4}\eta_{\uL}
  +\ui\bar{\eta}_{\uL}\bar{\sig}^{\mu}\partial_{\mu}\lam_{\uR}
  -\bar{\eta}_{\uL}\partial_{4}\bar{\lam}_{\uL}
  ,\notag \\
  \delta_{\eta_{\uL}}X^{3}=&
  \bar{\eta}_{\uL}\bar{\sig}^{\mu}\partial_{\mu}\lam_{\uL}
  -\ui\bar{\eta}_{\uL}\partial_{4}\bar{\lam}_{\uR}
  -\eta_{\uL}\sig^{\mu}\partial_{\mu}
  \bar{\lam}_{\uR}
  +\ui\eta_{\uL}\partial_{4}\lam_{\uR},\notag\\
  \delta_{\eta_{\uL}}\lam_{\uL}=&
  F_{\mu\nu}\sig^{\mu\nu}
  \eta_{\uL}
  +\ui\bigl(X^{3}-\partial_{4}\Sig\bigr)\eta_{\uL},\notag\\
  \delta_{\eta_{\uL}}\lam_{\uR}=&
  \ui F_{\mu 4}\sig^{\mu}\bar{\eta}_{\uL}
  +\partial_{\mu}\Sig\sig^{\mu}\bar{\eta}_{\uL}
  -\ui\bigl(X^{1}+\ui X^{2}\bigr)\eta_{\uL}.\notag
\end{align}
It was recognized in~\cite{Mirabelli:1997aj,*Hebecker:2001ke} that
these are the transformation rules for the components of a chiral
superfield $\Phi$ and a vector superfield $\cV$ in the Wess-Zumino
gauge under the combination of a supersymmetry transformation and a
complexified gauge transformation required to maintain the Wess-Zumino
condition.  The superfields take the form
\begin{subequations}
\begin{align}
  \Phi=&\phi+\sqrt{2}\theta\psi
  + \theta^{2}F+\ui\theta\sig^{\mu}\bar{\theta}\partial_{\mu}\phi
  +\frac{\ui}{\sqrt{2}}\theta^{2}\bar{\theta}\bar{\sig}^{\mu}
  \partial_{\mu}\psi
  +\frac{1}{4}\theta^{2}\bar{\theta}^{2}\partial^{2}\phi,\\
  \cV=&
  -\theta\sig^{\mu}\bar{\theta}A_{\mu}
  +\ui\theta^{2}\bar{\theta}\bar{\lambda}
  -\ui\bar{\theta}^{2}\theta\lambda
  +\frac{1}{2}\theta^{2}\bar{\theta}^{2}D,
\end{align}
in which $\partial^{2}=\eta^{\mu\nu}\partial_{\mu}\partial_{\nu}$. The
curl superfield $\cW_{\alpha}=-\frac{1}{4}D^{2}\bar{D}_{\alpha}\cV$
takes the standard form
\begin{align}
  \cW_{\alpha}=&
  -\ui\lambda_{\alpha}+\theta_{\alpha}D
  -\frac{\ui}{2}\left(\sig^{\mu}\bar{\sig}^{\nu}\right)_{\alpha}^{\ \beta}
  \theta_{\beta}F_{\mu\nu}
  +\theta^{2}
  \sig^{\mu}_{\alpha\dot{\alpha}}\partial_{\mu}
  \bar{\lambda}^{\dot{\alpha}}
  +\theta\sig^{\mu}\bar{\theta}
  \partial_{\mu}\lambda_{\alpha}\notag\\
  &+\ui\theta_{\alpha}\theta\sig^{\mu}\bar{\theta}
  \partial_{\mu}D
  +\frac{1}{2}\left(\sig^{\mu}\bar{\sig}^{\nu}\right)_{\alpha}^{\ \beta}
  \theta_{\beta}
  \theta\sig^{\kappa}\bar{\theta}
  \partial_{\kappa}F_{\mu\nu}
  -\frac{\ui}{4}\theta^{2}\bar{\theta}^{2}
  \partial^{2}\lambda_{\alpha}.
\end{align}
\end{subequations}
The component fields are related to the usual 5d fields through
\begin{equation}
  \label{eq:5d_in_4d}
  \phi=\Sig+\ui A_{4},\quad
  \psi=\ui\sqrt{2}\lambda_{\uR},\quad
  F=X^{1}+\ui X^{2},\qquad
  \lambda=\lambda_{\uL},\quad
  D=X^{3}-\partial_{4}\Sig,
\end{equation}
where $\lambda_{\uL}$ and $\lambda_{\uR}$ are the left- and
right-handed components of $\lam^{1}$ defined as
in~\eqref{eq:decompose_sm_spinors}.  Under the complexified gauge
transformation
\begin{equation}
  \cV\to \cV+\frac{1}{2}\bigl(\Lambda+\Lambda^{\ast}\bigr),
\end{equation}
where $\Lambda$ is a chiral superfield, $\Phi$ transforms as
\begin{equation}
  \Phi\to \Phi+\partial_{4}\Lambda.
\end{equation}
Then, the action~\eqref{eq:offshell_maxwell_action} can be written in
the language of $\cN_{4}=1$ superspace
\begin{equation}
  \label{eq:flatspace_off_shell_theory}
  S=\frac{1}{g_{5}^{2}}\int_{R^{4,1}}\nn\nn\nn\ud^{5}x
  \biggl\{
  \frac{1}{4}\int\ud^{2}\theta\, \cW^{\alpha}\cW_{\alpha}
  +\frac{1}{4}\int\ud^{2}\bar{\theta}\,
  \overline{\cW}_{\dot{\alpha}}\overline{\cW}^{\dot{\alpha}}
  +\int\ud^{4}\theta
  \left(
  \frac{1}{2}
  \bigl(\Phi+\Phi\bigr)^{\ast}-\partial_{4}\cV\right)^{2}\biggr\},
\end{equation}
where the coefficients are chosen to recover the normalization of the
action in~\eqref{eq:offshell_maxwell_action}.  We can couple this
$\cN_{5}=1$ theory to an $\cN_{4}=1$ theory localized at some point
$x^{4}=x_{0}^{4}$ by introducing an action
\begin{equation}
  \label{eq:couple_4d_to_5d}
  2\int_{x^{4}=x_{0}^{4}}\nn\nn\nn\nn\nn\nn\ud^{4}x\,
  \int\ud^{4}\theta\,
  \cV\cJ
  =\int_{x^{4}=x_{0}^{4}}\nn\nn\nn\nn\nn\nn\ud^{4}x\,
  \bigl\{JD-\lambda j -\bar{\lambda}\bar{j}-A^{\mu}j_{\mu}\bigr\},
\end{equation}
where $\cJ$ is a current
superfield~\eqref{eq:define_current_superfield}.

Given this flat space off-shell theory, we can produce the off-shell
theory for the vector supermultiplet on
$AdS^{5}$~\cite{Gregoire:2004nn, *Bagger:2006hm, *Bagger:2011na,
  McGarrie:2010yk}.  We can apply the same techniques to deduce the
off-shell action for the $\cN_{5}=1$ theory resulting from dimensional
reduction onto a supersymmetric probe $\uD 7$-brane in $AdS^{5}\times
X^{5}$ where $X^{5}$ is an Einstein-Sasaki manifold\footnote{A similar
  process should in fact work for any of the $\cN_{4}\ge 1$
  compactifications of~\cite{Giddings:2001yu}.  However, once
  supersymmetry is broken by the background, one would have to
  consider the supergravity multiplet as well.}.  The procedure is to
begin by identifying the Killing spinor for the 5d theory.  We can
begin by considering the Killing spinors of the 10d theory.  For this
geometry, the Killing spinor equations are
\begin{equation}
  \cD_{M}\ve=\nabla_{M}\ve+\frac{g_{\us}}{16}\slashed{F}^{\left(5\right)}
  \Ga_{M}\bigl(\ui\sig^{2}\bigr)\ve,
\end{equation}
where $\ve$ is a Majorana-Weyl bispinor
\begin{equation}
  \ve=\begin{pmatrix}\ve_{1} \\ \ve_{2}\end{pmatrix}.
\end{equation}
Taking $\ve_{2}=-\ui\ve_{1}$, this becomes
\begin{equation}
  0=\nabla_{M}\ve_{1}+\frac{\ui g_{\us}}{16}\slashed{F}^{\left(5\right)}
  \Ga_{M}\ve_{1}.
\end{equation}
Writing
\begin{equation}
  \ve=\begin{pmatrix}1 \\ 0\end{pmatrix}
  \otimes\ep\otimes\beta-\ui\begin{pmatrix}1 \\ 0\end{pmatrix}
  \otimes\tilde{B}_{5}\eta^{\ast}\otimes\hat{B}_{5}\beta^{\ast},
\end{equation}
where $\ep$ is an $\SO{4,1}$ spinor and $\beta$ is an $\SO{5}$ spinor,
this equation becomes
\begin{equation}
  \label{eq:5d_Killing_spinors}
  \hat{\nabla}_{\phi}\beta=-\frac{\ui}{2}\hat{\ga}_{\phi}\beta,\quad
  \tilde{\nabla}_{m}\ep=+\frac{1}{2L}\tilde{\ga}_{m}\ep,
\end{equation}
where $\hat{\nabla}_{\phi}$ is the covariant derivative built from the
metric on $X^{5}$.  The covariant derivative can be easily deduced
from the spin-connection on the $\uD 7$
worldvolume~\eqref{eq:D7_spin_connection}.  Since $X^{5}$ is an
Einstein-Sasaki space, the first of~\eqref{eq:5d_Killing_spinors} has
a solution.  The second does as well, the explicit form of
which~\cite{Breitenlohner:1982jf,Lu:1996rhb,Shuster:1999zf} will be
useful for us.  Since
$\left\{\mathbb{I}_{4},\tilde{\ga}_{\ul{m}},\tilde{\ga}_{\ul{mn}}\right\}$
form a basis for $4\times 4$ matrices, we can take the ansatz
\begin{equation}
  \ep=\left(a+b^{\ul{\mu}}\tilde{\ga}_{\ul{\mu}}
    +c^{\ul{\mu\nu}}\tilde{\ga}_{\ul{\mu\nu}}
    \right)\eta^{+}
    +\left(d+e^{\ul{\mu}}\tilde{\ga}_{\ul{\mu}}
    +c^{\ul{\mu\nu}}\tilde{\ga}_{\ul{\mu\nu}}
    \right)\eta^{-},
\end{equation}
where $\eta^{\pm}$ are constant Dirac spinors satisfying
$\ga_{\left(4\right)}\eta^{\pm}=\pm\eta^{\pm}$.  The $r$-component of
the Killing spinor equation is
\begin{equation}
  \partial_{r}\ep=-\frac{1}{2}\ga_{\left(4\right)}\ep,
\end{equation}
which implies that $c^{\ul{\mu}\ul{\nu}}=0$ while
\begin{equation}
  a=\sqrt{\frac{L}{r}}\bar{a},\quad
  b^{\ul{\mu}}=\sqrt{\frac{r}{L}}\bar{b}^{\ul{\mu}},\quad
  d=\sqrt{\frac{r}{L}}\bar{d},\quad
  e^{\ul{\mu}}=\sqrt{\frac{L}{r}}\bar{e}^{\ul{\mu}},
\end{equation}
where $\bar{a}$, $\bar{b}^{\ul{\mu}}$, $\bar{d}$, and
$\bar{e}^{\ul{\mu}}$ are all functions of the $R^{3,1}$ coordinates
$x^{\mu}$.
The $\mu$ components are
\begin{equation}
  \partial_{\mu}\ep=-\frac{r}{2L^{2}}
  \delta_{\mu}^{\ul{\mu}}\bigl(1+\ga_{\left(4\right)}\bigr)\ep,
\end{equation}
which are solved by
\begin{equation}
  \bar{a}=1,\quad \bar{b}^{\ul{\mu}}=-\frac{1}{L}\delta^{\ul{\mu}}_{\mu}x^{\mu},
  \quad
  \bar{d}=1,\quad\bar{e}^{\ul{\mu}}=0.
\end{equation}
Hence, the Killing spinor takes the
form~\cite{Breitenlohner:1982jf,Lu:1996rhb,Shuster:1999zf}
\begin{equation}
  \ep=
  \left(\frac{r}{L}\right)^{-\ga_{\left(4\right)}/2}
  \left(1-\frac{x^{\mu}}{L}\tilde{\ga}_{\mu}
    \left(1+\ga_{\left(4\right)}\right)
  \right)\eta
  =\begin{pmatrix}
    \sqrt{\frac{r}{L}}
    \left(\eta_{\uL\alpha}
      -\frac{\ui r}{L^{2}}x^{\mu}\delta_{\mu}^{\ul{\mu}}
      \bar{\sig}^{\ul{\mu}}_{\alpha\dot{\alpha}}
      \bar{\eta}_{\uR}^{\dot{\alpha}}\right) \\
    \ui\sqrt{\frac{L}{r}}\bar{\eta}_{\uR}^{\dot{\alpha}}
  \end{pmatrix}.
\end{equation}
where $\eta=\eta^{+}+\eta^{-}$.  Upon dimensional reduction, $\ep$
generates SUSY transformations for the 5d effective field theory on
$\cM$ where $r$ is to be treated as a function of $\rho$.

The off-shell theory for the $\cN_{5}=1$ vector multiplet can be
determined by again considering the restricted supersymmetry
transformations characterized by $\bar{\eta}_{\uR}=0$.  The
remaining components parametrize a rigid $\cN_{4}=1$ transformation.
An $\cN_{5}=1$ transformation is induced by
\begin{equation}
  \bar{\ep}_{i}\cR^{i}=
  \ui\ep_{\uL}\cR_{\uR}-\ui\bar{\ep}_{\uL}\bar{\cR}_{\uR}
  +\ui\bar{\ep}_{\uR}\bar{\cR}_{\uL}-\ui\ep_{\uR}\cR_{\uL}
  =\ui\sqrt{\frac{r}{L}}\eta_{\uL}\cR_{\uR}
  -\ui\sqrt{\frac{r}{L}}\bar{\eta}_{\uL}\bar{\cR}_{\uR},
\end{equation}
where after the second equality we have set $\eta_{\uR}=0$.  If we
identify $\eta_{\uL}$ as characterizing a rigid $\cN_{4}=1$
transformation, then the corresponding generator is
\begin{equation}
  Q=\sqrt{\frac{r}{L}}\cQ=\ui\sqrt{\frac{r}{L}}\cR_{\uR}.
\end{equation}
$Q$ induces translations in superspace
\begin{equation}
  Q_{\alpha}=
  \frac{\partial}{\partial\theta^{\alpha}}
  -\ui\sig^{\ul{\mu}}_{\alpha\dot{\alpha}}
  \bar{\theta}^{\dot{\alpha}}\delta_{\ul{\mu}}^{\ {\mu}}\partial_{\mu},
\end{equation}
and the restricted $\cN_{5}=1$ transformations induce translations
through a warped $\cN_{4}=1$ superspace
\begin{equation}
  \cQ_{\alpha}=
  \frac{\partial}{\partial\vt^{\alpha}}
  -\ui\sig^{\ul{\mu}}_{\alpha\dot{\alpha}}
  \bar{\vt}^{\dot{\alpha}}\tilde{\ep}_{\ul{\mu}}^{\ {\mu}}\partial_{\mu},
\end{equation}
where
\begin{equation}
  \vt^{\alpha}=\sqrt{\frac{r}{L}}\theta^{\alpha}=\ue^{A/2}\theta^{\alpha}.
\end{equation}

The $\cN_{5}=1$ super-Maxwell theory on $AdS^{5}$ can be recovered
from~\eqref{eq:flatspace_off_shell_theory} by
writing~\cite{Gregoire:2004nn, Bagger:2006hm, McGarrie:2010yk,
  Bagger:2011na}
\begin{subequations}
\begin{align}
  \Phi=&\phi+\sqrt{2}\vt\psi
  + \vt^{2}F+\ui\vt\sig^{\ul{\mu}}\bar{\vt}
  \tilde{e}_{\ul{\mu}}^{\phantom{\mu}\mu}\partial_{\mu}\phi
  +\frac{\ui}{\sqrt{2}}\vt^{2}\bar{\vt}\bar{\sig}^{\ul{\mu}}
  \tilde{e}_{\ul{\mu}}^{\phantom{\mu}\mu}
  \partial_{\mu}\psi
  +\frac{1}{4}\vt^{2}\bar{\vt}^{2}
  \tilde{g}^{\mu\nu}\partial_{\mu}\partial_{\nu}
  \phi\notag\\
  =&\phi+\sqrt{2}\ue^{A/2}\theta\psi
  + \ue^{A}\theta^{2}F+
  \ui\theta\sig^{\ul{\mu}}\bar{\theta}
  \delta_{\ul{\mu}}^{\phantom{\mu}\mu}\partial_{\mu}\phi
  +\frac{\ui}{\sqrt{2}}\ue^{A/2}\theta^{2}\bar{\theta}\bar{\sig}^{\ul{\mu}}
  \delta_{\ul{\mu}}^{\phantom{\mu}\mu}
  \partial_{\mu}\psi
  +\frac{1}{4}\theta^{2}\bar{\theta}^{2}\partial^{2}\phi,\\
  \cV=&
  -\vt\sig^{\ul{\mu}}\bar{\vt}\tilde{e}_{\ul{\mu}}^{\ \mu}
  A_{\mu}
  +\ui\vt^{2}\bar{\vt}\bar{\lambda}
  -\ui\bar{\vt}^{2}\vt\lambda
  +\frac{1}{2}\vt^{2}\bar{\vt}^{2}D\notag\\
  =&-\theta\sig^{\ul{\mu}}\bar{\theta}\delta_{\ul{\mu}}^{\phantom{\mu}\mu}A_{\mu}
  +\ui\ue^{3A/2}\theta^{2}\bar{\theta}\bar{\lambda}
  -\ui\ue^{3A/2}\bar{\theta}^{2}\theta\lambda
  +\frac{1}{2}\ue^{2A}\theta^{2}\bar{\theta}^{2}D.
\end{align}
The curl chiral superfield is likewise
\begin{align}
  \cW_{\alpha}=&-\frac{1}{4}\cD^{2}\bar{\cD}_{\alpha}\cV\notag\\
  =&
  -\ui\lambda_{\alpha}+\vt_{\alpha}D
  -\frac{\ui}{2}\left(\sig^{\ul{\mu}}\bar{\sig}^{\ul{\nu}}
  \right)_{\alpha}^{\ \beta}
  \vt_{\beta}\tilde{e}_{\ul{\mu}}^{\phantom{\mu}\mu}
  \tilde{e}_{\ul{\nu}}^{\phantom{\nu}\nu}
  F_{\mu\nu}
  +\vt^{2}
  \sig^{\ul{\mu}}_{\alpha\dot{\alpha}}
  \tilde{e}_{\ul{\mu}}^{\phantom{\mu}\mu}\partial_{\mu}
  \bar{\lambda}^{\dot{\alpha}}
  +\vt\sig^{\ul{\mu}}\bar{\vt}\tilde{e}_{\ul{\mu}}^{\phantom{\mu}\mu}
  \partial_{\mu}\lambda_{\alpha}\notag\\
  &+\ui\vt_{\alpha}\vt\sig^{\ul{\mu}}\bar{\vt}
  \tilde{e}_{\ul{\mu}}^{\phantom{\mu}\mu}
  \partial_{\mu}D
  +\frac{1}{2}\left(\sig^{\ul{\mu}}\bar{\sig}^{\ul{\nu}}\right)_{\alpha}^{\ \beta}
  \vt_{\beta}
  \vt\sig^{\ul{\kappa}}\bar{\vt}
  \tilde{e}_{\ul{\mu}}^{\phantom{\mu}\mu}
  \tilde{e}_{\ul{\nu}}^{\phantom{\nu}\nu}
  \tilde{e}_{\ul{\kappa}}^{\phantom{\kappa}\kappa}
  \partial_{\kappa}F_{\mu\nu}
  -\frac{\ui}{4}\vt^{2}\bar{\vt}^{2}
  \tilde{g}^{\mu\nu}
  \partial_{\mu}\partial_{\nu}
  \lambda_{\alpha}\notag\\
  =&
  -\ui\lambda_{\alpha}+\ue^{A/2}\theta_{\alpha}D
  -\frac{\ui}{2}\ue^{-3A/2}
  \left(\sig^{\ul{\mu}}\bar{\sig}^{\ul{\nu}}\right)_{\alpha}^{\ \beta}
  \delta_{\ul{\mu}}^{\phantom{\mu}\mu}
  \delta_{\ul{\nu}}^{\phantom{\nu}\nu}
  \theta_{\beta}F_{\mu\nu}
  +\theta^{2}
  \sig^{\ul{\mu}}_{\alpha\dot{\alpha}}
  \delta_{\ul{\mu}}^{\phantom{\mu}\mu}
  \partial_{\mu}
  \bar{\lambda}^{\dot{\alpha}}
  +\theta\sig^{\ul{\mu}}\bar{\theta}
  \delta_{\ul{\mu}}^{\phantom{\mu}\mu}
  \partial_{\mu}\lambda_{\alpha}\notag\\
  &+\ui\ue^{A/2}\theta_{\alpha}\theta\sig^{\ul{\mu}}\bar{\theta}
  \delta_{\ul{\mu}}^{\phantom{\mu}\mu}
  \partial_{\mu}D
  +\frac{1}{2}\ue^{-3A/2}\left(\sig^{\ul{\mu}}
    \bar{\sig}^{\ul{\nu}}\right)_{\alpha}^{\ \beta}
  \theta_{\beta}
  \theta\sig^{\ul{\kappa}}\bar{\theta}
  \delta_{\ul{\mu}}^{\phantom{\mu}\mu}
  \delta_{\ul{\nu}}^{\phantom{\nu}\nu}
  \delta_{\ul{\kappa}}^{\phantom{\kappa}\kappa}
  \partial_{\kappa}F_{\mu\nu}
  -\frac{\ui}{4}\theta^{2}\bar{\theta}^{2}
  \partial^{2}\lambda_{\alpha},
\end{align}
\end{subequations}
where $\cD_{\alpha}=\ue^{-A/2}D_{\alpha}$.  The same result holds for
$\cM$ with $A\to A\left(\rho\right)$. The components of the chiral
superfield are as in~\eqref{eq:5d_in_4d} except with
\begin{subequations}
\begin{equation}
  \phi=\Sigma+\ui A_{\ul{4}}=\Sigma+\ui \ue^{A}A_{\rho}.
\end{equation}
Similarly for the vector superfield,
\begin{equation}
  D=X^{3}-\partial_{\ul{4}}\Sigma
  =X^{3}-\ue^{A}\partial_{\rho}\Sigma,
\end{equation}
where here we have used that the $\uD 7$ is on $AdS^{5}\times X^{5}$ so
that $A=\bar{B}$.
\end{subequations}
The off-shell action follows from~\eqref{eq:flatspace_off_shell_theory}
\begin{equation}
  S=\frac{1}{g_{5}^{2}}\int_{\cM}\nn\!\ud^{5}x\, \sqrt{\tilde{g}}\ue^{-3B}
  \biggl\{\frac{1}{4}
  \int\ud^{2}\vt\,\cW^{\alpha}\cW_{\alpha}
  +\frac{1}{4}
  \int\ud^{2}\bar{\vt}\,\overline{\cW}_{\dot{\alpha}}
  \overline{\cW}^{\dot{\alpha}}
  +\int\ud^{4}\vt
  \biggl(\frac{1}{2}\bigl(\Phi+\Phi\bigr)^{\ast}-\ue^{A}
  \partial_{\rho}\cV\biggr)^{2}\biggr\},
\end{equation}
where the additional factor of $\ue^{3B}$ in the measure is introduced
to match with the gauge kinetic term of~\eqref{eq:on_shell_action}. In
terms of components,
\begin{align}
  S=-\frac{1}{g_{5}^{2}}\int_{\cM}\nn\!
  \ud^{5}x\,\sqrt{\tilde{g}}\frac{\rho^{3}}{L^{3}}\ue^{-3A}
  \biggl\{&
  \frac{1}{4}F_{mn}F^{mn}
  +\frac{\ui}{2}\bar{\lam}_{i}\tilde{\ga}^{m}\nabla_{m}\lam^{i}
  +\frac{1}{2}\partial_{m}\Sig\partial^{m}\Sig\notag\\
  & +\frac{\ui}{2}m_{\lam}
  \bar{\lam}_{i}\bigl(\sig^{3}\bigr)^{i}_{\ j}\lam^{j}
  +\frac{1}{2}\bar{m}_{\Sig}^{2}\Sig^{2}
  -\frac{1}{2}X^{I}X^{I}
  -\beta\Sig X^{3}\biggr\},
\end{align}
in which
\begin{align}
  \bar{m}_{\Sig}^{2}=&
  \frac{1}{\rho^{2}}\ue^{2A}\bigl(\rho^{2}A''
  +2\rho^{2}A'^{2}-3\rho A'-6\bigr),\notag\\
  m_{\lam}=&\frac{1}{2\rho}\ue^{A}\bigl(3-2\rho A'\bigr),\notag\\
  \beta=&\frac{1}{\rho}\ue^{A}\bigl(3-\rho A'\bigr),\notag
\end{align}
where we have used that in the supersymmetric case
$B=A+\log{L}/{\rho}$.  In the $AdS^{5}$ case, which can be
recovered by taking $\mu\to 0$,
\begin{equation}
  \bar{m}_{\Sig}^{2}=-\frac{8}{L^{2}},\quad
  m_{\lam}=\frac{1}{2L},\quad
  \beta=\frac{2}{L}.
\end{equation}

The auxiliary fields can be easily integrated out,
\begin{equation}
  X^{1,2}=0,\quad X^{3}=-\beta\Sig.
\end{equation}
Then the action can be written as~\eqref{eq:on_shell_action} with
\begin{align}
  m_{\Sig}^{2}=&\frac{1}{\rho^{2}}\ue^{2A}\bigl(\rho^{2}A''+
  3\rho^{2}A'^{2}-9\rho A'+3\bigr)\notag\\
  m_{\lam}=&\frac{1}{2\rho}\ue^{A}\bigl(3-2\rho A'\bigr),\notag\\
  \alpha=&\frac{3\ue^{A}}{\rho}\bigl(1-\rho A'\bigr),\notag
\end{align}
in which we have written $\lam=\lam_{1}$.  Note that this agrees with
the result in the previous subsection in supersymmetric limit
$B=A+\log L/p$.  In the $AdS$ case, we recover
$Lm_{\lam}={1}/{2}$, $L^{2}m_{\Sig}^{2}=-4$, and $\alpha=0$.

\section{\label{sec:SUSY_currents}Supersymmetric current-current correlators}

We now turn to the calculation of the current-current correlators
which, as discussed in section~\ref{sec:ggm}, can be used to calculate
visible-sector soft terms.  These two-point functions can be
calculated by considering the dual classical gravity solution.  We
will make use of the method of holographic
renormalization~\cite{Henningson:1998gx,*Henningson:1998ey,*de_Haro:2000xn,*Bianchi:2001de,Bianchi:2001kw,Skenderis:2002wp}.
The first step in the procedure is to solve the equations of motion
near the boundary of the 5d spacetime $\cM$.  The resulting on-shell
action will be divergent but can be regularized by cutting off the
spacetime at finite $\rho=R$.  The divergences can be subtracted by
the addition of an appropriate boundary term action and then the
renormalized action is defined in the $R\to \infty$ limit.  The
solution to the field equation for a particular field $\Phi$ that is
dual to a operator $\cO$ will be given in terms of two coefficients
that are set by boundary conditions.  One of these coefficients can be
fixed by determining a boundary condition at $\rho=\infty$ while the
other requires a boundary condition at small $\rho$.  The former
coefficient gives the leading behavior at large $\rho$ and corresponds
to a source for $\cO$ while the second corresponds to the resulting
point-point function for $\cO$.  Higher-point functions can then be
determined by differentiation of the one-point with respect to the
source.

To employ these methods, it is convenient to define a coordinate $u$
by
\begin{equation}
  \label{eq:define_u}
  \frac{\ud u}{\ud \rho}=-\frac{2u}{\rho}\ue^{-B}.
\end{equation}
Then the metric for $\cM$ is written as
\begin{equation}
  \label{eq:metric_in_u}
  \ud s_{5}^{2}=\ue^{2A}\eta_{\mu\nu}\ud x^{\mu}\ud x^{\nu}
  +\frac{L^{2}}{4u^{2}}\ud u^{2}.
\end{equation}

\subsection{\label{eq:massless_susy}Massless messengers}

The gravity fields dual to the field theory operators can be inferred
from~\eqref{eq:couple_4d_to_5d}.  In order to fix the normalizations,
we will first consider the case of $AdS^{5}$ with $\mu=0$.  Since the
spacetimes that we are considering are asymptotically anti-de Sitter,
the normalizations will apply also in these other cases.  Many of the
results of this subsection have been presented previously in the
literature.

\subsubsection{Scalar current}

The interaction Lagrangian contains the term $JD$ where $D$ is the
auxiliary component of the $\cN_{4}=1$ vector multiplet and $J$ is the
scalar component of the current superfield.  The action for $D$ was
determined only in the supersymmetric case in
section~\ref{sec:off_shell}.  The action after integrating it out in
either the SUSY or non-SUSY case was determined in
section~\ref{sec:on_shell} and since we are interested in only the
on-shell action, this action is sufficient for calculating two-point
functions.  However, the off-shell action is required to determine the
leading behavior of $D$ and thus the duality between the boundary
operator and the bulk field.  Since the spacetimes we are considering
are asymptotically AdS, we can apply the result of the $AdS^{5}$ case
to the other spacetimes.  The analysis of scalar correlators were
performed early in the stages of
AdS/CFT~\cite{Gubser:1998bc,Witten:1998qj,Muck:1998rr,Klebanov:1999tb}
though the analysis here follows the methods of holographic
renormalization (see, e.g.~\cite{Skenderis:2002wp}) with the slight
difference that we consider an auxiliary field.

In the pure-AdS case, $B=0$ and so~\eqref{eq:define_u} is solved by
\begin{equation}
  u=\frac{L^{2}}{\rho^{2}},
\end{equation}
and the action for the scalar $\Sig$ is
\begin{equation}
  S=-\frac{1}{2}\int_{AdS^{5}}\nn\nn\nn\nn\ud^{5}x\,
  \sqrt{\tilde{g}}
  \biggl\{\tilde{g}^{mn}\partial_{m}\Sig\partial_{n}\Sig
  +m^{2}\Sig^{2}\biggr\},
\end{equation}
in which $m^{2}=-4/L^{2}$ and we have redefined the field to absorb
a factor of $g_{5}$.  The resulting equation of motion, using the
coordinate $u$ is
\begin{equation}
  0=4u^{2}\Sig''-4u\Sig'+4\Sig-u \ka^{2}\Sig,
\end{equation}
where we have performed a Fourier transformation on the Minkwoski
spacetime and have defined $\ka^{2}=L^{2}k^{2}$ where $k^{2}$ is the
momentum.  This is solved by the series expansion
\begin{equation}
  \Sig=u\sum_{n=0}^{\infty}\biggl\{
  \sig_{\left(2n\right)}+\tilde{\sig}_{\left(2n\right)}\log u
  \biggr\}u^{n},
\end{equation}
in which $\sig_{\left(0\right)}$ and $\tilde{\sig}_{\left(0\right)}$
are undetermined while for $n>0$,
\begin{align}
  0=&4n^{2}\sig_{\left(2n\right)}+8n\tilde{\sig}_{\left(2n\right)}
  -\kappa^{2}\sig_{\left(2n-2\right)},\notag\\
  0=&4n^{2}\tilde{\sig}_{\left(2n\right)}
  -\kappa^{2}\tilde{\sig}_{\left(2n-2\right)}.
\end{align}
The scalar current $J$ is dual to the auxiliary field $D$
\begin{equation}
  D=X^{3}+\frac{2u}{L}\partial_{u}\Sig
  =\frac{2u}{L^{2}}\tilde{\sig}_{\left(0\right)}+\cdots,
\end{equation}
where we have used the fact that on-shell $X^{3}=-\frac{2}{L}\Sig$ and
have made a small $u$ expansion.  Since
$\tilde{\sig}_{\left(0\right)}$ is leading order term, we identify it
as the source for the field theory operator $J$.  The other
undetermined coefficient ${\sig}_{\left(0\right)}$ should then
be identified with the response.

The regulated action is defined by cutting off the integral at some
small $u=\ep$.  On-shell, this gives
\begin{equation}
  S_{\mathrm{reg}}
  =-\frac{1}{2}
  \int_{u\ge\ep}\nn\nn\ud^{5}x\,\sqrt{\tilde{g}}
  \biggl\{\tilde{g}^{mn}\partial_{m}\Sig\partial_{n}\Sig
  +m^{2}\Sig^{2}\biggr\}
  =\frac{1}{2}
  \int_{u=\ep}\nn\nn\ud^{4}x\,\sqrt{h}
  \frac{2u}{L}\Sig\partial_{u}\Sig,
\end{equation}
where after the second equality we have integrated by parts, applied
the equation of motion and have written the boundary metric as
\begin{equation}
  \ud s_{4}^{2}=h_{\mu\nu}\ud x^{\mu}\ud x^{\nu}.
\end{equation}
Inserting in the above solution, we find
\begin{equation}
  S_{\mathrm{reg}}=\frac{1}{L}\int\frac{\ud^{4}k}
  {\left(2\pi\right)^{4}}
  \biggl\{
  \tilde{\sig}_{\left(0\right)}^{2}\left(\log \ep\right)^{2}+
  2\sig_{\left(0\right)}\tilde{\sig}_{\left(0\right)}
  \log \ep 
  +\tilde{\sig}_{\left(0\right)}^{2}\log \ep
  +\cdots\bigr\},
\end{equation}
where $\cdots$ indicates those terms that are
finite or vanishing as $\ep\to 0$.  The action diverges in the limit
$\ep\to 0$, but the divergence can be removed by adding a
counterterm action
\begin{equation}
  S_{\mathrm{ct}}=-\frac{1+\left(\log \ep\right)^{-1}}
  {L}\int_{u=\ep}\nn\nn\ud^{4}x\,
  \sqrt{h}\Sig^{2}.
\end{equation}
The subtracted action
\begin{equation}
  S_{\mathrm{sub}}=S_{\mathrm{reg}}+S_{\mathrm{ct}},
\end{equation}
is then finite as $\ep\to 0$.  Defining the renormalized action
\begin{equation}
  \label{eq:define_renormalized_action}
  S_{\mathrm{ren}}=\lim_{\ep\to 0}S_{\mathrm{sub}},
\end{equation}
the response of the current $J$ to the source
$\tilde{\sig}_{\left(0\right)}$ is
\begin{equation}
  \left\langle J\right\rangle_{\mathrm{s}}=
  \frac{1}{\sqrt{\det\left(\eta_{\mu\nu}\right)}}
  \frac{\delta S_{\mathrm{ren}}}{\delta \tilde{\sig}_{\left(0\right)}}.
\end{equation}
Using~\eqref{eq:define_renormalized_action}
\begin{equation}
  \left\langle J\right\rangle_{\mathrm{s}}
  =\lim_{\ep\to 0}
  \frac{1}{\ep^{2}\sqrt{h}}
  {\ep\log \ep }
  \frac{\delta S_{\mathrm{sub}}}{\delta \Sig},
\end{equation}
where we have used the fact that for small $u$
\begin{equation}
  \Sig=\tilde{\sig}_{\left(0\right)}u\log u +\cdots.
\end{equation}
Under $\Sig\to\Sig+\delta \Sig$ at the boundary
\begin{equation}
  \delta S_{\mathrm{sub}}
  =
  \frac{2}{L}\int_{u=\ep}\nn\nn{\ud^{4}x}\sqrt{h}
  \biggl\{u\partial_{u}\Sig-\left(1+\left(\log \ep\right)^{-1}\right)
  \Sig\biggr\}\delta\Sig.
\end{equation}
Inserting in the above solution the divergent parts cancel and we get
\begin{equation}
  \bigl<J\bigr>_{\us}
  =-\frac{2}{L}\sig_{\left(0\right)}.
\end{equation}

The two-point function is then
\begin{equation}
  \bigl\langle J\bigl(k\bigr)J\bigl(-k\bigr)\bigr>
  =-\left.\frac{\delta\left\langle J\left(k\right)\right\rangle_{\us}}
    {\delta \tilde{\sig}_{\left(0\right)}\left(-k\right)}\right
  \rvert_{\tilde{\sig}_{\left(0\right)}=0}
  \to\frac{2D_{0}}{L}
  \left.\frac{\delta\sig_{\left(0\right)}}
    {\delta\tilde{\sig}_{\left(0\right)}}
  \right\rvert_{\tilde{\sig}_{\left(0\right)}=0},
\end{equation}
where the arrow indicates that have introduced a constant $D_{0}$ to
account a possible normalization.  In the $AdS^{5}$ case, it is
possible to solve the equation of motion exactly and we get
\begin{equation}
  \Sig=N_{0}u I_{0}\bigl(\kappa\sqrt{u}\bigr)
  +M_{0}u K_{0}\bigl(\kappa\sqrt{u}\bigr),
\end{equation}
where $I_{\nu}$ and $K_{\nu}$ are modified Bessel functions of the
first and second kinds and $\kappa=\sqrt{\kappa^{2}}$ and in this
expression, $\kappa$ is taken to be the Euclidean momentum.  Demanding
that $\Sig\to 0$ as $u\to\infty$ gives the condition $N_{0}=0$.  Then
a small $u$ expansion gives
\begin{equation}
  \Sig=u\biggl[-M_{0}\left(\ga+\log \kappa +\log 2
    \right)-\frac{M_{0}}{2}\log u +\cdots\biggr],
\end{equation}
where $\ga$ is the Euler-Mascheroni constant.  From this expression we
read off
\begin{equation}
  \sig_{\left(0\right)}\left(k\right)
  =\tilde{\sig}_{\left(0\right)}
  \biggl[\log \kappa^{2} +2\ga-\log 4 \biggr],
\end{equation}
giving
\begin{equation}
  \label{eq:massless_scalar_current}
  \bigl\langle J\left(k\right)J\left(-k\right)\bigr\rangle
  =\frac{2D_{0}}{L}
  \biggl(\log \kappa^{2} +2\ga-\log 4 \biggr).
\end{equation}
The non-analytic behavior in $k$ is completely determined by conformal
invariance and has the expected form.  Additionally, we have
suppressed a factor of $\left(2\pi\right)^{4}\delta^{4}\left(0\right)$
resulting from momentum conservation.

The correlator for the operator dual to a BF scalar in $AdS^{d+1}$ was
first calculated in~\cite{Klebanov:1999tb} and in position space
\begin{equation}
  \bigl\langle\cO\bigl(x\bigr)\cO\bigl(0\bigr)\bigr\rangle
  =\frac{2}{\pi^{2}}\frac{1}{x^{4}}+\left(\text{contact terms}\right).
\end{equation}
Note that the dual operator $J$ that we consider here is not precisely
dual to the BF scalar $\Sig$ but is instead dual to the auxiliary
field $D$. However, since $D$ latter is closely related to the BF
scalar we will use the result of~\cite{Klebanov:1999tb} to fix the
normalization.  Fourier transforming and comparing
to~\eqref{eq:massless_scalar_current}, we get $D_{0}=-L$.

\subsubsection{Vector current}

A similar analysis applies for the vector
correlators~\cite{Mueck:1998iz,*l'Yi:1998yt} though again we apply the
method of holographic renormalization as in, e.g.,
~\cite{Bianchi:2001kw}.  The bulk field due to the vector current
$j_{\mu}$ are the components $A_{\mu}$ of the 5d vector field.  The
action for the vector field is
\begin{equation}
  S=-\frac{1}{4}\int_{AdS^{5}}\nn\nn\nn\nn\ud^{5}x\,\sqrt{\tilde{g}}
  \tilde{g}^{mn}\tilde{g}^{st}F_{ms}F_{nt},
\end{equation}
resulting in the equations of motion
\begin{align}
  0=&4u^{2}A_{\mu}+uL^{2}\partial^{2}A_{\mu}-uL^{2}\partial_{\mu}
  \bigl(\partial\cdot A\bigr)-4u^{2}\partial_{\mu}
  \bigl(\partial_{u}A_{u}\bigr),\notag\\
  0=&u\partial^{2}A_{u}-u\partial_{u}\bigl(\partial\cdot A\bigr),
\end{align}
where $\partial\cdot A:=\eta^{\mu\nu}\partial_{\mu}A_{\nu}$ and as
before $\partial^{2}=\eta^{\mu\nu}\partial_{\mu}\partial_{\nu}$.  In
section~\ref{sec:on_shell}, we imposed the 5d Lorenz condition
\begin{equation}
  \tilde{\nabla}^{m}A_{m}=0.
\end{equation}
For the metric~\eqref{eq:metric_in_u}, the non-vanishing Christoffel
symbols are
\begin{equation}
  \tilde{\Ga}^{u}_{\phantom{u}uu}=-\frac{1}{u},\quad
  \tilde{\Ga}^{u}_{\phantom{u}\mu\nu}=\frac{2}{L^{2}}\eta_{\mu\nu},\quad
  \tilde{\Ga}^{\mu}_{\phantom{\mu}u\nu}=-\frac{1}{2u}\delta^{\mu}_{\nu},
\end{equation}
and so the Lorenz condition becomes
\begin{equation}
  0=uL^{2}\partial\cdot A+4u^{2}\partial_{u}A_{u}-4uA_{u}.
\end{equation}
Then the equations of motion become
\begin{subequations}
\begin{align}
  \label{eq:minkowski_component_gauge_eom}
  4\ui u k_{\mu}A_{u}=&4u^{2}\partial_{u}^{2}A_{\mu}-u\kappa^{2}A_{\mu},\\
  \label{eq:radial_component_gauge_eom}
  0=&4u^{2}\partial_{u}^{2}A_{u}-u\kappa^{2}A_{u}.
\end{align}
\end{subequations}

The solution to~\eqref{eq:radial_component_gauge_eom} is
\begin{equation}
  A_{u}=\sum_{n=0}^{\infty}\bigl\{a_{{\left(2n\right)}}+
  \tilde{a}_{\left(2n\right)}\log u \bigr\}u^{n},
\end{equation}
in which $\tilde{a}_{\left(0\right)}=0$ while $a_{\left(0\right)}$ and
$a_{\left(2\right)}$ are undetermined.  For $n>0$,
\begin{align}
  0=&4n\left(n-1\right)a_{\left(2n\right)}
  +4\left(2n-1\right)\tilde{a}_{\left(2n\right)}
  -\kappa^{2}a_{\left(2n-2\right)},\notag\\
  0=&4n\left(n-1\right)\tilde{a}_{\left(2n\right)}
  -\kappa^{2}\tilde{a}_{\left(2n-2\right)}.
\end{align}
The homogeneous part of~\eqref{eq:minkowski_component_gauge_eom} has a
similar solution
\begin{equation}
  A_{\mu}^{\left(\uH\right)}
  =\sum_{n=0}^{\infty}\bigl\{a_{\mu\left(2n\right)}+\tilde{a}_{\mu\left(2n\right)}
  \log u \bigr\}u^{n},
\end{equation}
with $\tilde{a}_{\mu\left(0\right)}=0$, $a_{\mu\left(0\right)}$ and
$a_{\mu\left(2\right)}$ undetermined, and for $n>0$,
\begin{align}
  0=&4n\left(n-1\right)a_{\mu\left(2n\right)}
  +4\left(2n-1\right)\tilde{a}_{\mu\left(2n\right)}
  -\kappa^{2}a_{\mu\left(2n-2\right)},\notag\\
  0=&4n\left(n-1\right)\tilde{a}_{\mu\left(2n\right)}
  -\kappa^{2}\tilde{a}_{\mu\left(2n-2\right)}.
\end{align}
For the inhomogeneous part, we write
\begin{equation}
  A_{\mu}^{\left(\uI\right)}
  =-\ui k_{\mu}\sum_{n=0}^{\infty}
  \bigl\{\alpha_{\left(2n\right)}+\tilde{\alpha}_{\left(2n\right)}
  \log u \bigr\}u^{n}.
\end{equation}
This leads to $\tilde{\alpha}_{\left(0\right)}=0$. For $n>0$,
\begin{align}
  0=&4n\left(n-1\right)\alpha_{\left(2n\right)}
  +4\left(2n-1\right)\tilde{\alpha}_{\left(2n\right)}
  +4a_{\left(2n-2\right)}-\kappa^{2}\alpha_{\left(2n\right)},\notag\\
  0=&4n\left(n-1\right)\tilde{\alpha}_{\left(2n\right)}
  +4\tilde{a}_{\left(2n-2\right)}
  -\kappa^{2}\tilde{\alpha}_{\left(2n-2\right)}.
\end{align}
We can add some of the homogeneous solution to set
$\tilde{\alpha}_{\left(2\right)}=0$.

The gauge-fixing condition in Fourier space is
\begin{equation}
  0=4u^{2}\partial_{u}-4uA_{u}+\ui L u\kappa^{\mu}A_{\mu},
\end{equation}
where $\kappa^{\mu}:=\eta^{\mu\nu}\kappa_{\nu}$.  This imposes the relations
\begin{align}
  0=&\ui L\ka^{\mu}a_{\mu\left(2n\right)}
  +4\left(n-1\right)a_{\left(2n\right)}+4\tilde{a}_{\left(2n\right)}
 -\kappa^{2}\alpha_{\left(2n\right)},\notag\\
  0=&\ui L\ka^{\mu}\tilde{a}_{\mu\left(2n\right)}
  +4\left(n-1\right)\tilde{a}_{\left(2n\right)}-
  \kappa^{2}\tilde{\alpha}_{\left(2n\right)}.
\end{align}
Note that the latter implies
$\kappa^{\mu}\tilde{a}_{\mu\left(2\right)}=\kappa^{\mu}a_{\mu\left(0\right)}=0$.

With this solution, the regulated action is
\begin{equation}
  S_{\mathrm{reg}}=\frac{1}{L}\int\frac{\ud^{4}k}{\left(2\pi\right)^{4}}
  \eta^{\mu\nu}\frac{\kappa^{2}}{4}a_{\mu\left(0\right)}a_{\nu\left(0\right)}
  \log \ep +\cdots.
\end{equation}
The divergence can be cancelled by adding the counterterm action
\begin{equation}
  S_{\mathrm{ct}}=-\frac{L\log \ep }{8}
  \int_{u=\ep}\nn\nn\nn\ud^{4}x\,
  \sqrt{h}h^{\mu\nu}h^{\sig\tau}
  F_{\mu\sig}F_{\nu\tau}.
\end{equation}

Writing
\begin{equation}
  A_{\mu}=\sum_{n=0}\bigl\{A_{\mu\left(n\right)}
  +\tilde{A}_{\mu\left(n\right)}\log \ep \bigr\}u^{n},
\end{equation}
the leading behavior of $A_{\mu}$ is
\begin{equation}
  A_{\mu}=A^{\left(0\right)}_{\mu}+\cO\bigl(u\bigr),
\end{equation}
and so the response function in the dual field theory is
\begin{equation}
  \left<j_{\mu}\right>_{\us}
  =\frac{1}{\sqrt{\det\left(\eta_{\mu\nu}\right)}}
  \frac{\delta S_{\mathrm{ren}}}{\delta A_{\left(0\right)}^{\mu}}\notag\\
  =\lim_{\ep\to 0}\frac{1}{\ep^{2}\sqrt{h}}
  \frac{\delta S_{\mathrm{sub}}}{\delta A^{\mu}},
\end{equation}
where as before, $S_{\mathrm{sub}}=S_{\mathrm{reg}}+S_{\mathrm{sub}}$
and $S_{\mathrm{ren}}$ is the limit of this sum as $\ep\to 0$ and the
$\mu$ index has been raised with $\eta^{\mu\nu}$.  Under $A_{\mu}\to
A_{\mu}+\delta A_{\mu}$ at the boundary,
\begin{equation}
  \delta S_{\mathrm{sub}}
  =\frac{2}{L}\int_{u=\ep}\nn\nn\nn\ud^{4}x\,\sqrt{h}\ep^{2}
  \bigl\{F_{u\mu}+\frac{L^{2}\log \ep }{8}
  \eta^{\sig\kappa}\partial_{\sig}F_{\kappa\mu}\bigr\}A^{\mu}.
\end{equation}
Inserting in the above solution gives
\begin{equation}
  \bigl<j_{\mu}\bigr>_{\us}=\frac{2}{L}
  \biggl(a_{\mu\left(2\right)}-\frac{k_{\mu}k^{\nu}}{k^{2}}a_{\nu\left(2\right)}
  +\frac{\kappa^{2}}{4}a_{\mu\left(0\right)}\biggr).
\end{equation}
Using that $k^{\mu}a_{\mu\left(0\right)}=0$, this gives
\begin{align}
  \bigl<j_{\mu}\bigr>_{\us}=&\frac{2}{L}\bigl(\delta^{\nu}_{\mu}
  -\frac{k_{\mu}k^{\nu}}{k^{2}}\bigr)
  \biggl(a_{\nu\left(2\right)}+\frac{\kappa^{2}}{4}a_{\nu\left(0\right)}\biggr).
\end{align}
Note that the longitudinal part is projected out, as expected for a
conserved current.  Since the solution to the inhomogeneous equation
is transverse, we have
\begin{equation}
  \label{eq:general_vector_two_point}
  \bigl<j_{\mu}\bigl(k\bigr)j_{\nu}\bigl(-k\bigr)\bigr>
  =-\left.\frac{\delta\bigl<j_{\mu}\bigl(k\bigr)\bigr>_{\us}}
    {\delta A_{\left(0\right)}^{\nu}
    \bigl(-k\bigr)}\right\rvert_{a_{\nu\left(0\right)\kappa}=0}
  \to-\frac{2D_{1}}{L}
  \bigl(\delta_{\mu}^{\lambda}-\frac{k_{\mu}k^{\lambda}}{k^{2}}\bigr)
  \biggl(\frac{\delta a_{\lambda\left(2\right)}}{\delta a^{\nu}_{\left(0\right)}}
  +\frac{\kappa^{2}}{4}\eta_{\lambda\nu}\biggr).
\end{equation}
In the $AdS^{5}$ case, we can again solve the equations of motion
exactly, and the solution to the homogeneous part
of~\eqref{eq:minkowski_component_gauge_eom} is
\begin{equation}
  A_{\mu}^{\left(H\right)}=N_{\mu}\sqrt{u}I_{1}\bigl(\kappa\sqrt{u}\bigr)
  + M_{\mu}\sqrt{u}K_{1}\bigl(\kappa\sqrt{u}\bigr).
\end{equation}
Demanding that $A_{\mu}\to 0$ as $u\to \infty$ sets $N_{\mu}=0$.  Then
expanding for small $u$,
\begin{equation}
  A_{\mu}^{\left(\uH\right)}
  =\frac{M_{\mu}}{\kappa^{}}
  +\frac{\kappa M_{\mu}}{4}u\bigl(\log \kappa^{2} 
  +2\ga-\log 4-1\bigr)
  +\frac{\kappa M_{\mu}}{4}u\log u+\cdots,
\end{equation}
so that
\begin{equation}
  a_{\mu\left(2\right)}=\frac{\kappa^{2}a_{\mu\left(0\right)}}{4}
    \biggl(\log \kappa^{2}+2\ga-\log 4-1\biggr),
\end{equation}
giving
\begin{equation}
  \bigl\langle j_{\mu}\bigl(k\bigr)j_{\nu}\bigl(-k\bigr)
  \bigr\rangle
  =-\frac{D_{1}L}{2}\bigl(
  k^{2}\delta_{\mu\nu}-k_{\mu}k_{\nu}\bigr)
  \biggl(\log \kappa^{2}
  +2\ga-\log 4 \biggr),
\end{equation}
where we have again moved into Euclidean space.  Note that this
satisfies
\begin{equation}
  \bigl\langle j_{\mu}\bigl(k\bigr)j_{\nu}\bigl(-k\bigr)\bigr\rangle
  \propto -\bigl(k^{2}\delta_{\mu\nu}-k_{\mu}k_{\nu}\bigr)
  \bigl\langle J\bigl(k\bigr)J\bigl(-k\bigr)\bigr\rangle,
\end{equation}
as expected from supersymmetry.
  
\subsubsection{Spinor current}  

We now turn to the analysis of the spinor
current~\cite{Henningson:1998cd,*Mueck:1998iz,*Ghezelbash:1998pf},
making use of holographic renormalization techniques as
in~\cite{Ammon:2010pg}.  It was argued in~\cite{Henneaux:1998ch,
  *Contino:2004vy} that the action must be supplemented by a particular
boundary term,
\begin{equation}
  S=-\ui\int_{AdS^{5}}\nn\nn\nn\nn\ud^{5}x\,\sqrt{\tilde{g}}
  \biggl\{\bar{\lambda}\tilde{\ga}^{m}\tilde{\nabla}_{m}\lambda
  +m\bar{\lambda\lambda}\biggr\}
  +\frac{\ui}{2}\int_{\delta AdS^{5}}\nn\nn\nn\nn\nn\ud^{4}x\,\sqrt{h}
  \bar{\lambda}\lambda,
  \label{eq:supplemented_Dirac}
\end{equation}
in which $m=1/2L$.  The non-vanishing components of spin connection are
\begin{equation}
  \tilde{\omega}_{\mu}^{\phantom{\mu}\ul{4}\ul{\pi}}=\frac{2}{L\sqrt{u}}
  \delta_{\mu}^{\ul{\pi}},
\end{equation}
and so the equation of motion resulting from this action is
\begin{equation}
  \label{eq:AdS_Dirac}
  0=\ui\kappa_{\mu}\sqrt{u}\tilde{\ga}^{\ul{\mu}}
  \delta_{\ul{\mu}}^{\ \mu}\lambda
  -2u\ga_{\left(4\right)}\partial_{u}\lambda
  +2\ga_{\left(4\right)}\lambda
  +\frac{1}{2}\lambda.
\end{equation}
Following~\cite{Mueck:1998iz}, we apply $\tilde{\ga}^{m}\partial_{m}$,
giving
\begin{equation}
  0=4u^{2}\partial_{u}^{2}\lambda
  -6u\partial_{u}\lambda
  +\frac{1}{2}\ga_{\left(4\right)}\lambda+\frac{23}{4}\lambda
  -u\kappa^{2}\lambda.
\end{equation}
Writing $\lambda$ as
\begin{equation}
  \lambda=\begin{pmatrix}\lambda_{\uL} \\ \ui\bar{\lambda}_{\uR}
    \end{pmatrix},
\end{equation}
we get the solutions
\begin{align}
  \lambda_{\uL}=&u^{3/4}\sum_{n=0}^{\infty}
  \bigl\{\lambda_{\uL\left(2n\right)}+\tilde{\lambda}_{\uL\left(2n\right)}
  \log u \bigr\}u^{n},\notag\\
  \lambda_{\uR}=&u^{5/4}\sum_{n=0}^{\infty}
  \bigl\{\lambda_{\uR\left(2n\right)}+\tilde{\lambda}_{\uR\left(2n\right)}
  \log u \bigr\}u^{n},
  \label{eq:LH_fermion_eq}
\end{align}
in which $\lambda_{\uL\left(0\right)}$ and
$\lambda_{\uL\left(2\right)}$ are undetermined,
$\tilde{\lambda}_{\uL\left(0\right)}=0$ and for $n>0$,
\begin{align}
  0=&4n\left(n-1\right)\lambda_{\uL\left(2n\right)}
  +4\left(2n-1\right)\tilde{\lambda}_{\uL\left(2n\right)}
  -\kappa^{2}\lam_{\uL\left(2n-2\right)},\notag\\
  0=&4n\left(n-1\right)\tilde{\lambda}_{\uL\left(2n\right)}
  -\kappa^{2}\tilde{\lambda}_{\uL\left(2n-2\right)}.
\end{align}
Similarly $\lambda_{\uR\left(0\right)}$ and
$\tilde{\lambda}_{\uR\left(0\right)}$ are (for the moment) unfixed and
\begin{align}
  0=&4n^{2}\lambda_{\uR\left(2n\right)}
  +8n\tilde{\lambda}_{\uR\left(2n\right)}
  -\kappa^{2}{\lambda}_{\uR\left(2n-2\right)},\notag\\
  0=&4n^{2}\tilde{\lam}_{\uR\left(2n\right)}
  -\ka^{2}\tilde{\lambda}_{\uR\left(2n-2\right)}.
\end{align}
Writing $\lambda_{\uL}=u^{3/4}\chi_{\uL}$ and
$\lambda_{\uR}=u^{5/4}\chi_{\uR}$, the Dirac
equation~\eqref{eq:AdS_Dirac} gives the (not independent) relations
\begin{align}
  0=&2\chi_{\uL}'-\ka_{\mu}\sig^{\ul{\mu}}\delta_{\ul{\mu}}^{\ \mu}
  \bar{\chi}_{\uR},\notag\\
  0=&2u\bar{\chi}_{\uR}'+\ka_{\mu}\bar{\sig}^{\ul{\mu}}\delta_{\ul{\mu}}^{\ \mu}
  \chi_{\uL}.
\end{align}
Matching coefficients
\begin{align}
  0=&2n\lam_{\uL\left(2n\right)}+2\tilde{\lam}_{\uL\left(2n\right)}
  -\ka_{\mu}\delta_{\ul{\mu}}^{\ \mu}\sig^{\ul{\mu}}
  \bar{\lam}_{\uR\left(2n-2\right)},\notag\\
  0=&2n\lam_{\uL\left(2n\right)}
  -\ka_{\mu}\delta^{\ \mu}_{\ul{\mu}}\sig^{\ul{\mu}}
  \bar{\tilde{\lam}}_{\uR\left(2n-2\right)},\notag\\
  0=&2n\bar{\lam}_{\uR\left(2n\right)}
  +2\bar{\tilde{\lam}}_{\uR\left(2n\right)}
  +\ka_{\mu}\delta^{\ \mu}_{\ul{\mu}}
  \bar{\sig}^{\ul{\mu}}
  \lam_{\uL\left(2n\right)},
  \label{eq:Dirac_coefficients}\\
  0=&2n\bar{\tilde{\lam}}_{\uR\left(2n\right)}
  +\ka_{\mu}\delta^{\ \mu}_{\ul{\mu}}
  \bar{\sig}^{\ul{\mu}}
  \tilde{\lam}_{\uL\left(2n\right)}.\notag
\end{align}
The spinor current $j$ on the boundary couples to $\lambda_{\uL}$.  We have
\begin{equation}
  \lambda_{\uL}=\lambda_{\uL\left(0\right)}+\cO\bigl(u\bigr),
\end{equation}
and so $\lambda_{\uL\left(0\right)}$ is the source for the dual
current $j$ and $\lambda_{\uL\left(2\right)}$ is the response.

On-shell, the bulk part of the action~\eqref{eq:supplemented_Dirac}
vanishes, and so the regulated action comes only from the boundary
term
\begin{equation}
  S_{\mathrm{reg}}=
  \frac{1}{2}\int\frac{\ud^{4}k}{\left(2\pi\right)^{4}}\,
  \bigl\{\tilde{\lam}_{\uR\left(0\right)}\lam_{\uL\left(0\right)}
  +\bar{\tilde{\lam}}_{\uR\left(0\right)}\bar{\lam}_{\uL\left(0\right)}
  \bigr\}\log \ep +\cdots,
\end{equation}
where we have used the fact that $\tilde{\lam}_{\uL\left(0\right)}=0$.
Making use of~\eqref{eq:Dirac_coefficients}, this is
\begin{equation}
  S_{\mathrm{reg}}
  =-\frac{1}{2}\int\frac{\ud^{4}k}{\left(2\pi\right)^{4}}
  \bar{\lam}_{\uL\left(0\right)}\bar{\sig}^{\ul{\mu}}
  \delta^{\ \mu}_{\ul{\mu}}\ka_{\mu}\lam_{\uL\left(0\right)}\log \ep
  +\cdots.
\end{equation}
The divergences can be canceled by the counterterm
\begin{equation}
  S_{\mathrm{ct}}=-\frac{\ui L\log \ep }{2}
  \int_{u=\ep}\nn\nn\nn\nn\ud^{4}x\, \sqrt{h}
  \bar{\lam}\tilde{\ga}^{\mu}\tilde{\nabla}_{\mu}
 \frac{1}{2}\bigl(1-\ga_{\left(4\right)}\bigr)\lam.
\end{equation}

The response function is
\begin{equation}
  \bigl\langle j_{\alpha}\bigr\rangle
  =\frac{1}{\sqrt{\det\left(\eta_{\mu\nu}\right)}}
  \frac{\delta S_{\mathrm{ren}}}{\delta \lam_{\uL\left(0\right)}^{\alpha}}
  =\lim_{\ep\to 0}\frac{\ep^{3/4}}{\ep^{2}\sqrt{h}}
  \frac{\delta S_{\mathrm{sub}}}{\delta \lam_{\uL}^{\alpha}}.
\end{equation}
This gives
\begin{equation}
  \bigl<j_{\alpha}\bigr>_{\us}
  =\frac{1}{2}\lim_{\ep\to 0}
  \ep^{-5/4}\biggl\{
  \lam_{\uR\alpha}
  +\bar{\lam}_{\uL\dot{\alpha}}
  \frac{\delta\bar{\lam}_{\uR}^{\dot{\alpha}}}
  {\delta\lam_{\uL}^{\alpha}}
  +\ep^{1/2}\log \ep 
  \bar{\lam}_{\uL\dot{\beta}}\ka_{\mu}\delta^{\ \ul{\mu}}_{\ul{\mu}}
  \bar{\sig}^{\ul{\mu}\dot{\beta}\ga}\ep_{\ga\alpha}\biggr\},
\end{equation}
where we have used the fact that on shell, $\bar{\lam}_{\uR}$ and
$\lam_{\uL}$ are not independent and indeed at small $u$
\begin{equation}
  \bar{\lam}_{\uR}=-\frac{1}{2}u^{1/2}\log u 
  \ka_{\mu}\delta^{\ \mu}_{\ul{\mu}}
  \bar{\sig}^{\ul{\mu}}\lam_{\uL}+\cdots.
\end{equation}
Thus,
\begin{equation}
  \label{eq:fermionic_one_point}
  \bigl\langle j_{\alpha}\bigr\rangle_{\us}
  =\frac{1}{2}\lim_{\ep\to 0}\ep^{-5/4}\bigl\{\lam_{\uR\alpha}
  +\frac{1}{2}\ep^{1/2}\log \ep 
  \bar{\lam}_{\uL\dot{\beta}}\ka_{\mu}\delta^{\ \ul{\mu}}_{\ul{\mu}}
  \bar{\sig}^{\ul{\mu}\dot{\beta}\ga}\ep_{\ga\alpha}\bigr\}
  =\frac{1}{2}\lam_{\uR\alpha\left(0\right)}.
\end{equation}
Making use of~\eqref{eq:Dirac_coefficients}
and~\eqref{eq:LH_fermion_eq}, this gives
\begin{equation}
  \bigl\langle j_{\alpha}\bigr\rangle_{\us}
  =-\ka_{\mu}\delta^{\ {\mu}}_{\ul{\mu}}
  \sig^{\ul{\mu}}_{\alpha\dot{\alpha}}
  \biggl\{\frac{1}{\kappa^{2}}
  \bar{{\lam}}_{\uL\left(2\right)}^{\dot{\alpha}}
  +\frac{1}{4}\bar{{\lam}}_{\uL\left(0\right)}
  ^{\dot{\alpha}}\biggr\}.
\end{equation}

The two point functions are
\begin{align}
  \bigl\langle j_{\alpha}\bigl(k\bigr)
  j_{\beta}\bigl(-k\bigr)\bigr\rangle
  =&\left.\frac{\delta \left\langle j_{\alpha}\left(k\right)\right\rangle}
    {\delta \lambda^{\beta}_{\uL\left(0\right)}\left(-k\right)}
  \right\rvert_{\lam_{\uL\left(0\right)}=0},\notag\\
  \bigl\langle j_{\alpha}\bigl(k\bigr)
  \bar{j}_{\dot{\beta}}\bigl(-k\bigr)\bigr\rangle
  =&\left.\frac{\delta \left\langle j_{\alpha}\left(k\right)\right\rangle}
    {\delta \bar{\lambda}^{\dot{\beta}}_{\uL\left(0\right)}\left(-k\right)}
  \right\rvert_{\bar{\lam}_{\uL\left(0\right)}=0}.
\end{align}
So in this case,
\begin{equation}
  \label{eq:general_spinor_correlator}
  \bigl\langle j_{\alpha}\bigl(k\bigr)j_{\beta}\bigl(-k\bigr)\bigr\rangle
  =0,\qquad
  \bigl\langle j_{\alpha}\bigl(k\bigr)
  \bar{j}_{\dot{\beta}}\bigl(-k\bigr)\rangle
  \to-D_{1/2}\ka_{\mu}\delta^{\ \mu}_{\ul{\mu}}
  \sig^{\ul{\mu}}_{\alpha\dot{\alpha}}
  \biggl\{
  \frac{1}{\kappa^{2}}
  \frac{\delta\bar{{\lam}}^{\dot{\alpha}}_{\uL\left(2\right)}}
  {\delta\bar{\lam}^{\dot{\beta}}_{\uL\left(0\right)}}
  +\frac{1}{4}\delta^{\dot{\alpha}}_{\dot{\beta}}\biggr\}.
\end{equation}

In the $AdS^{5}$ case, we have the exact solution
\begin{equation}
  \lam_{\uL\alpha}=N_{\alpha}u^{5/4}I_{1}\bigl(\ka\sqrt{u}\bigr)
  +M_{\alpha}u^{5/4}K_{1}\bigl(\ka\sqrt{u}\bigr)
\end{equation}
Again imposing that $\lam_{\uL\to 0}$ as $u\to 0$ sets $N_{\alpha}=0$,
and so for small $u$
\begin{equation}
  \lam_{\uL\alpha}=M_{\alpha}u^{3/4}\biggl[\frac{1}{\kappa}
  +\frac{\kappa}{4}\big(\log \kappa^{2} +2\ga-\log 4
  -1\bigr)u
  +\frac{\kappa}{4}u\log u +\cdots\biggr],
\end{equation}
from which we read off
\begin{equation}
  \lam_{\uL\left(2\right)}=\frac{\kappa^{2}\lam_{\uL\left(0\right)}}{4}
  \biggl(\log \ka^{2} +2\ga-\log 4 -1\biggr),
\end{equation}
and so
\begin{equation}
  \bigl\langle j_{\alpha}\bigl(k\bigr)\bar{j}_{\dot{\beta}}\bigl(-k\bigr)
  \bigr\rangle
  =-\frac{D_{1/2}}{4L}k_{\mu}
  \delta^{\ \mu}_{\ul{\mu}}
  \sig^{\ul{\mu}}_{\alpha\dot{\beta}}
  \biggl(\log \ka^{2}+2\ga-\log 4 \biggr).
\end{equation}
The correlators satisfy the relations
\begin{align}
  \bigl\langle j_{\alpha}\bigl(k\bigr)j_{\beta}\bigl(-k\bigr)\bigr\rangle
  =&0\\
  \bigl\langle j_{\alpha}\bigl(k\bigr)\bar{j}_{\dot{\beta}}\bigl(-k\bigr)
  \bigr\rangle
  \propto&
  -k_{\mu}
  \delta^{\ \mu}_{\ul{\mu}}
  \sig^{\ul{\mu}}_{\alpha\dot{\beta}}
  \bigl\langle J\bigl(k\bigr)J\bigl(-k\bigr)\bigr\rangle,
\end{align}
as again expected from SUSY.
  
\subsection{Massive messengers}

The messengers can be made massive by taking $\mu>0$.  In this case,
it turns out to be easier to work with the $\rho$ coordinate.  This
system was considered also in~\cite{Kruczenski:2003be} and we follow
that work closely, but here we go beyond just determination of the
spectrum to find the full two-point functions. We have
\begin{equation}
  A=\frac{1}{2}\log\biggl(\frac{\rho^{2}+\mu^{2}}{L^{2}}\biggr),\quad
  B=\frac{1}{2}\log\biggl(\frac{\rho^{2}+\mu^{2}}{\rho^{2}}\biggr).
\end{equation}

From~\eqref{eq:on_shell_action}, the
equation of motion for the scalar is
\begin{align}
  0=&
  \frac{\ue^{3B}}{\sqrt{\tilde{g}}}\bigl[\sqrt{\tilde{g}}\ue^{-3B}\tilde{g}^{mn}
  \partial_{m}\Sig\bigr]-m_{\Sig}^{2}\Sig\notag\\
  \label{eq:general_scalar_eom}
  =&\ue^{-2A+4B}{\rho}
  \partial_{\rho}\bigl[\ue^{4A-2B}{\rho}\partial_{\rho}\Sig\bigr]
  -\ue^{2A}L^{2}m_{\Sig}^{2}\Sig-\kappa^{2}\Sig.
\end{align}
Writing $\rho=\mu p$,
\begin{equation}
  0=\frac{1+p^{2}}{p^{3}}\frac{\ud}{\ud p}
  \biggl[p^{3}\bigl(1+p^{2}\bigr)\frac{\ud\Sig}{\ud p}\biggr]
  -\nu^{2}\Sig
  -\frac{3-2p^{2}-4p^{4}}{p^{2}}\Sig,
\end{equation}
in which $\nu^{2}=L^{4}k^{2}/\mu^{2}$.  Taking
\begin{equation}
  \Sig=p^{m}\bigl(1+p^{2}\bigr)^{n}P\bigl(p\bigr),
\end{equation}
This equation becomes
\begin{multline}
  0=\bigl(1+p^{2}\bigr)
  \frac{\ud^{2}P}{\ud p^{2}}
  +\frac{1}{p}
  \bigl[\bigl(5+2m+4n\bigr)p^{2}
  +\bigl(3+2m\bigr)\bigr]
  \frac{\ud P}{\ud p}\\
  +\frac{1}{p^{2}\left(1+p^{2}\right)}
  \bigl[\bigl(2+m+2n\bigr)^{2}p^{4}
  +\bigl(2+8n+2m\bigl(3+m+2n\bigr)-\nu^{2}\bigr)p^{2}
  +\bigl(m+3\bigr)\bigl(m-1\bigr)\bigr]P.
\end{multline}
Defining $y=-p^{2}$,
\begin{multline}
  0=y\bigl(1-y\bigr)\frac{\ud^{2}P}{\ud y^{2}}
  +\bigl[-\bigl(3+m+2n\bigr)y+\bigl(2+m\bigr)\bigr]
  \frac{\ud P}{\ud y}\\
  -\frac{1}{4\left(y-1\right)}
  \bigl[\bigl(2+m+2n\bigr)^{2}y
  -\bigl(2+8n+2m\bigl(3+m+2n\bigr)-\nu^{2}\bigr)
  +\frac{1}{y}\bigl(3+m\bigr)\bigl(m-1\bigr)\bigr]P.
\end{multline}
This can be cast as a hypergeometric differential equation
\begin{equation}
  0=y\bigl(y-1\bigr)\frac{\ud^{2} P}{\ud y^{2}}
  +\bigl[-\bigl(a+b+1\bigr)y+c\bigr]\frac{\ud P}{\ud y}
  -ab\,P,
\end{equation}
by taking
\begin{equation}
  m=1,\quad n=-\frac{1}{2}\sqrt{1-\nu^{2}}=:\eta.
\end{equation}
Then the full solution is
\begin{align}
  \Sig = p\bigl(1+p^{2}\bigr)^{\eta} \biggl\{&
  M_{0}\,F\biggl(\frac{3}{2}+\eta, \frac{3}{2}+\eta;
  3;-p^{2}\biggr)
  + N_{0}\, p^{-4}F\biggl(-\frac{1}{2}+\eta,-\frac{1}{2}+\eta;
  -1;-p^{2}\biggr) \biggr\},
\end{align}
in which $F\bigl(a,b;c;y\bigr)$ is the hypergeometric function.
Demanding that $\Sig$ be regular at $y=0$ sets $N_{0}=0$.  The
solution to~\eqref{eq:define_u}
is\footnote{\label{foot:normalization}There is an overall
  multiplicative factor arising as an integration constant.  A choice
  different from the one here would change the leading term of
  $\ue^{2A}$ by a multiplicative constant, and the result of
  section~\ref{eq:massless_susy} would have to be re-performed.  The
  end of result of course would be the same.  With this choice of the
  integration constant, naive application of the expressions for the
  dual stress-energy tensor presented
  in~\cite{deHaro:2000xn,*Skenderis:2000in} would imply that
  $\left\langle T_{\mu\nu}\right\rangle\neq 0$.  However, since the
  analysis leading to those results makes use of the Einstein
  equations for the 5d metric and the metric above doesn't satisfy
  such equations (that is, the pullback of the 10d metric onto the
  worldvolume of course satisfies the pullback of the Einstein
  equations, but does not satisfy Einstein equations built from the
  pullback metric alone since generally the pullback of curvature
  tensors are different from the curvature of the pullback of the
  metric), the analysis does not apply.}
\begin{equation}
  \label{eq:massive_u}
  \rho=\frac{L}{\sqrt{u}}\biggl(1-\frac{\mu^{2}u}{4L^{2}}\biggr),\quad
  u\in\bigl[0,4L^{2}/\mu^{2}\bigr],
\end{equation}
and a small $u$ expansion gives
\begin{equation}
  \Sig=\frac{-8 M_{0} \mu^{2}}{L^{2}\pi\nu^{2}}
  \cos\bigl(\pi\eta\bigr)u
  \biggl[
  \psi\biggl(\frac{3}{2}+\eta\biggr)
  +\psi\biggl(\frac{3}{2}-\eta\biggr)
  +\log\biggl(\frac{\mu^{2}}{L^{2}}\biggr)
  +2\ga
  +\log u \biggr]+\cdots,
\end{equation}
in which $\psi$ is the digamma function.  This gives the correlator
function (c.f.~\eqref{eq:correlator_functions}).
\begin{equation}
  \label{eq:massive_correlator_function}
  C_{0}=\frac{2D_{0}}{L}
  \biggl[\psi\biggl(\frac{3}{2}+\eta\biggr)+\psi\biggl(\frac{3}{2}-\eta\biggr)
  +\log\biggl(\frac{\mu^{2}}{L^{2}}\biggr)+2\ga\biggr],
\end{equation}
in which again
\begin{equation}
  \nu^{2}=\frac{L^{4}k^{2}}{\mu^{2}},\quad
  \eta=-\frac{1}{2}\sqrt{1-\nu^{2}}.
\end{equation}
In the limit where $\mu\to 0$, this agrees with the result in the
conformal case.

The digamma functions have poles corresponding to resonances located
at
\begin{equation}
  k^{2}=-\frac{4\mu^{2}}{L^{4}}\bigl(\ell+1\bigr)\bigl(\ell+2\bigr),\quad
  \ell\in\mathbb{N},
\end{equation}
agreeing with the analysis of~\cite{Kruczenski:2003be}.

For the vectors, the equation of motion is
\begin{equation}
  0=\frac{\ue^{3B}}{\sqrt{\tilde{g}}}
  \partial_{m}\bigl[\sqrt{\tilde{g}}\ue^{-3B}\tilde{g}^{mn}\tilde{g}^{st}
  F_{nt}\bigr].
\end{equation}
So,
\begin{align}
  0=&\ue^{4B}\rho\,\partial_{\rho}\bigl[\ue^{2A-2B}
  \rho\bigl(\partial_{\rho}A_{\mu}-\partial_{\mu}A_{\rho}\bigr)\bigr]
  +L^{2}\partial^{2}A_{\mu}-L^{2}\partial_{\mu}\bigl(\partial\cdot A\bigr),\\
  0=&\partial^{2}A_{\rho}-\partial_{\rho}\bigl(\partial\cdot A\bigr)\notag.
\end{align}

The Christoffel symbols are
\begin{equation}
  \tilde{\Ga}^{\rho}_{\phantom{\rho}\rho\rho}=-\frac{1}{\rho}
  \bigl(1+\rho B'\bigr),\quad
  \tilde{\Ga}^{\rho}_{\phantom{\rho}\mu\nu}
  =-\frac{\rho^{2}A'}{L^{2}}\ue^{2A+2B}\eta_{\mu\nu},\quad
  \tilde{\Ga}^{\mu}_{\phantom{\mu}\rho\nu}
  =A'\delta^{\mu}_{\nu},
\end{equation}
and so the gauge-fixing condition becomes
\begin{equation}
  0=\tilde{g}^{mn}\tilde{\nabla}_{m}A_{n}
  =\ue^{-2A}\partial\cdot A
  +\frac{\rho}{L^{2}}\ue^{2B}
  \bigl(4\rho A'+\rho B'+1\bigr)A_{\rho}
  +\frac{\rho^{2}}{L^{2}}\ue^{2B}\partial_{\rho}A_{\rho}.
\end{equation}
With this condition, the equation of motion for $A_{\mu}$ is
\begin{equation}
  \label{eq:external_eq}
  -\rho\,\ue^{2A+2B}\bigl(2\rho A'+3\rho B'\bigr)\ui k_{\mu}A_{\rho}=
  \rho\,\ue^{4B}\partial_{\rho}\bigl[\rho\,\ue^{2A-2B}\partial_{\rho}A_{\mu}\bigr]
  -\ka^{2}A_{\mu},
\end{equation}
and that for $A_{\rho}$ is
\begin{multline}
  0=\rho^{2}\ue^{2A+2B}\partial_{\rho}^{2}A_{\rho}
  +\bigl(6\rho A'+3\rho B'+3\bigr)\ue^{2A+2B}\rho\partial_{\rho}A_{\rho}\\
  +\bigl(4\rho^{2} A''+\rho^{2}B''
  +8\rho^{2}A'^{2}+2\rho^{2}B'^{2}
  +10\rho^{2}A'B'
  +10\rho A'+4\rho B'+1\bigr)\ue^{2A+2B}A_{\rho}.
\end{multline}
The homogeneous part of~\eqref{eq:external_eq} is
\begin{equation}
  0=\frac{\bigl(1+p^{2}\bigr)^{2}}{p^{3}}
  \frac{\ud}{\ud p}\biggl[p^{3}\frac{\ud A_{\mu}}{\ud p}\biggr]
  -\nu^{2}A_{\mu},
\end{equation}
the general solution to which is
\begin{equation}
  A_{\mu}=\bigl(1+p^{2}\bigr)^{\frac{1}{2}+\eta}\biggl\{
  \biggl(M_{\mu}F\biggl(\frac{3}{2}+\eta,\frac{1}{2}+\eta;2;-p^{2}\biggr)
  +N_{\mu}\,p^{-2}F\biggl(\frac{1}{2}-\eta,-\frac{1}{2}-\eta;0;-p^{2}\biggr)
  \biggr\}.
\end{equation}
Regularity imposes $N_{\mu}=0$ and at small $u$,
\begin{align}
  A_{\mu}=\frac{M_{\mu}\cos\bigl(\pi\eta\bigr)}{\pi\nu^{2}}
  \biggl\{4+\frac{\mu^{2}\nu^{2}}{L^{2}}
  \biggl[\psi\biggl(\frac{3}{2}+\eta\biggr)
  +\psi\biggl(\frac{3}{2}-\eta\biggr)+
  \log\biggl(\frac{\mu^{2}}{L^{2}}\biggr)+2\ga-1\biggr]u
  \biggr\}+\cdots.
\end{align}
From this and~\eqref{eq:general_vector_two_point}, we find
$C_{1}=C_{0}$ where $C_{0}$ is as in~\eqref{eq:massive_correlator_function}.

The Dirac equation is
\begin{align}
  0=&\bigl(\tilde{\slashed{\nabla}}+m_{\lam}
  +\alpha\ga_{\left(4\right)}\bigr)\lam \\
  0=&\biggl\{\ue^{-A}\tilde{\ga}^{\ul{\mu}}\delta_{\ul{\mu}}^{\mu}\partial_{\mu}
  +\frac{\rho}{L}\ue^{B}\ga_{\left(4\right)}\partial_{\rho}+
  \biggl(\alpha+\frac{2A'\rho}{L}\ue^{B}\biggr)\ga_{\left(4\right)}
  +m_{\lam}\biggr\}\lam.
\end{align}
Again applying $\tilde{\ga}^{m}\partial_{m}$,
\begin{align}
  &0=\rho^{2}\ue^{2B+2A}\partial_{\rho}^{2}\lam
  +\bigl(5\rho A'+\rho B'+1+2\alpha L\ue^{-B}\bigr)\rho\,\ue^{2B+2A}
  \partial_{\rho}\lam\notag\\
  &+\bigl[2\rho^{2}\bigl(A''+3A'^{2}+A'B'\bigr)
  +\rho\bigl(2A'
  +\alpha' L\ue^{-B}
  +5 A'\alpha L\ue^{-B}\bigr)
  +\bigl(\alpha^{2}-m_{\lam}^{2}\bigr)L^{2}\ue^{-2B}
  \bigr]\ue^{2B+2A}\lam\notag\\
  &+\bigl(\rho A' m_{\lam}L+\rho m_{\lam}'L\bigr)\ue^{B+2A}\ga_{\left(4\right)}
  \lam
  -\ka^{2}\lam.
\end{align}
For the left-handed spinor,
\begin{equation}
  0=\bigl(1+p^{2}\bigr)^{2}\frac{\ud^{2}\lam_{\uL}}{\ud p^{2}}
  +\frac{3\left(1+p^{2}\right)(1+2p^{2}\bigr)}{p}
  \frac{\ud\lam_{\uL}}{\ud p}
  +\frac{3\left(8+7p^{2}\right)}{4}\lam_{\uL}-\nu^{2}\lam_{\uL}.
\end{equation}
This is solved by
\begin{equation}
  \lam_{\uL}=\bigl(1+p^{2}\bigr)^{\eta-\frac{1}{4}}
  \biggl\{M_{\uL}\,F\biggl(\frac{3}{2}+\eta,\frac{1}{2}+\eta;2;-p^{2}\biggr)
  +N_{\uL}\,p^{-2}F\biggl(\frac{1}{2}-\eta,-\frac{1}{2}-\eta;0;-p^{2}\biggr)
  \biggr\}.
\end{equation}
Regularity again imposes that $N_{\uL}=0$, and then at small $u$,
\begin{align}
  \lam_{\uL}=\frac{M_{\uL}\mu^{3/2}\cos\bigl(\pi\eta\bigr)u^{3/4}}{
    \pi\nu^{2}L^{3/2}}
  \biggl\{4+\frac{\mu^{2}\nu^{2}}{L^{2}}
  \biggl[\psi\biggl(\frac{3}{2}+\eta\biggr)
  +\psi\biggl(\frac{3}{2}-\eta\biggr)+
  \log\biggl(\frac{\mu^{2}}{L^{2}}\biggr)+2\ga-1\biggr]u
  \biggr\}+\cdots,
\end{align}
and so, using~\eqref{eq:general_spinor_correlator}, we get
$C_{1/2}=C_{1}=C_{0}$ and $B_{1/2}=0$.

\section{\label{sec:non_SUSY_currents}Non-supersymmetric correlators}

We now turn to the analysis of non-supersymmetric cases which is our
main interest in this work.  In particular, we consider a normalizable
non-supersymmetric perturbation to the above geometry and recalculate
the correlators of section~\ref{sec:SUSY_currents} in this perturbed
background.  As discussed in section~\ref{sec:geometry}, these classical
two-point functions correspond to the current-current correlators of
the hidden sector after the latter has obtained a non-supersymmetric
state and so calculation of these functions is tantamount to
calculation of visible-sector soft terms resulting from the gauge
mediation of supersymmetry breaking.

Inspired by the well-studied case in
Klebanov-Strassler~\cite{DeWolfe:2008zy, Bena:2009xk,
  McGuirk:2009xx,*Bena:2010ze,*Dymarsky:2011pm,*Bena:2011wh}, the toy
case that we consider is the addition of $p$ $\uD 3$-$\overline{\uD
  3}$ pairs to $AdS^{5}\times X^{5}$.  The geometry was considered
in~\cite{DeWolfe:2008zy} and (as argued in~\cite{DeWolfe:2008zy}) can
be found by taking a near-horizon limit of geometries considered
in~\cite{Zhou:1999nm,*Brax:2000cf}.  In our notation, the solution is
\begin{align}
  \label{eq:DKM}
  A=\log\biggl(\frac{r}{L}\biggr)-\frac{L^{8}\cS}{5r^{8}},\quad
  \bar{B}=\log\biggl(\frac{r}{L}\biggr)-\frac{L^{8}\cS}{10r^{8}},\quad
  C=\log\biggl(\frac{r}{L}\biggr)+\frac{3L^{8}\cS}{10 r^{8}},
\end{align}
in which we have taken $\cS=\frac{p}{N}$ to be a small parameter.  It
was pointed out in~\cite{DeWolfe:2008zy} that the perturbations are
such that $\left\langle T_{\mu\nu}\right\rangle=0$ and so the dual
supersymmetry, which is rigid, is unbroken.  This should be reflected
in the correlators resulting from this background.  Nevertheless, we
find below that the messenger spectrum is split and therefore the
correlators do not respect the relationships expected from
supersymmetry.  Presumably, this is related to the fact that a finite
messenger mass spoils the conformal behavior of the dual theory and
taking into account the backreaction of the $\uD 7$s should result in
a finite vacuum energy.

To leading order in $\cS$, the equation of motion for the scalar field
takes the form
\begin{equation}
  \label{eq:DKM_eom}
  0=\biggl\{
  \frac{1+p^{2}}{p^{3}}\frac{\ud}{\ud p}
  \biggl[p^{3}\bigl(1+p^{2}\bigr)
  \biggl(1-\frac{\delta}{\left(1+p^{2}\right)^{4}}\biggr)
  \frac{\ud}{\ud p}\biggr]
  -\frac{3-2p^{2}-4p^{4}}{p^{2}}
  \biggl(1-\frac{\delta}{\left(1+p^{2}\right)^{4}}\biggr)
  -\nu^{2}\biggr\}\Sig,
\end{equation}
in which
\begin{equation}
  \delta=\frac{3L^{8}}{5\mu^{8}}\cS.
\end{equation}
This is a Sturm-Liouville problem and so for a set of boundary
conditions the solutions will be orthogonal with respect to the inner
product
\begin{equation}
  \label{eq:scalar_inner_product}
  \bigl(\Sig_{\ell},\Sig_{\ell'}\bigr)_{0}
  =\int_{0}^{\infty}\ud p\,\frac{p^{3}}{1+p^{2}}
  \Sig_{\ell}\Sig_{\ell'}\propto\delta_{\ell\ell'}.
\end{equation}
Instead of attempting to solve~\eqref{eq:DKM_eom} directly, we can
apply perturbation theory.  When $\delta=0$,~\eqref{eq:DKM_eom} is
again a Sturm-Liouville problem and so the solutions are orthogonal
with respect to the same inner product.  The solutions that are
regular were found in the last section, and the resulting correlation
function~\eqref{eq:massive_correlator_function} can be written as
\begin{equation}
  \label{eq:massive_correlator_resonances}
  C_{0}=\frac{2D_{0}}{L}
  \sum_{\ell=0}^{\infty}\frac{-4\left(3+2\ell\right)}{\nu^{2}+4\left(\ell+1\right)
    \left(\ell+2\right)}+\cdots,
\end{equation}
where we have omitted the contact terms that were specified by
holographic renormalization.  The simple structure of this correlator
is a consequence of the fact that the messengers form mesonic bound
states which are free in the large $N$ limit~\cite{'tHooft:1973jz,
  *'tHooft:1974hx,*Witten:1979kh}.

The correlator in the perturbed geometry should be
expressible in a similar way
\begin{equation}
  C_{0}=\frac{2D_{0}}{L}\sum_{\ell=0}^{\infty}\frac{Z_{\ell}}{\nu^{2}+
    {L^{4}m_{\ell}^{2}}/{\mu^{2}}}
  +\cdots.
\end{equation}
The precise form of this requires the precise solution which we will
not attempt to find.  However, we can find the spectrum perturbatively
by writing the equation of motion~\eqref{eq:DKM_eom} for the modes
corresponding to these poles as
\begin{equation}
  \cH_{0}\Sig_{\ell}=\lam_{\ell}\Sig_{\ell},
\end{equation}
with $\lam_{\ell}=-L^{4}m_{\ell}^{2}/\mu^{2}$.
When $\delta=0$,
\begin{equation}
  \cH_{0}=\cH_{0}^{\left(0\right)}=
  \frac{1+p^{2}}{p^{3}}\frac{\ud}{\ud p}\biggl[p^{3}\bigl(1+p^{2}\bigr)
  \frac{\ud}{\ud p}\biggr]
  -\frac{3-2p^{2}-4p^{4}}{p^{3}}.
\end{equation}
For $\delta\neq 0$, we write
$\cH_{0}=\cH_{0}^{\left(0\right)}+\delta\cH_{0}^{\left(1\right)}$ with
\begin{equation}
  \cH_{0}^{\left(1\right)}=\frac{8p}{\left(1+p^{2}\right)^{3}}\frac{\ud}{\ud p}
  -\frac{1}{\left(1+p^{2}\right)^{4}}\cH^{\left(0\right)}_{0}.
\end{equation}
The unperturbed spectrum is
$\lam^{\left(0\right)}_{\ell}=-4\left(\ell+1\right)\left(\ell+2\right)$
and the corresponding eigenfunctions,
\begin{equation}
  \Sig_{\ell}^{\left(0\right)}=M_{\ell}\,p\bigl(1+p^{2}\bigr)^{-\frac{3}{2}-\ell}
  F\bigl(-\ell,-\ell;3;-p^{2}\bigr),
\end{equation}
form a complete set of functions that vanish as $p\to\infty$ and, by
appropriate choice of $M_{\ell}$, are orthonormal with respect
to~\eqref{eq:scalar_inner_product}.  We similarly expand
$\lam_{\ell}^{\left(0\right)}+\delta\lambda^{\left(1\right)}_{\ell}$
and
$\Sig_{\ell}=\Sig_{\ell}^{\left(0\right)}+\delta\Sig_{\ell}^{\left(1\right)}$
with
\begin{equation}
  \Sig_{\ell}^{\left(1\right)}=\sum_{\ell'}c_{\ell\ell'}\Sig_{\ell'}^{\left(0\right)}.
\end{equation}
Demanding that $\Sig_{\ell}$ is normalized sets $c_{\ell\ell}=0$.  It
follows then
\begin{equation}
  \lambda_{\ell}^{\left(1\right)}=
  {\left(\Sig_{\ell}^{\left(0\right)},\cH_{0}^{\left(1\right)}
    \Sig_{\ell}^{\left(0\right)}\right)_{0}}.
\end{equation}
Since $\ell$ is an integer, the solutions are a polynomial of order
$\ell$ and so these integrals can be easily performed.  The resulting
spectrum appears in table~\ref{table:spectrum}.  Meanwhile, for
$\ell'\neq\ell$,
\begin{equation}
  c_{\ell\ell'}=\frac{\left(\Sig_{\ell'}^{\left(0\right)},\cH_{0}^{\left(1\right)}
      \Sig_{\ell}^{\left(0\right)}\right)_{0}}
  {\lam_{\ell}^{\left(0\right)}-\lam_{\ell'}^{\left(0\right)}}.
\end{equation}

The homogeneous equation for $A_{\mu}=a_{\mu}\cA$ takes the form
\begin{equation}
  0=\biggl\{\frac{\left(1+p^{2}\right)^{2}}{p^{3}}
  \biggl(1-\frac{2}{3}\frac{\delta}{\left(1+p^{2}\right)^{4}}\biggr)
  \frac{\ud}{\ud p}\biggl[
  p^{3}\biggl(1-\frac{1}{3}\frac{\delta}{\left(1+p^{2}\right)^{4}}\biggr)
  \frac{\ud}{\ud p}\biggr]-\nu^{2}\biggr\}\cA.
\end{equation}
Here the appropriate inner product is
\begin{equation}
  \bigl(\cA_{\ell},\cA_{\ell'}\bigr)_{1}
  =\int_{0}^{\infty}\ud p\frac{p^{3}}{\left(1+p^{2}\right)^{2}}
  \biggl(1+\frac{2}{3}\frac{\delta}{\left(1+p^{2}\right)^{4}}\biggr)
  \cA_{\ell}\cA_{\ell'}.
\end{equation}
The equation of motion takes the form
$\cH_{1}\cA_{\ell}=\lam_{\ell}\cA_{\ell}$ with
\begin{align}
  \cH_{1}=&\cH_{1}^{\left(0\right)}+\delta\cH_{1}^{\left(1\right)},\notag\\
  \cH_{1}^{\left(0\right)}=&\frac{\left(1+p^{2}\right)^{2}}{p^{3}}
  \frac{\ud}{\ud p}
  \biggl[p^{3}\frac{\ud}{\ud p}\biggr],\\
  \cH_{1}^{\left(1\right)}=&
  \frac{8p}{3\left(1+p^{2}\right)^{3}}\frac{\ud}{\ud p}
  -\frac{1}{\left(1+p^{2}\right)^{4}}\cH_{1}^{\left(0\right)}.\notag
\end{align}
The solutions to the $\delta=0$ equation are
\begin{equation}
  \cA_{\ell}^{\left(0\right)}
  =M_{\ell}\,\bigl(1+p^{2}\bigr)^{-1-n}
  F\bigl(-n,-1-n;2;-p^{2}\bigr).
\end{equation}
These are orthonormal with respect to
\begin{equation}
  \bigl\langle\cA_{\ell}^{\left(0\right)},\cA^{\left(0\right)}_{\ell'}\bigr\rangle_{1}
  =\int_{0}^{\infty}\ud p\,\frac{p^{3}}{\left(1+p^{2}\right)^{2}}
  \cA_{\ell}\cA_{\ell'}.
\end{equation}
Write
$\cA_{\ell}=\cA_{\ell}^{\left(0\right)}+\delta\cA_{\ell}^{\left(1\right)}$ with
\begin{equation}
  \cA_{\ell}^{\left(1\right)}=\sum_{\ell'}c_{\ell\ell'}\cA_{\ell'},
\end{equation}
then imposing $\bigl(\cA_{\ell},\cA_{\ell}\bigr)_{1}=1$ sets
\begin{equation}
  c_{\ell\ell}=-\frac{1}{3}\int_{0}^{\infty}\ud p\,
  \frac{p^{3}}{\left(1+p^{2}\right)^{2}}
  \frac{1}{\left(1+p^{2}\right)^{4}}
  \cA_{\ell}^{\left(0\right)}\cA_{\ell}^{\left(0\right)}.
\end{equation}
While as with the scalar,
\begin{equation}
  \lam_{\ell}^{\left(1\right)}=
  \bigl\langle\cA_{\ell}^{\left(0\right)},\cH_{1}^{\left(1\right)}
  \cA_{\ell}^{\left(0\right)}\bigr\rangle_{1},
\end{equation}
and for $\ell'\neq\ell$,
\begin{equation}
  c_{\ell\ell'}=\frac{\bigl\langle
    \cA_{\ell'}^{\left(0\right)},\cH_{1}^{\left(0\right)}
    \cA_{\ell}^{\left(0\right)}\bigr\rangle_{1}}
  {\lam_{\ell}^{\left(0\right)}-\lam_{\ell'}^{\left(0\right)}}.
\end{equation}

Writing $\lam^{\alpha}_{\uL}=a_{\uL}^{\alpha}\psi$, the equation of
motion for the spinor takes the form
\begin{equation}
  \label{eq:spinor_eom}
  0=\biggl\{f\bigl(p\bigr)\frac{\ud^{2}}{\ud p^{2}}
  +g\bigl(p\bigr)\frac{\ud}{\ud p}
  +h\bigl(p\bigr)
  -\nu^{2}\biggr\}\psi,
\end{equation}
with
\begin{align}
  f\bigl(p\bigr)=&\bigl(1+p^{2}\bigr)^{2}
  \biggl(1-\frac{\delta}{\left(1+p^{2}\right)^{4}}\biggr),\notag\\
  g\bigl(p\bigr)=&\frac{3\left(1+p^{2}\right)\left(1+2p^{2}\right)}{p}
  \biggl(1+\frac{\delta\left(-9+14p^{2}\right)}{9\left(1+p^{2}\right)^{4}
    \left(1+2p^{2}\right)}\biggr),\\
  h\bigl(g\bigr)=&\frac{3\left(8+7p^{2}\right)}{4}
  \biggl(1+\frac{7\delta\left(8-9p^{2}\right)}
  {9\left(1+p^{2}\right)^{4}\left(8+7p^{2}\right)}\biggr).
\end{align}
We can cast~\eqref{eq:spinor_eom} as a Sturm-Liouville problem
\begin{equation}
  0=\biggl\{f\ue^{-\zeta}\frac{\ud}{\ud p}
  \biggl[\ue^{\zeta}\frac{\ud}{\ud p}\biggr]
  +h-\nu^{2}\biggr\}\psi,
\end{equation}
in which
\begin{equation}
  \zeta\bigl(p\bigr)=\int^{p}\ud p'\,\frac{g\left(p'\right)}{f\left(p'\right)}.
\end{equation}
Then,
\begin{multline}
  0=\biggl\{\frac{\sqrt{1+p^{2}}}{p^{3}}
  \biggl(1+\frac{\delta}{3\left(1+p^{2}\right)^{4}}\biggr)
  \frac{\ud}{\ud p}\biggl[p^{3}\bigl(1+p^{2}\bigr)^{3/2}
  \biggl(1-\frac{4}{3}\frac{\delta}{\left(1+p^{2}\right)^{4}}\biggr)
  \frac{\ud}{\ud p}\biggr]\\
  +\frac{3\left(8+7p^{2}\right)}{4}
  \biggl(1+\frac{7}{9}\frac{\delta\left(8-9p^{2}\right)}
  {\left(1+p^{2}\right)^{4}\left(8+7p^{2}\right)}\biggr)
  -\nu^{2}\biggr\}\psi.
\end{multline}
The inner product is
\begin{equation}
  \bigl(\psi_{\ell},\psi_{\ell'}\bigr)_{1/2}=
  \int_{0}^{\infty}\ud p\,\frac{p^{3}}{\sqrt{1+p^{2}}}
  \biggl(1-\frac{\delta}{3\left(1+p^{2}\right)^{4}}\biggr)
  \psi_{\ell}\psi_{\ell'}.
\end{equation}
The equation of motion is $\cH_{1/2}\psi_{\ell}=\lam_{\ell}\psi_{\ell}$ with
\begin{align}
  \cH_{1/2}=&\cH_{1/2}^{\left(0\right)}+\delta\cH_{1}^{\left(1\right)},\notag\\
  \cH_{1/2}^{\left(0\right)}=&
  \frac{\sqrt{1+p^{2}}}{p^{3}}
  \frac{\ud}{\ud p}\biggl[p^{3}\bigl(1+p^{2}\bigr)^{3/2}
  \frac{\ud}{\ud p}\biggr]
  +\frac{3\left(8+7p^{2}\right)}{4},\\
  \cH_{1/2}^{\left(1\right)}
  =&\frac{32p}{3\left(1+p^{2}\right)^{3}}
  \frac{\ud}{\ud p}
  +\frac{7\left(8-9p^{2}\right)}{12\left(1+p^{2}\right)^{4}}
  -\frac{1}{\left(1+p^{2}\right)^{4}}\cH_{1/2}^{\left(0\right)}.\notag
\end{align}
For $\delta=0$, the solutions take the form
\begin{equation}
  \psi_{\ell}^{\left(0\right)}=M_{\ell}
  \left(1+p^{2}\right)^{-7/4-\ell}F\bigl(-\ell,-1-\ell;2;-p^{2}\bigr),
\end{equation}
and are orthonormal with respect to the inner product
\begin{equation}
  \bigl\langle
  \psi_{\ell}^{\left(0\right)},\psi_{\ell'}^{\left(0\right)}\bigr\rangle_{1/2}=
  \int_{0}^{\infty}\ud p\,\frac{p^{3}}{\sqrt{1+p^{2}}}
  \psi_{\ell}\psi_{\ell'}
\end{equation}
Writing again
\begin{equation}
  \psi_{\ell}^{\left(1\right)}=\sum_{\ell'}c_{\ell\ell'}\psi_{\ell'}^{\left(0\right)},
\end{equation}
and imposing $\left(\psi_{\ell},\psi_{\ell}\right)_{1/2}=1$ sets
\begin{equation}
  c_{\ell\ell}=
  \frac{1}{6}\int_{0}^{\infty}\ud p\,\frac{p^{3}}{\sqrt{1+p^{2}}}
  \frac{1}{\left(1+p^{2}\right)^{4}}
  \psi_{\ell}^{\left(0\right)}\psi_{\ell}^{\left(0\right)}.
\end{equation}
Once again the perturbations to the masses are
\begin{equation}
  \lam_{\ell}^{\left(1\right)}=
  \left\langle\psi_{\ell}^{\left(0\right)},\cH_{1/2}^{\left(1\right)}
    \psi_{\ell}^{\left(0\right)}\right\rangle_{1/2},
\end{equation}
while for $\ell\neq\ell'$,
\begin{equation}
  c_{\ell\ell'}=
  \frac{\bigl\langle
    \psi_{\ell'}^{\left(0\right)},\cH_{1/2}^{\left(0\right)}
    \psi_{\ell}^{\left(0\right)}\bigr\rangle_{1/2}}
  {\lam_{\ell}^{\left(0\right)}-\lam_{\ell'}^{\left(0\right)}}.
\end{equation}

The residues $Z_{\ell}$ can be obtained as follows (see~\cite{Berg:2006xy}).
At large $p$, the solution for the scalar mode takes the
form\footnote{Note that this takes a different from
  from~\eqref{eq:general_asymptote} since here we have $\Delta=2$.}
\begin{equation}
  \Sig=\sig_{1}p^{-2}+\sig_{2}p^{-2}\log p +\cdots.
\end{equation}
When the $4$-momentum is on a resonance, $\sig_{2}=0$ which can be seen
by expanding~\eqref{eq:massive_correlator_function} for
$\nu^{2}=-4\left(\ell+1\right)\left(\ell+2\right)$.  Then, when
$\cS=0$ and the solution is normalized according
to~\eqref{eq:scalar_inner_product}, we have
\begin{equation}
  Z_{\ell}\sim \sig_{\ell,1}^{2}.
\end{equation}
As a check, the normalized $\ell=0$ solution is
\begin{equation}
  \Sig_{0}^{\left(0\right)}=\frac{\sqrt{6}p}{\left(1+p^{2}\right)^{3/2}}
  =\frac{\sqrt{6}}{p^{2}}+\cdots,
\end{equation}
comparing to~\eqref{eq:massive_correlator_resonances} $Z_{\ell}$, we
get
\begin{equation}
  Z_{\ell}=-2\sig_{\ell,1}^{2}.
\end{equation}
It is straightforward to check that this holds for higher $\ell$ as
well\footnote{Alternatively, since the resonant solutions are
  polynomials of finite order, this should be possible to check
  for general $\ell$.  Note that the relationship presented
  in~\cite{Berg:2006xy} between the residues $Z_{\ell}$ and the
  coefficient of the sub-dominant solutions must be modified for
  scalars satisfying the BF bound.}.  When $\delta>0$, we have
$\sig_{\ell,1}=\sig_{\ell,1}^{\left(0\right)}+\delta\sig_{\ell,1}^{\left(1\right)}$ with
\begin{equation}
  \sig_{\ell,1}^{\left(1\right)}
  =\sum_{\ell'}c_{\ell\ell'}\sig_{\ell}^{\left(0\right)}.
\end{equation}
The residue is then $Z_{\ell}=Z_{\ell}^{\left(0\right)}+\delta
Z_{\ell}^{\left(1\right)}$ with
\begin{equation}
  \label{eq:corrected_vector_residue}
  Z_{\ell}^{\left(0\right)}=-2\bigl(\sig_{\ell,1}^{\left(0\right)}\bigr)^{2},
  \qquad
  Z_{\ell}^{\left(1\right)}=-4\sig_{\ell,1}^{\left(0\right)}
  \sig_{\ell,1}^{\left(1\right)}.
\end{equation}
The residues are also presented in table~\ref{table:spectrum}.  Since
$c_{\ell,\ell'}$ vanishes if $\left\lvert\ell-\ell'\right\rvert\le
4$, the correction to the residue can be calculated explicitly.

For the vectors, at large $p$,
\begin{equation}
  \cA=a_{1}p^{-2}+a_{2}\bigl(1+ a\bigl(k\bigr)p^{-2}\log p\bigr)+\cdots.
\end{equation}
For the mass eigenstates, $a_{2}=0$ and
\begin{equation}
  Z_{\ell}=\frac{8}{\nu^{2}}a_{\ell,1}^{2}.
\end{equation}
Finally for the spinors, at large $p$
\begin{equation}
  \psi=s_{1}p^{-2}+s_{2}\bigl(1+ a\bigl(k\bigr)p^{-2}\log p\bigr)+\cdots,
\end{equation}
and for mass eigenstates
\begin{equation}
  Z_{\ell}=\frac{8}{\nu^{2}}s_{\ell,1}^{2}.
\end{equation}
We then find an expression for the perturbed residue that is similar
to~\eqref{eq:corrected_vector_residue}.

\begin{table}
  \begin{center}
    \begin{tabular}{|c c c|c c|c c|c c|}
      \hline
      & & &
      \multicolumn{2}{c|}{Scalar} &
      \multicolumn{2}{c|}{Vector} &
      \multicolumn{2}{c|}{Spinor} \\
      $\ell$ &
      $-\lam_{\ell}^{\left(0\right)}$ & 
      $Z^{\left(0\right)}_{\ell}$ &
      $-\lam_{\ell}^{\left(1\right)}/\lam_{\ell}^{\left(0\right)}$ &
      $Z_{\ell}$ &
      $-\lam_{\ell}^{\left(1\right)}/\lam_{\ell}^{\left(0\right)}$ &
      $Z_{\ell}$ &
      $-\lam_{\ell}^{\left(1\right)}/\lam_{\ell}^{\left(0\right)}$ &
      $Z_{\ell}$  \\
      \hline
      $0$ & $8$ & $-12$ &
      $0.0143$ & $0.474$ &
      $0.104$ & $2.33$ &
      $-0.0780$ & $0.538$ \\
      $1$ & $24$ & $-20$ &
      $0.0767$ & $3.36$ & 
      $0.201$ & $5.37$ & 
      $0.0941$ & $0.491$ \\
      $2$ & $48$ & $-28$ &
      $0.145$ & $6.72$ & 
      $0.236$ & $7.62$ &
      $0.178$ & $7.50$ \\
      $3$ & $80$ & $-36$ &
      $0.188$ & $9.34$ & 
      $0.251$ & $9.83$ &
      $0.215$ & $9.78$ \\
      $4$ & $120$ & $-44$ &
      $0.214$ & $11.8$ & 
      $0.258$ & $12.0$  & 
      $0.234$ & $12.0$ \\
      $5$ & $168$ & $-52$ &
      $0.223$ & $14.1$ & 
      $0.263$ & $14.2$ &
      $0.245$ & $14.2$ \\
      $6$ & $224$ & $-60$ &
      $0.234$ & $16.3$ & 
      $0.265$ & $16.4$ &
      $0.252$ & $16.4$ \\
      $7$ & $288$ & $-68$ &
      $0.247$ & $18.5$ &
      $0.267$ & $18.6$ &
      $0.257$ & $18.6$ \\
      $8$ & $360$ & $-76$ &
      $0.252$ & $20.7$ & 
      $0.268$ & $20.8$ & 
      $0.260$ & $20.8$ \\
      $9$ & $440$ & $-84$ &
      $0.256$ & $22.9$ & 
      $0.269$ & $23.0$ &
      $0.262$ & $23.0$ \\
      $10$ & $528$ & $-92$ &
      $0.259$ & $25.1$ & 
      $0.270$ & $25.2$ &
      $0.264$ & $25.2$ \\      
      $20$ & $1848$ & $-172$ &
      $0.269$ & $47.0$ & 
      $0.272$ & $47.0$ &
      $0.271$ & $47.0$ \\
      $50$ & $10608$ & $-412$ &
      $0.273$ & $113.$ & 
      $0.273$ & $113.$ &
      $0.273$ & $113.$ \\
      $100$ & $41208$ & $-812$ &
      $0.273$ & $222.$ & 
      $0.273$ & $222.$ &
      $0.273$ & $222.$ \\
      \hline
    \end{tabular}
    \caption{\label{table:spectrum}Perturbed spectrum of mesonic
      messengers in the theory and state dual to the
      geometry~\eqref{eq:DKM}.  Note that although higher modes
      contribute increasingly large amounts to the correlators, the
      spectra also become degenerate as $\ell$ increases and so cancel
      out of~\eqref{eq:sfermion_mass}.  All of the entries in this
      table are approximations to rational numbers that can be
      determined for any $\ell$, though the general expression is not
      simple.}
  \end{center}
\end{table}

The analytic terms in the correlators $C_{a}$ correspond to contact
terms and must cancel in~\eqref{eq:sfermion_mass}.  The non-analytic
parts can be easily integrated.  We have,
\begin{equation}
  \int_{0}^{\Lam^{2}} \ud \nu^{2}\sum_{\ell=0}\frac{Z_{\ell}}{\nu^{2}-\lam_{\ell}}
  \approx \sum_{\ell}^{\ell_{\um}}Z_{\ell}\bigl\{\log\bigl(-\lam_{\ell}\bigr)
  -\log\bigl(-\lam_{\ell}-\lam_{\ell_{\mathrm{max}}}\bigr)
  \bigr\},
\end{equation}
where $\ell_{\um}$ is the largest $\ell$ satisfying
$\lam_{\ell}<\Lam^{2}$.  \eqref{eq:sfermion_mass} then gives
(suppressing the group index)
\begin{equation}
  \Ga=\frac{-\delta\mu^{2}}{16\pi^{2}L^{4}}
  \biggl(3\Ga^{\left(1\right)}_{0}-4\Ga^{\left(1\right)}_{1/2}+
  \Ga^{\left(1\right)}_{1}\biggr),
\end{equation}
in which for any particular spin,
\begin{equation}
  \Ga_{a}^{\left(0\right)}=\sum_{\ell}^{\ell_{\um}}
  \biggl\{Z_{\ell}^{\left(1\right)}
  \log\biggl[\frac{\lam_{\ell}^{\left(0\right)}-\lam^{\left(0\right)}_{\ell_{\um}}}
  {\lam_{\ell}^{\left(0\right)}}\biggr]
  -Z_{\ell}^{\left(1\right)}\lam_{\ell}^{\left(1\right)}
  \biggl(\frac{1}
  {\lam_{\ell}^{\left(0\right)}+\lam_{\ell_{\um}}^{\left(0\right)}}
  -\frac{1}{\lam_{\ell}^{\left(0\right)}}\biggr)\biggr\}.   
\end{equation}
As we observe from table~\ref{table:spectrum} and as expected from
supersymmetry, the residues and masses become degenerate as $\ell$.
The result is
\begin{equation}
  \Ga=\frac{3 L^{4}\cS}{80 \pi^{2}\mu^{6}} c,
\end{equation}
in which $c\approx 90$.  We would like to be able to express this in
terms of quantities on the gauge theory side, for example the
messenger mass and the scale of supersymmetry breaking.  The messenger
mass is related to the $\uD 7$-position by
$m_{\mu}=\mu\ell_{\us}^{-2}$.  However, the dual scale of
supersymmetry breaking is not immediately obvious in this set up.  The
analysis of~\cite{DeWolfe:2008zy} indicates that (at least without the
$\uD 7$-brane) the dual state has no vacuum energy and supersymmetry
is not broken.  However, since the current-current
correlators for the component fields of $\cJ$ do not satisfy
supersymmetric relations, supersymmetry must be broken at least after
the introduction of the flavor branes\footnote{This indicates that the
  mediation of supersymmetry breaking is no longer precisely
  semi-direct.}.  Relating the parameters of the gravitational theory
to the gauge theory may then require calculation of the backreaction
of the $\uD 7$-brane, an analysis that we leave to future work.

\section{\label{sec:conclusions}Conclusions}

In this work, we considered models of supersymmetry breaking and
mediation, using the language and techniques of the gauge/gravity
correspondence.  In this sense, this work is very much along the lines
of~\cite{Benini:2009ff, McGuirk:2009am}.  However, the approach used
here differed from those works in that in the latter soft terms (in
particular, gaugino masses) were inferred directly from dimensional
reduction.  In contrast, here we used the correspondence to calculate
correlation functions in the gauge theory, much in the spirit of the
some of the foundational works on AdS/CFT.  Such correlators can be
related to visible-sector soft terms via the formalism of general
gauge mediation.  The downside to this technique however is that it
requires more explicit knowledge of solutions of the classical
solutions of the equations of motion.  In particular, the fact that
some of the fields of interest had non-trivial angular dependence
required us to consider particularly simple geometries.

One surprising result of the analysis performed concerns the geometry
resulting from adding a small number of $\uD 3$-$\overline{\uD
  3}$-branes to $AdS^{5}\times S^{5}$.  Although naively such a
construction is not supersymmetric, it was argued in~\cite{DeWolfe:2008zy}
that supersymmetry was preserved in the state realized by the dual
theory.  However, we found that when a $\U{1}$ flavor group and a
massive quark is added to the theory, the resulting spectrum of mesons
is not supersymmetric.  A possibility is that supersymmetry is broken
only after the addition of the quarks.  This could be confirmed by
calculating the backreaction of the $\uD 7$ on the geometry.

The advantage of the technique used here is that soft terms for chiral
matter fields, which were not directly calculable from holography in
the setups of~\cite{Benini:2009ff,McGuirk:2009am}, are readily
obtainable here.  However, because of the large amount of symmetry,
the gaugino in the dual gauge theory remains massless.  That is, the
response function for the spinor component of the dual current
superfield took the form~\eqref{eq:fermionic_one_point} and so the
correlator function $B_{1/2}$ given in~\eqref{eq:correlator_functions}
vanish, leading to a vanishing $m_{1/2}$.  This is directly related to
the fact that the geometry does contain any $3$-form flux which is
necessary to give rise to gaugino masses in this class of
constructions~\cite{Lust:2004fi,*Camara:2004jj,*Lust:2008zd,
  Benini:2009ff, McGuirk:2009am}.  It would be of interest to extend
the techniques considered here to geometries that are supported by
fluxes such as~\cite{Klebanov:2000nc, Klebanov:2000hb} and their
non-supersymmetric
perturbations~\cite{DeWolfe:2008zy,McGuirk:2009xx,*Bena:2010ze,*Dymarsky:2011pm,
  Bena:2011wh}, especially since such constructions are perturbatively
stable.  However, in addition to the complications presented by an
angular space\footnote{See~\cite{Gherghetta:2006yq} for discussions on
  this point.}, such theories do not, strictly speaking, flow from a
conformal fixed point but are instead cascading theories.  Although
the methods of holographic renormalization have been discussed for
such theories~\cite{Aharony:2005zr}, the geometry is such that the
correlators cannot be explicitly calculated.  Although this was also
the case for the non-supersymmetric cases considered above, for
asymptotically AdS spaces it is known how to infer the two-point
functions from the $1$-point functions and the spectrum.  For
cascading theories, this is less clear~\cite{Berg:2006xy}.

\section*{Acknowledgments}

It is a pleasure to thank G. Shiu, Y. Sumitomo, and P. Ouyang for
useful discussions and suggestions.  I also thank the Hong Kong
Institute for Advanced Study at the Hong Kong University of Science
and Technology for hospitality and support.  This work was supported
by the US National Science Foundation under grant PHY-0757868, the US
Department of Energy under contract DE-FG-02-95ER40896, a String
Vacuum Project Graduate Fellowship funded through the US National
Science Foundation grant PHY-0918807, and a Cottrell Scholar Award
from Research Corporation.

\appendix

\section{\label{app:conv}Conventions}

The index $\mu$ indicates an $R^{3,1}$ coordinate $x^{0,1,2,3}$ while
$m$ runs over all five non-compact coordinates (i.e. $R^{3,1}$ and the
holographic direction).  $\alpha$ runs over the coordinates of the
probe $\uD 7$-brane discussed in the text, with $a,b,c,d$ running over
coordinates transverse to $R^{3,1}$ and $i,j$ running transverse to
the brane. $\phi, \psi$ denote angular directions, either those in the
full 10d space or along the worldvolume. $M, N$ run over all 10
directions.  Underlined indices are used to denote locally orthogonal
non-coordinate bases for the (co-)vector spaces.  The index $\alpha$
is also used to denote spinor indices, but context should allow these
to be distinguished from worldvolume indices.

\subsection{\label{subsec:fermions}Fermion conventions}

\subsubsection{SO(3,1) spinors}

We make use of the dotted and undotted notation of~\cite{Wess:1992cp}.
Such indices are raised and lowered with $\ep^{12}=\ep_{21}=1$ and we
define $\theta^{2}=\theta\theta$.  An $\SO{3,1}$ Dirac spinor takes
the form
\begin{equation}
  \psi=\begin{pmatrix}
    \psi_{\uL\alpha} \\ \ui\bar{\psi}^{\dot{\alpha}}_{\uR}
  \end{pmatrix},
\end{equation}
where we have chosen a Weyl basis for the $\SO{3,1}$ $\ga$-matrices
\begin{equation}
  \ga^{\ul{\mu}}=\begin{pmatrix}
    0 & \sig^{\ul{\mu}} \\ -\bar{\sig}^{\ul{\mu}} & 0
  \end{pmatrix},
\end{equation}
in which
\begin{equation}
  \sig^{\ul{\mu}}=\bigl(-\II_{2},\bs{\sig}\bigr),\quad
  \bar{\sig}^{\ul{\mu}}=\bigl(-\II_{2},-\bs{\sig}\bigr),
\end{equation}
where $\bar{\sig}$ are the usual Pauli matrices
\begin{equation}
  \sig^{1}=\begin{pmatrix} 0 & 1 \\ 1 & 0\end{pmatrix},\quad
  \sig^{2}=\begin{pmatrix} 0 & -\ui \\ \ui & 0\end{pmatrix},\quad
  \sig^{3}=\begin{pmatrix} 1 & 0 \\ 0 & -1\end{pmatrix},
\end{equation}
and so the $\ga$-matrices satisfy the Clifford algebra
\begin{equation}
  \bigl\{\ga^{\ul{\mu}},\ga^{\ul{\nu}}\bigr\}=2\eta^{\ul{\mu}\ul{\nu}}.
\end{equation}
The 4d chirality operator is
\begin{equation}
  \ga_{\left(4\right)}=-\ui \ga^{\ul{0}}\ga^{\ul{1}}\ga^{\ul{2}}\ga^{\ul{3}}
  =\begin{pmatrix} -1 & 0 \\ 0 & 1\end{pmatrix}.
\end{equation}

\subsubsection{SO(4,1) spinors} For $\SO{4,1}$, we take
\begin{equation}
  \tilde{\ga}^{\ul{\mu}}={\ga}^{\ul{\mu}},\quad
  \tilde{\ga}^{\ul{4}}=\ga_{\left(4\right)}.
\end{equation}
These then satisfy
$\left\{\tilde{\ga}^{\ul{m}},\tilde{\ga}^{\ul{n}}\right\}=2\tilde{\eta}^{\ul{m}\ul{n}}$.  The Majorana matrix is
\begin{equation}
  \tilde{B}_{5}=\tilde{\ga}^{\ul{2}},
\end{equation}
which is imaginary, satisfies $\tilde{B}_{5}\tilde{B}_{5}^{\ast}=-1$ and
$\tilde{B}_{5}^{-1}\tilde{\ga}^{\ul{m}}\tilde{B}_{5}=-\tilde{\ga}^{\ul{m}\ast}$.
The charge-conjugation operator is
\begin{equation}
  \tilde{C}_{5}=\tilde{B}_{5}\tilde{\ga}^{\ul{0}}
  =\begin{pmatrix} \sig^{\ul{2}} & 0 \\ 0 & -\sig^{\ul{2}}\end{pmatrix},
\end{equation}
which satisfies
$\tilde{C}_{5}\tilde{\ga}^{\ul{m}}\tilde{C}_{5}^{-1}=\tilde{\ga}^{\ul{m}\uT}$.
$\cN_{5}=1$ is usefully parametrized in terms of symplectic-Majorana
spinors.  A pair of symplectic-Majorana spinors are Dirac spinors
$\psi^{i=1,2}$ satisfying the property
\begin{equation}
  \psi^{i}=\ep^{ij}\tilde{C}_{5}\bar{\psi}_{j}^{\uT},
\end{equation}
where $\psi_{j}^{\ast}:=\bigl(\psi^{j}\bigr)^{\ast}$ and again
$\ep^{12}=\ep_{21}=1$.  Such spinors satisfy the identities
\begin{subequations}
\begin{align}
  \bar{\chi}_{i}\tilde{\ga}^{\ul{m}_{1}}\cdots
  \tilde{\ga}^{\ul{m}_{n}}\psi^{j}=&
  \ep_{il}\ep^{jk}
  \bar{\psi}_{k}
  \tilde{\ga}^{\ul{m_{n}}}\cdots\tilde{\ga}^{\ul{m_{1}}}
  \chi^{l},\\
  \bigl(\bar{\chi}_{i}\tilde{\ga}^{\ul{m}_{1}}\cdots
  \tilde{\ga}^{\ul{m}_{n}}\psi^{j}\bigr)^{\ast}=&
  \bigl(-1\bigr)^{n+1}
  \ep^{ik}\ep_{jl}
  \bar{\chi}_{k}\tilde{\ga}^{\ul{m}_{1}}\cdots
  \tilde{\ga}^{\ul{m}_{n}}
  \psi^{l}.
\end{align}
\end{subequations}
For a pair of symplectic-Majorana spinors, we may write
\begin{equation}
  \label{eq:decompose_sm_spinors}
  \psi^{1}=\begin{pmatrix}
    \psi_{\uL\alpha} \\
    \ui\bar{\psi}_{\uR}^{\dot{\alpha}}
  \end{pmatrix},\qquad
  \psi^{2}=\begin{pmatrix}
    -\psi_{\uR\alpha} \\ \ui\bar{\psi}^{\dot{\alpha}}_{\uL}
  \end{pmatrix},\qquad
  \bar{\psi}_{1}=\begin{pmatrix}
    -\ui\psi_{\uR}^{\alpha} \\
    -\bar{\psi}_{\uL\dot{\alpha}}\end{pmatrix}^{\uT},\qquad
  \bar{\psi}_{2}=\begin{pmatrix}
    -\ui\psi_{\uL}^{\alpha} \\
    \bar{\psi}_{\uR\dot{\alpha}}\end{pmatrix}^{\uT}.
\end{equation}

\subsubsection{SO(5) spinors}

For $\SO{5}$ we choose a basis that is useful for the decomposition
$\SO{5}\to\SO{2}\times\SO{3}$,
\begin{equation}
  \hat{\ga}^{\ul{1}}=\sig^{\ul{1}}\otimes\II_{2},\quad
  \hat{\ga}^{\ul{2}}=\sig^{\ul{2}}\otimes\II_{2},\quad
  \hat{\ga}^{\ul{3}}=\sig^{\ul{3}}\otimes\sig^{\ul{1}},\quad
  \hat{\ga}^{\ul{4}}=\sig^{\ul{3}}\otimes\sig^{\ul{2}},\quad
  \hat{\ga}^{\ul{5}}=\sig^{\ul{3}}\otimes\sig^{\ul{3}}.
\end{equation}
These satisfy
\begin{equation}
  \bigl\{\hat{\ga}^{\ul{\phi}},\hat{\ga}^{\ul{\psi}}\bigr\}=
  2\delta^{\ul{\phi}\ul{\psi}}.
\end{equation}
The Majorana matrix is
\begin{equation}
  \hat{B}_{5}=\hat{\ga}^{\ul{2}}\hat{\ga}^{\ul{4}}
  =\sig^{\ul{1}}\otimes\ui \sig^{\ul{2}}.
\end{equation}
It satisfies $\hat{B}_{5}=\hat{B}_{5}^{\ast}$, $\hat{B}_{5}^{2}=-1$, and
\begin{equation}
  \hat{B}_{5}^{-1}\hat{\ga}^{\ul{\phi}}\hat{B}_{5}=
  \hat{\ga}^{\ul{\phi}\ast}.
\end{equation}

\subsubsection{SO(9,1) spinors} On $R^{9,1}$ we take
\begin{equation}
  \Ga^{\ul{m}}=\sig^{\ul{1}}\otimes\tilde{\ga}^{\ul{m}}\otimes\II_{4},\quad
  \Ga^{\ul{\phi}}=\sig^{\ul{2}}\otimes\II_{4}\otimes
  \hat{\ga}^{\ul{\phi}},
\end{equation}
where the second equality should be understood to mean
$\Ga^{\ul{5}}=\sig^{\ul{2}}\otimes\II_{4}\otimes\hat{\ga}^{\ul{1}}$,
etc.  The $\Ga$-matrices satisfy
$\left\{\Ga^{\ul{M}},\Ga^{\ul{N}}\right\}=2\eta^{\ul{M}\ul{N}}$.  The
10d chirality operator is
\begin{equation}
  \Ga_{\left(10\right)}=\ui\Ga^{\ul{0}}\Ga^{\ul{1}}\cdots\Ga^{\ul{9}}
  =\sig^{\ul{3}}\otimes\II_{4}\otimes\II_{4},
\end{equation}
which satisfies $\Ga_{\left(10\right)}^{2}=1$ and anti-commutes with
all of the $\Ga^{\ul{M}}$.  The 10d Majorana matrix is
\begin{equation}
  B_{10}=\Ga^{\ul{2}}\Ga^{\ul{5}}\Ga^{\ul{7}}\Ga^{\ul{9}}
  =-\ui\sig^{3}\otimes\tilde{B}_{5}\otimes\hat{B}_{5},
\end{equation}
which satisfies $B_{10}=B_{10}^{\ast}$, $B_{10}B_{10}^{\ast}=1$ and
\begin{equation}
  B_{10}^{-1}\Ga^{\ul{M}}B_{10}=\Ga^{\ul{M}\ast}.
\end{equation}
In 10d, we generally work with Majorana-Weyl spinors satisfying
$\Ga_{\left(10\right)}\Psi=\Psi$, and $B_{10}\Psi=\Psi^{\ast}$.
Additionally, we work with bispinors which are combination of two 10d
spinors
\begin{equation}
  \Psi=\begin{pmatrix}\Psi_{1} \\ \Psi_{2}\end{pmatrix}.
\end{equation}
$\Gamma$-matrices act on bispinors as
\begin{equation}
  \Ga^{\ul{M}}\Psi=\begin{pmatrix}
    \Ga^{\ul{M}}\Psi_{1} \\ \Ga^{\ul{M}}\Psi_{2}
  \end{pmatrix}.
\end{equation}
    
\subsection{\label{app:typeIIB}Type-IIB supergravity}

Our conventions for the bosonic modes of type-IIB supergravity are
summarized by the pseudo-action which we write as
\begin{equation}
  S_{\mathrm{IIB}}=S_{\mathrm{IIB}}^{\mathrm{NS}}+S_{\mathrm{IIB}}^{\mathrm{R}}+
  S_{\mathrm{IIB}}^{\mathrm{CS}},
\end{equation}
in which
\begin{align}
  S_{\mathrm{IIB}}^{\mathrm{NS}}=&
  \frac{1}{2\ka_{10}^{2}}\int\ud^{10}x\,\sqrt{-\det\left(g\right)}
  \biggl\{R-\frac{1}{2}\partial_{M}\Phi\partial^{M}\Phi
  -\frac{g_{\us}}{2}\ue^{-\Phi}\bigl(H^{\left(3\right)}\bigr)^{2}\biggr\},\notag\\
  S_{\mathrm{IIB}}^{\mathrm{R}}=&
  -\frac{1}{4\ka_{10}^{2}}\int\ud^{10}x\,
  \sqrt{-\det\left(g\right)}
  \biggl\{\ue^{2\Phi}\bigl(F^{\left(1\right)}\bigr)^{2}
  +g_{\us}\ue^{\Phi}\bigl(F^{\left(3\right)}\bigr)^{2}+
  \frac{1}{2}g_{\us}^{2}\bigl(F^{\left(5\right)}\bigr)^{2}\biggr\},\\
  S_{\mathrm{IIB}}^{\mathrm{CS}}=&
  \frac{g_{\us}^{2}}{4\ka_{10}^{2}}
  \int C^{\left(4\right)}\wedge H^{\left(3\right)}
  \wedge \ud C^{\left(2\right)},
\end{align}
in which $g_{MN}$ is the 10d Einstein-frame metric, $\Phi$ is the
dilaton and the gauge-invariant field strengths of the NS-NS $2$-form
potential $B^{\left(2\right)}$ and R-R potentials
$C^{\left(0\right)}$, $C^{\left(2\right)}$, and $C^{\left(4\right)}$ are
\begin{align}
  H^{\left(3\right)}=&\ud B^{\left(2\right)},
  &F^{\left(1\right)}=&\ud C^{\left(0\right)},\notag\\
  F^{\left(3\right)}=&\ud C^{\left(2\right)}
  -C^{\left(0\right)}H^{\left(3\right)},
  &F^{\left(5\right)}=&\ud C^{\left(4\right)}
  +B^{\left(2\right)}\wedge\ud C^{\left(2\right)}.
\end{align}
The dilaton is normalized such that $\bigl<\Phi\bigr>=\log g_{\us}$
and for a $p$-form
\begin{equation}
  \bigl(\Omega^{\left(p\right)}\bigr)^{2}=\frac{1}{p!}
  \Omega_{M_{1}\cdots M_{p}}\Omega^{M_{1}\cdots M_{p}}.
\end{equation}

Type-IIB supergravity exhibits $\cN_{10}=2$ and so there are two
Majorana-Weyl gravitini and two Majorana-Weyl dilatini which can be
organized into bispinors
\begin{equation}
  \Psi_{M}=\begin{pmatrix}
    \Psi_{M}^{1} \\ \Psi_{M}^{2}
  \end{pmatrix},\quad
  \Lambda=\begin{pmatrix}
    \Lambda^{1}\\ \Lambda^{2}
  \end{pmatrix}.
\end{equation}
SUSY transformations are parametrized by a Majorana-Weyl bispinor
$\ep$.  For the fermionic fields
\begin{equation}
  \delta_{\ep}\Psi_{M}=\cD_{M}\ep,\quad
  \delta_{\ep}\Lambda=\Delta\ep,
\end{equation}
where, in the 10d Einstein frame,
\begin{subequations}
\label{eq:SUSY_vars}
\begin{align}
  \Delta=&\frac{1}{2}\slashed{\partial}\Phi
  -\frac{1}{2}\ue^{\Phi}\slashed{F}^{\left(1\right)}\bigl(\ui\sig^{2}\bigr)
  -\frac{1}{4}\bigl(g_{\us}\ue^{\Phi}\bigr)^{1/2}\cG_{3}^{+},\\
  \cD_{M}=&\nabla_{M}+\frac{1}{2}\ue^{\Phi}F_{M}\bigl(\ui\sig^{2}\bigr)
  +\frac{1}{16}g_{\us}\slashed{F}^{\left(5\right)}
  \Ga_{M}\bigl(\ui\sig^{2}\bigr)
  +\frac{1}{8}\bigl(g_{\us}\ue^{\Phi}\bigr)^{1/2}
  \bigl(\cG_{3}^{-}\Ga_{M}+\frac{1}{2}\Ga_{M}\cG_{3}^{-}\bigr),
\end{align}
\end{subequations}
where $\left\langle\Phi\right\rangle=\log g_{\us}$, $\nabla_{M}$ is
the covariant derivative built from the 10d metric, the Pauli matrices
rotate that spinors constituting the bispinor into each other, and
\begin{equation}
  \cG_{3}^{\pm}=\slashed{F}^{\left(3\right)}\sig^{1}
  \pm\ue^{-\Phi}\slashed{H}^{\left(3\right)}\sig^{3}.
\end{equation}
For a $p$-form $\Omega^{\left(p\right)}$, we have defined
\begin{equation}
  \slashed{\Omega}^{\left(p\right)}=\frac{1}{p!}\Omega_{M_{1}\cdots M_{p}}
  \Ga^{M_{1}\cdots M_{p}}.
\end{equation}


\bibliography{ghgm}

\end{document}